\documentclass[aps,prx,amsmath,amssymb,superscriptaddress,notitlepage,twocolumn,longbibliography,nofootinbib]{revtex4-1}

\usepackage{capt-of}%

\usepackage{amsmath}
\usepackage{amssymb}
\usepackage{latexsym}
\usepackage{graphicx}
\usepackage{times,psfrag}
\usepackage{enumerate}
\usepackage{amsmath}
\usepackage{dsfont}
\usepackage{dcolumn}
\usepackage{bm,bbm}
\usepackage{hyperref}
\usepackage{color}
\usepackage{latexsym,amsmath,amssymb,bm,euscript}
\usepackage{dsfont}
\usepackage{textcomp}
\usepackage{tabularx}
\usepackage{setspace}
\usepackage{sidecap}
\usepackage{placeins}
\usepackage{threeparttable}
\usepackage{multirow}
\usepackage{physics}
\usepackage{tikz-feynman}
\usepackage{pifont}
\usepackage{subcaption}
\usepackage{caption}

\let \oldbm \bm
\renewcommand{\vec}[1]{\oldbm{#1}}
\newcommand{\vechat}[1]{\hat{\oldbm{#1}}}

\def\bk{{\vec k}}

\def\bA{{\vec A}}

\def\bq{{\vec q}}

\def\bR{{\vec R}}
\def\bG{{\vec G}}

\def\bP{{\bf P}}

\def\bn{{\vec n}}
\def\bm{{\vec m}}

\def\br{{\vec r}}
\def\ad{\mathrm{ad}}
\def\Sp{\rm{Sp}}

\def\tr{\mathop{\mathrm{tr}}}
\def\Z{\mathds{Z}}

\def\T{\mathcal{T}}
\def\L{\mathcal{L}}
\def\D{\mathcal{D}}

\def\P{\mathcal{P}}
\def\S{\mathcal{S}}

\def\G{\mathcal{G}}

\def\H{\mathcal{H}}
\def\K{\mathcal{K}}

\def\diag{{\rm diag}}

\def\U{{\rm U}}

\def\SU{{\rm SU}}
\def\SO{{\rm SO}}
\def \tp{\text{pair}}
\def\R{\rho_{\rm ps}}

\def\ta{\tilde \alpha}
\def\tA{\tilde A}

\newcommand{\beq}{\begin{equation}}
\newcommand{\eeq}{\end{equation}}
\newcommand{\beqarray}{\begin{eqnarray}}
\newcommand{\eeqarray}{\end{eqnarray}}
\newcommand{\nn}{\nonumber \\}

\graphicspath{{Science_submit/}}

\date{\today}

\begin{document}

\title{Charged Skyrmions and Topological Origin of Superconductivity in Magic Angle Graphene}

\author{Eslam Khalaf}
\affiliation{Department of Physics, Harvard University, Cambridge, Massachusetts 02138, USA}

\author{Shubhayu Chatterjee}
\affiliation{Department of Physics, University of California, Berkeley, CA 94720, USA}

\author{Nick Bultinck}
\affiliation{Department of Physics, University of California, Berkeley, CA 94720, USA}
\affiliation{Department of Physics, Ghent university, 9000 Gent, Belgium}

\author{Michael P. Zaletel}
\affiliation{Department of Physics, University of California, Berkeley, CA 94720, USA}

\author{Ashvin Vishwanath}
\email[]{avishwanath@g.harvard.edu}
\affiliation{Department of Physics, Harvard University, Cambridge, Massachusetts 02138, USA}

\maketitle

\textbf{
Spontaneous symmetry breaking plays a pivotal role in many areas of physics, engendering a variety of excitations from sound modes in solids to  pions in nuclear physics. While these represent the  linear response of the symmetry breaking condensate,  equally important are  nonlinear configurations, solitons, which  enjoy exceptional stability when associated with a topological charge . A notable example is in the Skyrme model of nuclear forces, where the neutron and proton are interpreted as topological solitons of the pion condensate. We argue that  similar models may describe a remarkable new material - magic angle graphene - obtained on twisting and stacking two sheets of graphene. When the relative angle between the sheets is near the magic angle of $\sim 1^\circ$ 
degree - insulating  behavior resulting from ordering of electrons is observed, which gives way to superconductivity on changing the  electron  density.  Here we propose a unifying description of both the order underlying the insulating state as well as  the superconductivity. In our theory, while the symmetry breaking condensate leads to the ordered phase,  topological solitons in the condensate - skyrmions - are shown to be bosons that carry an electric charge $2e$. Condensation of skyrmions  leads to a  superconductor whose pairing strength, symmetry and other properties are inferred. More generally, we show how topological textures can mitigate  Coulomb repulsion to pair electrons and  provide a new route to superconductivity. Our mechanism potentially applies to much wider class of systems but crucially invokes certain key ingredient such as inversion symmetry present in magic angle graphene. We discuss how these  insights not only clarify why certain correlated moire materials do not superconduct, they also point to promising new platforms where robust superconductivity is anticipated.}

\section{Introduction}

Typically when charge is added to an insulator, electron or hole excitations are produced leading to charge conduction.  Can solitons play the role of charge carriers? This unusual scenario 
can be realized through electrically charged topological textures  which have been theoretically proposed in various contexts \cite{Frohlich,polyacetylene, AbanovWiegmann, GroverSenthil,KentaroRyuLee,Chamon2007,LeeSachdev}, although physical realizations are rare. In fact, the only experimentally established instance takes place in quantum Hall ferromagnets, where topological  textures of spin in the form of skyrmions acquire a charge due to the Landau level topology and are found to be the lowest energy charge excitations \cite{Sondhi,KaneLee,MoonMori,manfra1997skyrmions,Goerbig2011,ZhangSenthil,Chatterjee19}. On the other hand, finding situations where \textit{Cooper pairing} occurs  between charged topological textures, rather than between electrons, is even harder to come by. For example, in the aforementioned quantum Hall ferromagnets where charged  topological textures have been experimentally established, strong breaking of time-reversal symmetry makes superconductivity highly unlikely. In fact, obtaining robust superconductivity from topological textures typically requires simultaneously satisfying two conditions: (a) an unbroken time-reversal symmetry and (b) the existence of stable low energy charge $2e$ topological textures. The latter can be achieved if the fundamental defects have charge $2e$ or by pairing charge $e$ defects of the same electric charge. Here, we show that all these critera are satisfied in a simple model consisting of two time-reversal-related quantum Hall (or flat Chern band) ferromagnets with purely repulsive interactions, coupled via tunneling. Such a model, we point out, captures the essential physics of magic angle graphene making it a promising candidate for superconductivity arising from the pairing of topological textures, a mechanism that is fundamentally different from the conventional electron-phonon mechanism.

Let us start with a brief review of magic angle twisted bilayer graphene (MATBG). Two sheets of graphene twisted relative to one another generate a moir\'e pattern and correspondingly a reduced Brillouin zone in reciprocal space. Previous work demonstrated that the  reconstruction of the electronic bands by the moir\'e lattice leads to minibands with extremely narrow bandwidth near charge neutrality at a magic angle $\sim 1^o$ \cite{ MacDonald2011, Morell2010, Santos, EvaAndrei2010}, separated by a band gap from other bands. Recent  experiments \cite{PabloMott,PabloSC,Dean-Young,efetov} revealed dramatic new physics near the same magic angle. Band gaps arising from the moir\'e potential are  expected and observed at electron filling $\nu_T=\pm 4$ per moir\'e unit cell.  In addition, insulators at other integer filling  including $\nu_T=\pm 2$ \cite{PabloMott, Dean-Young, efetov} and in some experiments also at $\nu_T=0$ \cite{efetov, EfetovScreening} and at certain odd integer fillings have also been observed and are attributed to the effects of electron-electron interaction. Furthermore, superconductivity has been repeatedly observed in twisted bilayer graphene, though its precise relation to the correlated insulating phase remains to be determined. While early experiments observed the superconductivity in the vicinity of the $\nu_T=- 2$ correlated insulator, subsequently a wider extent of superconductivity has been observed \cite{EfetovScreening,YoungScreening}.

\section{Quantum Hall ferromagnetism model of magic angle graphene}
\begin{figure*}
    \centering
    \includegraphics[width=0.85\textwidth]{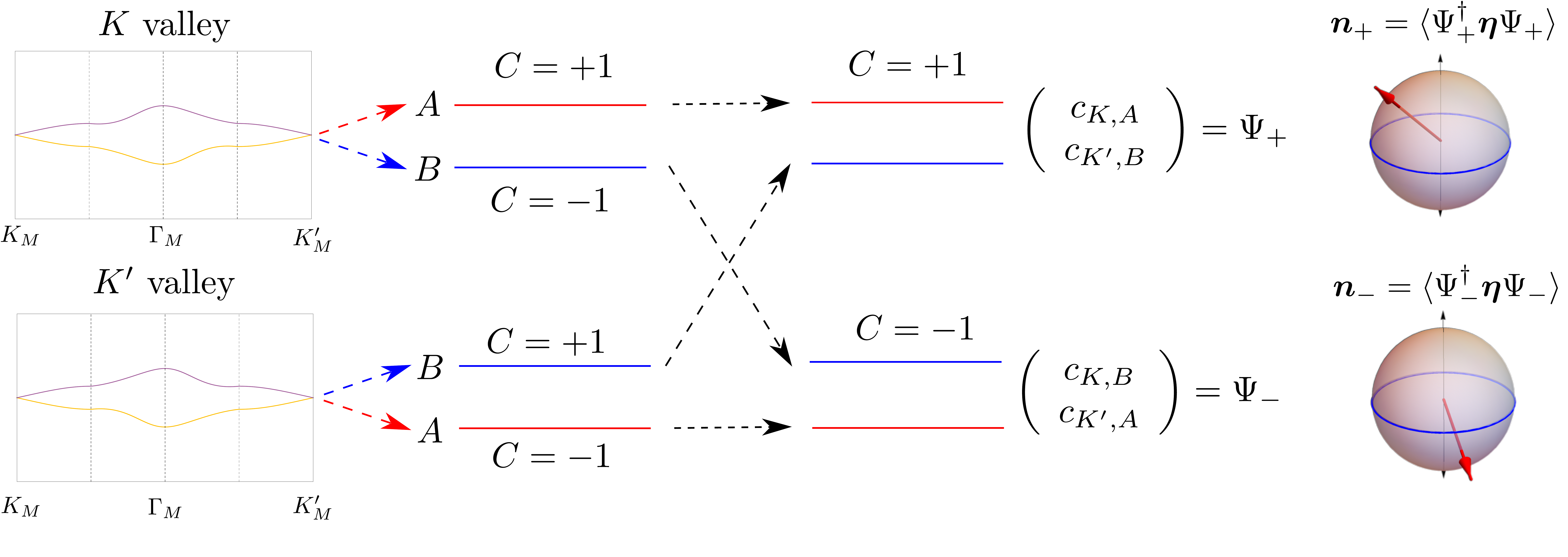}
    \caption{{\bf Defining Pseudospin}: Linear combinations of the nearly flat bands of twisted bilayer graphene (left) give rise to  pseudospin bands with opposite  Chern numbers. Two states with the same pseudospin that lie in opposite Chern sectors have identical valley wavefunctions but are distinguished by A-B sublattice polarization (red and blue, third column from left). In this figure we have omitted spin. Half filling of this spinless model with net Chern number $C=0$, at energies below the interaction scale,  correspond to quantum Hall pseudospin ferromagnets in each Chern band, described by unit vectors $\bn_\pm$.}
    \label{fig:QHF}
\end{figure*}

We begin by noting an important feature of the nearly flat bands near charge neutrality. Electrons in graphene carry both a  spin  ($s=\uparrow,\,\downarrow$) and a twofold valley degeneracy ($\tau=K,\,K'$) which would suggest that filling each moir\'e lattice site would  require four electrons. In reality, it takes twice as many electrons to  fill the nearly flat bands, since they consist of two connected bands as shown in Figure \ref{fig:QHF}. Conventionally, one distinguishes the two bands by their kinetic energy but in view of the narrow bandwidth, other linear combinations may be preferable. A different choice  is the sublattice basis - which separates bands on the basis of their weight on the two sites $\sigma=A,\,B$ of the honeycomb lattice of monolayer graphene. This basis arises naturally in the ideal chiral limit of Ref.~\cite{Tarnopolsky} where the sublattice polarization is complete, but can also be defined at   physical parameter values  as explained in Ref.~\cite{KIVCpaper}. Even for realistic parameters, the sublattice polarization is substantial, so a reasonable approximation is to ignore sublattice off diagonal terms in the density operator \cite{KIVCpaper}. This sublattice polarized (SP) approximation will be assumed throughout this paper.

A remarkable feature of the sublattice basis is that individual bands in this basis carry Chern number. This allows for a mapping of MATBG to a pair of time-reversal related copies of quantum-Hall like systems. The presence of 
flavors then relates our problem to quantum Hall {\em ferromagnetism}, as 
advocated in Ref.~\cite{KIVCpaper}.
This mapping to a generalized spin-valley ferromagnet is broadly consistent with the observation of a cascade of polarization transitions on varying electron density \cite{CascadeShahal,CascadeYazdani} and it acquires a particularly simple form in the vicinity of half-filling $\nu = \pm 2$ when the spin degree of freedom can be neglected. In this case, the low energy physics is dominated by four flat bands, each band being labelled by a valley and sublattice index. 

The valley-sublattice resolved flat bands are characterized by Chern number $C=\pm 1$. Labelling sublattice (valley) by $\sigma\left (\tau \right )$, we have $C=\sigma\tau$ which is opposite for opposite valleys due time-reversal symmetry $\T$ and also opposite for opposite sublattices within the same valley due to the combination of two-fold rotation and time-reversal $C_2 \T$. Thus, the flat bands can be conveniently studied by introducing the pseudospin spinors $\Psi_+ = ( c_{K A} , c_{K'B})^T $ and  $\Psi_- =  ( c_{K B}, c_{K'A})^T $ for the $\pm$ Chern sectors as shown in Fig.~\ref{fig:QHF}. A summary of the two different basis: valley/sublattice and Chern sector/pseudospin and the relationship between them is provided in Table \ref{BasisSymm} together with the implementation of each symmetry in both basis. 

The Hamiltonian arises from projecting the Coulomb interaction $V(\br)$ into the flat bands, 
\beq
\H = \int d^2 \bk \, \Psi^\dagger_\bk h_\bk \Psi_\bk + \frac{1}{2} \int d^2 \br \,  \delta n(\br) V(\br - \br') \delta n(\br')
\eeq
where $\delta n(\br)$ is the deviation of the density $n(\br) = \sum_{\gamma = \pm} \Psi^\dagger_\gamma(\br) \Psi_{\gamma}(\br)$ from its average value. The Fourier components of $n(\br)$ depend on the detailed structure of the Bloch wavefunctions $u_{\gamma \eta;\bk}(\br)$, which can be conveniently encapsulated in the ``form'' factor expression for the density which can be written within the SP approximation as ~\cite{KIVCpaper}:
\beq
n_\bq = \sum_{\bk,\gamma = \pm} F_\bq(\bk) e^{i \gamma \Phi_\bq(\bk) } \Psi^\dagger_{\gamma ,\bk} \Psi_{\gamma,\bk+\bq} 
\eeq
Note that the form factor is diagonal in the Chern sectors with equal amplitude $F$ and opposite phase $\Phi$ for opposite Chern sectors. 
It follows that $n_\bq$ - and hence the interaction - is symmetric under \emph{independent} pseudo-spin rotations $\U_{+}(2) \times \U_{-}(2)$ in each $C = \pm$ sector.  

Remarkably, at half-filling the ground state of the interaction: $\H_C = \frac{1}{2} \sum_\bq \delta n_\bq V_\bq \delta n_{-\bq}$  can be determined exactly for any purely repulsive interaction, $V_\bq > 0$ $\forall \bq$. In this case, $H_C$ is a sum of positive definite terms, so any state which is annihilated by each $\delta n_\bq \ket{\Omega} = 0$ is a ground state. This is satisfied by\footnote{Actually, there is another possible set of ground states with Chern number $|C|=2$ obtained by filling only one of the two Chern sectors. These states, however, do not admit non-trivial topological pseudospin textures and strongly break time-reversal symmetry making them incompatible with superconductivity and, as a result, not relevant for our discussion.} a manifold of $C = 0$ states,  $|\Omega\rangle$, obtained by choosing a direction $\bn_\pm$ in pseudo-spin space for each of the $\pm$ Chern sectors and ``filling'' the pseudo-spin bands  independent of $\bk$ yielding a pseudospin ferromagnet in each Chern sector. Thus: 
\begin{equation}
    \langle \Omega | \Psi_{\pm}^\dagger \vec{\eta} \Psi_{\pm}|\Omega\rangle = \bn_\pm
\end{equation}
\begin{table}
\begin{centering}
\parbox{0.8 \linewidth}{
\centering
 From valley/sublattice to Chern /pseudospin\\ $(\tau,\sigma) \rightarrow (\gamma,\eta)$\\
$ {\boldsymbol \gamma} = (\gamma_x,\,\gamma_y,\,\gamma_z) = (\sigma_x, \sigma_y \tau_z, \sigma_z \tau_z)$ \\ ${\boldsymbol \eta} = (\eta_x,\,\eta_y,\,\eta_z) = (\sigma_x \tau_x, \sigma_x \tau_y, \tau_z)$} \vspace{0.4cm}

\centering
Basis\\
\begin{tabular}{c|c}
\hline \hline
     Valley & $\tau_z = K/K'$\\
     \hline
     Sublattice & $\sigma_z = A/B$ \\ 
     \hline
     Chern sector & $\gamma_z = \sigma_z\tau_z=+/-$ \\
     \hline
     Pseudospin & $\eta_z = \tau_z=\uparrow_{\rm ps}/\downarrow_{\rm ps}$ \\
     \hline \hline
\end{tabular} 
 \vspace{0.4cm}

\centering
Symmetries\\
\begin{tabular}{c|c|c}
\hline \hline
Symm & $(\tau, \sigma)$ basis & $(\gamma, \eta)$ basis \\
\hline
     $\T$ & $\tau_x \K$ & $\gamma_x \eta_x \K$ \\
     \hline
     $C_2$ & $\sigma_x \tau_x$ & $\eta_x$ \\
     \hline
     $\U_V(1)$ & $e^{i \phi \tau_z}$ & $e^{i \phi \eta_z}$ \\
     \hline \hline
\end{tabular} 

\caption{Definition of new variables - Chern Sector and pseudospin, from the valley and sublattice degrees of freedom. Action of symmetries on the internal indices are shown on the right, in both set of variables. Additionally, the independent pseudospin rotations in the Chern sectors are generated by the ${\boldsymbol \eta}P_\pm $ where the projectors $P_\pm = \frac12(1\pm\gamma_z)$ single out a specific Chern sector.}
\label{BasisSymm}
\end{centering}
\end{table}

Now let us reintroduce the single particle dispersion $h$ whose form is constrained by the pseudo-spin and $C_2 \mathcal{T}$ symmetry, 
\begin{align}
h(\bk) = \gamma^x h_x(\bk) + \gamma^y h_y(\bk)
\label{Eqn:h}
\end{align}
It is important to stress here that the form of $h(\bk)$ above is the most general form of the dispersion allowed by the symmetries of the chiral model and we do \emph{not} assume the flat band limit. Deviation from the chiral limit introduces an extra term proportional to $\eta_z \gamma_0$ (other terms are prohibited either by $C_2 \T$ or particle-hole symmetry) which was shown in Ref.~\cite{KIVCpaper} to be relatively small. This means that the main effect of deviation from the chiral limit is altering the values of $h_{x,y}(\bk)$ rather than introducing new terms in (\ref{Eqn:h}). Thus, (\ref{Eqn:h}) takes realistic dispersion fully into account.

The dispersion $h(\bk)$ acts as tunneling between states with the same pseudospin and momentum in opposite Chern sectors. Its main effect in the limit of strong interaction is to introduce a 'superexchange' term $J \sim h^2/U$, where $U$ is the typical interaction scale, which antiferromagnetically couples the pseudospins in opposite Chern sectors. This reduces the $\SU(2)\times \SU(2)$ symmetry of the interaction Hamiltonian, down to a single $\SU(2)$. The generation of such superexchange term is quite generic for any pair of opposite Chern number tunneling-coupled ferromagnets with  $\SU(2)$ spin rotation symmetry, as shown in supplemental material, and it plays a crucial role in the skyrmion superconductivity discussed below. {It also plays an important role in selecting the insulating ground state among the quantum Hall ferromagnetic states, favoring those for which the pseudospins in the two Chern sectors are anti-aligned, $\bn_+ = -\bn_- = \bn$. In the language of Ref.~\cite{KIVCpaper}, the resulting state corresponds to the Kramers-inter valley coherent state (K-IVC) state if $\bn$ lies in the XY plane. In this phase the conservation of valley charge is spontaneously broken, (hence inter valley coherence) which corresponds to a tripling of the unit cell on the graphene lattice scale.  On the other hand, if $\bn$ points along the Z direction, the valley charge is conserved and this order corresponds to the valley Hall state. Anisotropies  neglected here which arise on going beyond the SP approximation, prefer $\bn$ ordering in the $n_x,\,n_y$ plane and select the K-IVC order.}

\begin{figure}
    \centering
    \includegraphics[width = 0.43\textwidth]{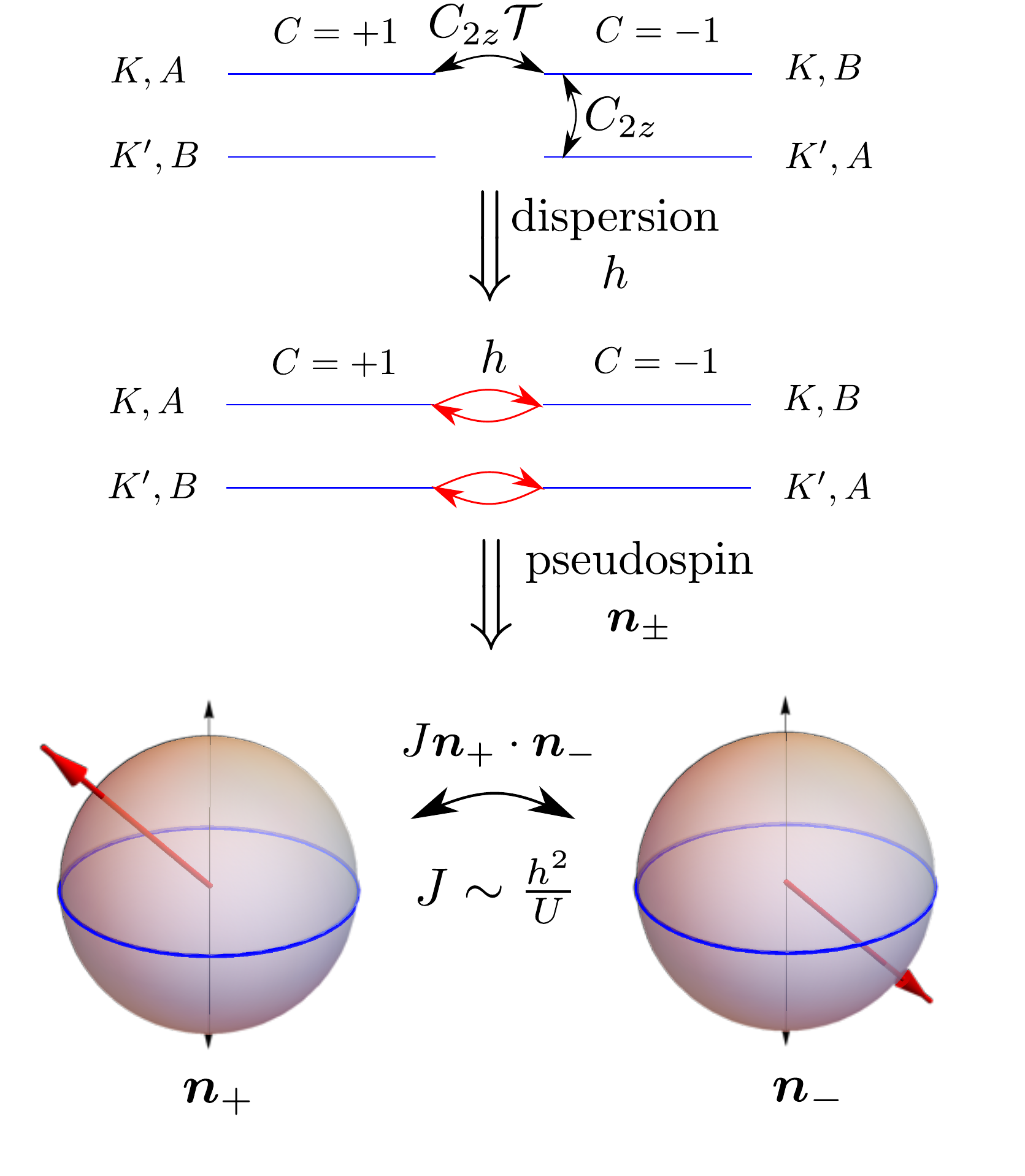}
    \caption{{\bf Tunneling-induced superexchange J}: (Top) Action of symetries in the Chern and pseudospin basis. (Middle) Starting from the two pseudospin ferromagnets that are  related by time reversal symmetry,  dispersion $h$ acts as tunneling between $C_{2z} \T$-related states (Bottom) leading  to a superexchange term $J \sim h^2/U$ that acts as an antiferromagnetic coupling between pseudospins in opposite Chern sectors.}
    \label{fig:Superexchange}
\end{figure}

\section{Charged Topological Textures and Effective Sigma Model}

 Due to the band topology of the two Chern sectors, pseudospin skyrmions in $\bn_\pm$ carry electric charge  \cite{Sondhi,KaneLee} of $\pm e$. The charge density  $\delta \rho(\br) = e \left ( q_+(\br) -q_-(\br) \right ) $ where $q_\pm(\br) = \frac{1}{4\pi} \bn_{\pm}\cdot (\partial_x \bn_{\pm} \times \partial_y \bn_{\pm})$ is the topological density in each Chern sector, which integrates to unity for a single skyrmion. Let us neglect the antiferromagnetic coupling $J$ between the two Chern sectors for now, postponing discussion of its effect to the next section. Then, we find that the size of a charge `$e$'  skyrmion in a single Chern sector is determined only by the Coulomb repulsion which prefers to spread it out over the entire system. In this case, its energy is given only by the elastic contribution $E_{\rm Sk}=4\pi \R$, where $\R$ is the pseudospin stiffness associated with the $\bn_\pm$ vector fields \cite{Sondhi,MoonMori}. 
 
\begin{figure}
    \centering
    \includegraphics[width=0.42 \textwidth]{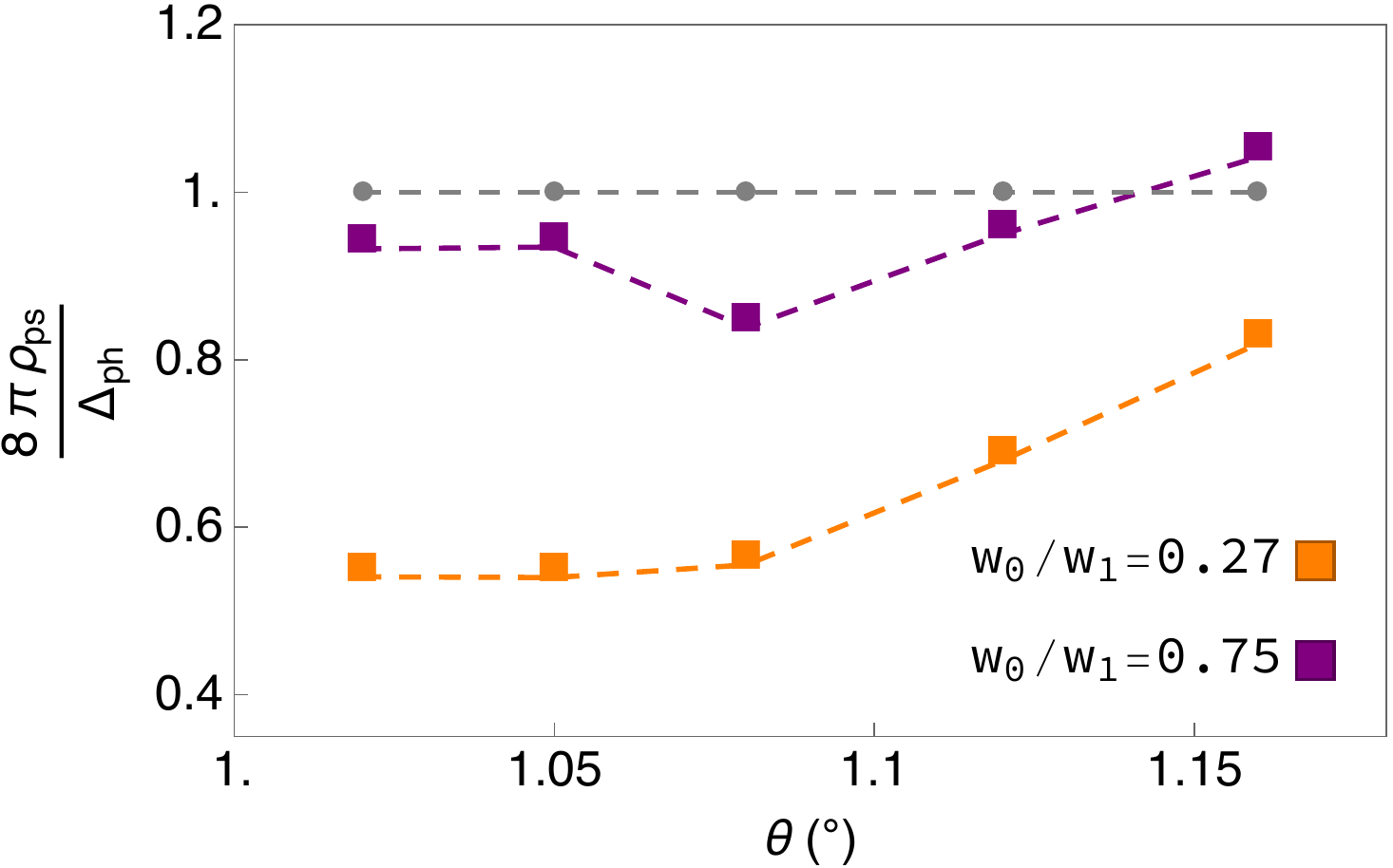}
    \caption{{\bf Energetics of charged excitations:} Ratio of the elastic energy of the $2e$ skyrmion to the particle-hole band gap in the nearly chiral (orange) and realistic (purple) limits, with stiffness and gap values extracted from self-consistent Hartree-Fock. Below the grey dashed line, the skyrmion energy lies within the particle-hole gap. Note that $w_0$ and $w_1$ are the interlayer hoppings in AA and AB-stacked regions (see Ref.~\cite{Tarnopolsky} for details)}
    \label{fig:E2skGapRatio}
\end{figure}

 To determine if  skyrmions play the role of  charge carriers, we must compare their energy  to that of other charge carriers in the system, in particular the particle-hole excitations. This question generally depends on system details - in quantum Hall systems at $\nu = 1$, the energy of a skyrmion-pair, $8 \pi \R$, is smaller than the particle-hole gap, $\Delta_{\rm ph}$ by a factor of 2. For MATBG, the ratio: $8 \pi \R/ \Delta_{\rm ph}$ can be computed numerically from the self-consistent Hartree-Fock calculation leading to the results shown in Fig.~\ref{fig:E2skGapRatio}. For model parameters close to the ideal chiral limit, the ratio is close to the quantum Hall value of $0.5$, whereas for realistic parameters it is about $0.8-0.9<1$  so  that, ignoring anisotropies,  
 skyrmions are indeed the lowest energy charged excitations. While anisotropies are absent in the SP approximation, we will leave a full  calculation of skyrmion energetics in MATBG including the most general kinds of ansiotropy, to future work.  It is sufficient to note that they remain low energy excitations,  and we will proceed based on the premise that skyrmion excitations are important to the charge physics. 
 
 To derive the effective field theory for skyrmions, we integrate out the electrons while retaining the charge degree of freedom in the topological textures. We thus derive an effective description of MATBG by taking the pseudospin variables in the two Chern sectors $\bn_\pm$ to be slowly varying fields leading to (see supplemental material for details)
\begin{multline}
\L[\bn_+, \bn_-] = \sum_{\gamma=\pm} \left(\frac{1}{2A_M} {\mathcal \bA}[\bn_\gamma]\cdot  \partial_\tau \bn_\gamma  + \frac{\R}{2} [\nabla \bn_\gamma]^2 \right) \\ + J \bn_+ \cdot \bn_-   - \mu e \delta \rho + \frac{1}{2} \int  d^2 \br' \delta \rho(\br) V(\br - \br') \delta \rho(\br')  
\label{eq:NLSM}
\end{multline}
where $A_M$ is the area of the moir\'e unit cell \footnote{given by $\frac{\sqrt{3} a^2}{2 \theta^2}$ (with $a = \sqrt{3} a_{\rm CC}$)}. The first term is the Berry phase term, the second represents the elastic energy of the non-uniform pseudospin configurations with $\R \sim 1$ meV, and the third term includes the effects of antiferromagnetic coupling via the `super-exchange'  $J \sim 0.5-1$ meV.  The chemical potential couples to the charge deviations from the background which is given by the skyrmion (antiskyrmion) topological charge in the $+$ ($-$) Chern sector. 

\begin{figure}
    \centering
    \includegraphics[width = 0.4 \textwidth]{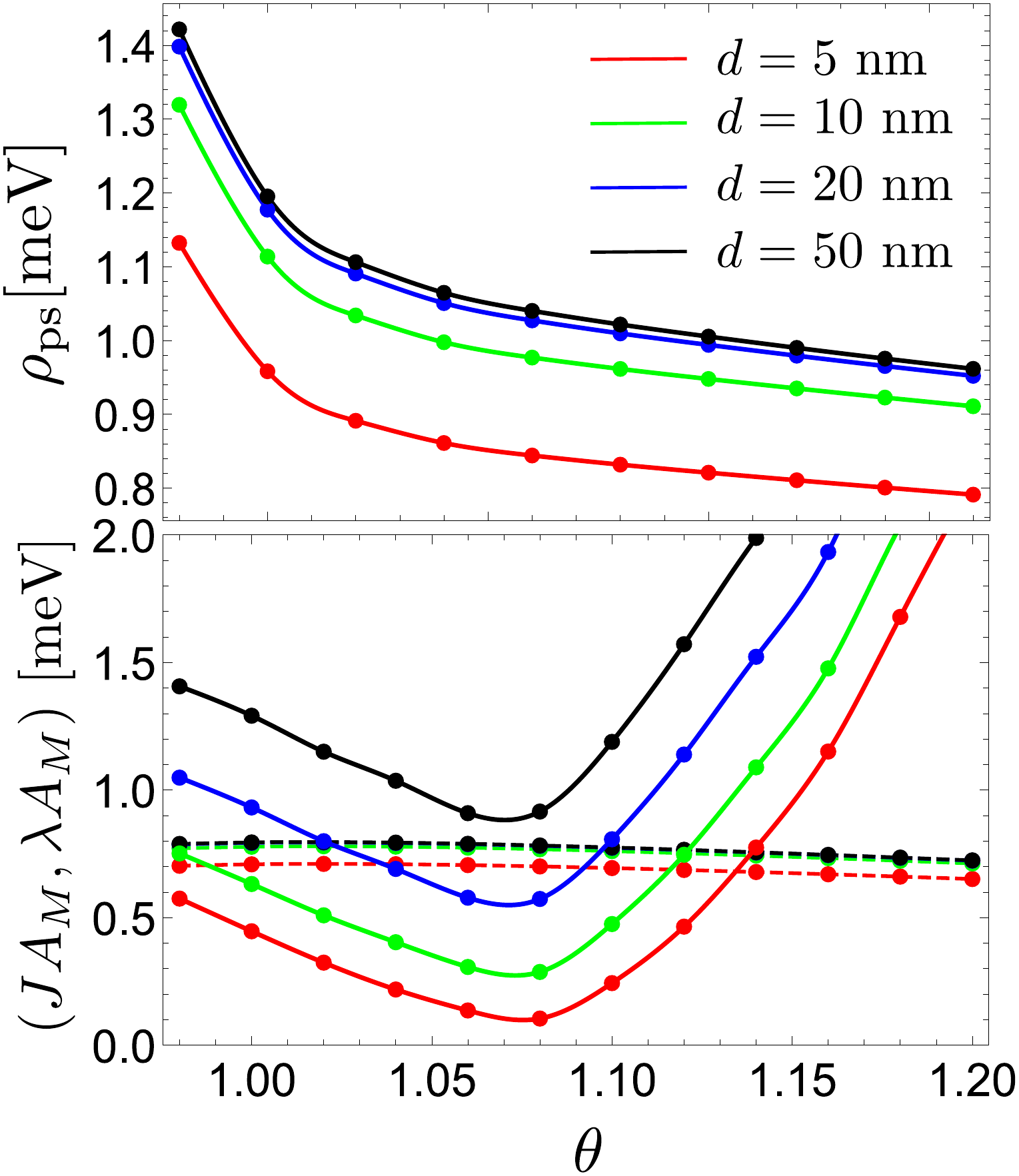}
    \caption{{\bf Sigma model parameters}: Pseudospin stiffness (upper panel), antiferromagnetic coupling $J$, and pseudospin easy-plane anisotropy $\lambda$ (lower panel) as a function of the twist angle $\theta$ for different values of the gate screening distance $d$ for dielectric constant $\epsilon = 9.5$. All curves are computed using the analytic expressions provided in the supplemental material using the band renormalization scheme of Ref.~\cite{Liu19}.}
    \label{fig:RhoJLambda}
\end{figure}

{We note here that the deviation from the perfect sublattice polarization assumed so far leads to an additional term $\lambda ( \bn_{+,xy} \cdot \bn_{-,xy} - \bn_{+,z} \cdot \bn_{-,z})$ with $\lambda \approx$ 0.5 meV for realistic MATBG parameters (cf.~Fig.~\ref{fig:RhoJLambda}). This term favors out-of-plane ferromagnetic coupling and in-plane anferromagnetic coupling, thus acting as an easy plane anisotropy which selects the XY pseudospin antiferromagnetic order (this is the Krammers intervalley coherent order discussed in Ref.~\cite{KIVCpaper}).} The dependence of the sigma  model parameter $J$ and $\lambda$ on the angle and interaction parameters is shown in Fig.~\ref{fig:RhoJLambda}.

\section{Superconductivity from Skyrmion Pairing}

The antiferromagnetic coupling $J$ between opposite Chern sectors leads to the binding of a skyrmion  \textit{anti}-skyrmion pair in the opposite Chern sectors.  
The net charge of this combined object is $Q = e(q_+-q_-)=2e$, i.e. the exchange $J$ has effectively resulted in an extended Cooper pair. Note the crucial interplay of anti-ferromagnetic coupling between opposite Chern-bands; if we had had $J < 0$, skyrmions would bind with skyrmions, leading to a charge-neutral object. Remarkably, despite the long-range Coulomb interaction, such a bound state will form \emph{no matter how small $J$ is}, in the absence of other anisotropies. Roughly speaking, an isolated charge $e$ skyrmion in a single Chern sector pays a ``Zeeman" energy due to coupling to the uniform ferromagnetic order in the opposite sector via the exhange term $J$. This energy cost scales with the size $R$ of the skyrmions as $\sim J R^2$. As in the case of quantum Hall skyrmions \cite{Sondhi}, the competition with the Coulomb repulsion $\sim U/R$ will lead to a finite size for such skyrmions and yields an extra energy penalty on top of the elastic contribution. On the other hand, a pair of antiferromagnetically locked charge $2e$ skyrmions does \textit{not} pay any exchange energy which enables it to evade Coulomb repulsion by becoming very large. Hence, the extended nature of the skyrmion allows for a pairing mechanism which 
evades the Coulomb repulsion while benefiting locally from the antiferromagnetic coupling.

\begin{figure}
    \centering
    \includegraphics[width=0.42\textwidth]{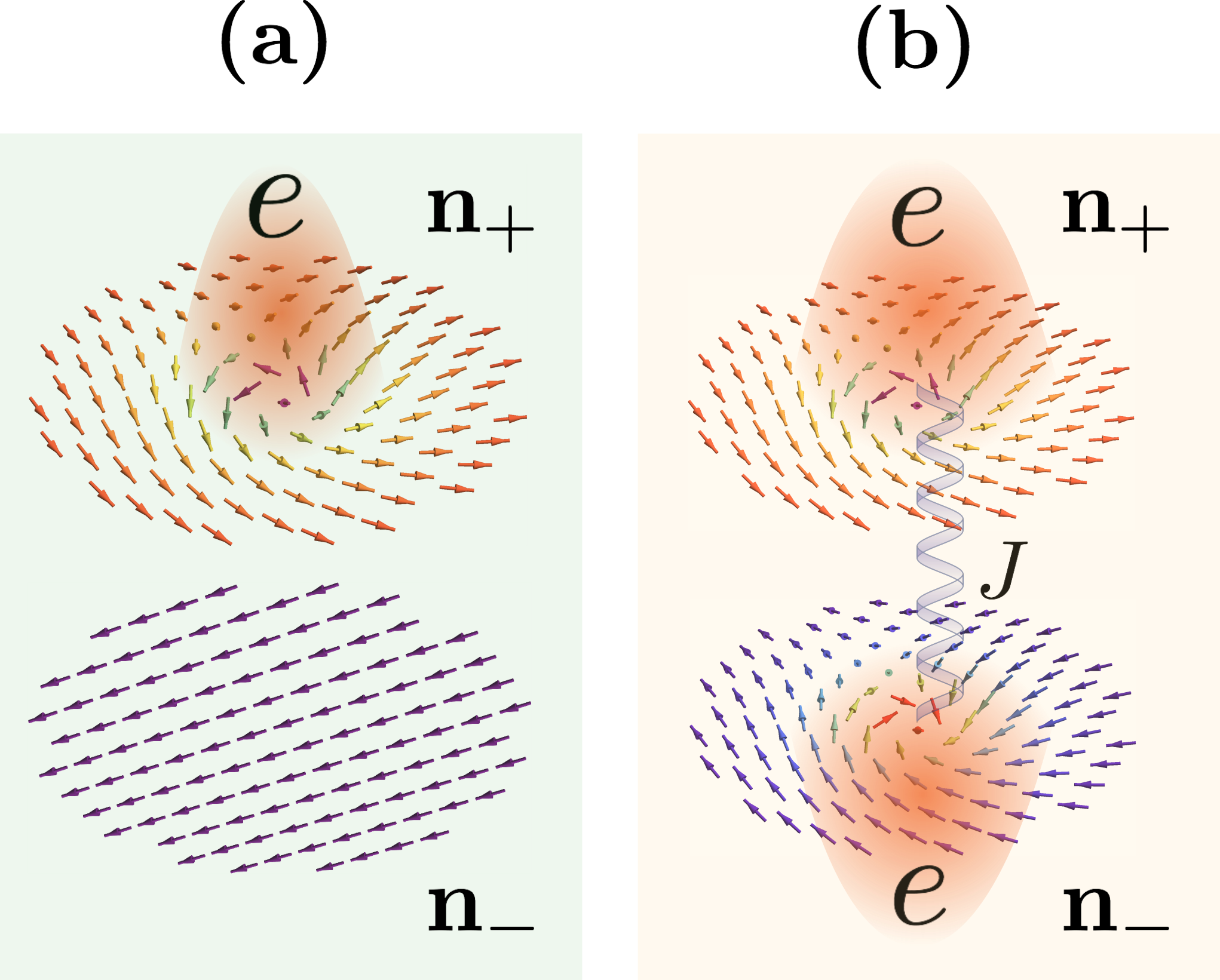}
    \caption{{\bf Pairing of Charged Topological textures}: (a) Single charge $e$ skyrmion in one of the Chern sector. Pseudospins in $\bn_+$ and $\bn_-$ not antiferromagnetically aligned in the skyrmion core. (b) skyrmion-antiskyrmion pair in $\bn_+$-$\bn_-$ binding together due to the antiferromagnetic coupling $J$ which favors local pseudospin anti-alignment forming a charge $2e$ object.}
    \label{fig:ChargedSkx}
\end{figure}

{The inclusion of the easy plane anisotropy $\lambda$ due to imperfect sublattice polarization affects the above scenario as follows. First, it leads to the deformation of skyrmions to a topologically equivalent configuration of a meron-antimeron pair to evade the penalty of out-of-plane pseudospin alignment. Second, it reduces the binding energy of anti-ferromagnetically locked skyrmion-antiskyrmion pair. This can be seen by noting that out-of-plane pseudospin only incurs an energy penalty for such skyrmion pairs but not for individual skyrmions (since the energy only has the term $n_{+,z} n_{-,z}$ but not $n_{+,z}^2$). As a result, this term reduces the binding energy of the charge $2e$  skyrmion-antiskyrmion pairs and eventually leads to their unbinding when it is sufficiently large. Evidence from a recent numerical study \cite{SkDMRG} suggests that the binding energy remains finite for the physically relevant parameter regime which we will assumed in what follows.}

Superconductivity does not follow from pairing alone. To establish a nonzero superfluid stiffness and transition temperature $T_c$, the condensing $2e$ bound state must have a finite effective mass despite the flat-band dispersion of charge $e$ skyrmions. Remarkably, this effective mass can be generated entirely by the Coulomb repulsion through the exchange scale $J$. This can be understood by noting that the skyrmion and antiskyrmion in opposite Chern sectors feel opposite effective magnetic fields $B_{\rm eff} = \frac{2\pi \hbar}{e A_M}$. This leads to a Lorentz force which tends to pull them apart when they move together. Given that a skyrmion-antiskyrmion pair are bound together by $J$ as shown in Figure \ref{fig:ChargedSkx}b, the restoring "spring" force balances the Lorentz force leading to an effective mass.  More precisely, writing the Lagrangian for a a skyrmion and an antiskyrmion at $\bR_+$ and $\bR_-$:
\beq
\L = \frac{e B_{\rm eff}}{2}(\dot{\bR_+}\times \bR_+ -\dot{\bR_-}\times \bR_-)- \frac{k}{2}(\bR_+-\bR_-)^2, \quad k =4 \pi J
\eeq
and eliminating the relative coordinate $\bR_+ -\bR_-$ in favor of the  center-of-mass coordinates: $\bR_s = \frac{\bR_+ + \bR_-}{2}$ yields $\L =  \frac{(e B_{\rm eff})^2}{2k} \dot{\bR}_s^2$ from which we can read off the effective mass:
\beq
M_{\rm pair} = \frac{(e B_{\rm eff})^2}{k} = \frac{\pi \hbar^2}{J A_M^2}
\label{eq:Mass}
\eeq
and the transition temperature \cite{NelsonKosterlitz}, related to the effective mass and stiffness through:

\beq
k_BT_c = \frac{\nu \pi \hbar^2}{2 A_M M_{\rm pair}} =\nu  \frac{ J A_M}{2} 
\label{eq:Tc}
\eeq
where $\nu$ is the skyrmion filling fraction. The effective mass sets the condensation scale of the composite objects. For MATBG, $J A_M \sim 1$ meV leading to the scale $T_c \sim 1-5$ K. We will verify these estimates using a field theoretical calculation of the  phase diagram with doping.

\section{Field Theory Description of the Skyrmion Superconductor at Finite Chemical Potential }

The antiferromagnetic coupling implies that the $ \bn_+$ and $\bn_-$ pseudospins are antiferromagnetically locked in the ground state. In addition, the lowest energy charge excitation are also bound states: of skyrmions in $\bn_+$ and of antiskyrmions in $\bn_-$ where $\bn_+ = -\bn_- = \bn$, which can be understood as  $\bn$-skyrmions which carry charge $2e$. Thus, we can integrate out the massive ferromagnetic fluctuations, $ \bn_+ +\bn_-$, whose mass is proportional to $J$. The resulting field theory is then written solely in terms of the $\SO(3)$ vector $\bn$. Furthermore, we can rewrite this in the well known CP${}^1$ representation by writing $\bn = z^\dagger {\boldsymbol \eta} z$, where we  introduced the bosonic spinon field $z = (z_1 , z_2)^T$ which  satisfies the constraint $z^\dagger z = 1$. \cite{CPNdAdda, CPNWitten, PolyakovBook}. The overall phase of $z$ is redundant which leads to the gauge field $a_\mu = - i  z^\dagger \partial_\mu z$. The topological density is given by  the flux of $a_\mu$, and is tied to the charge density. Thus: $\delta \rho = -\frac{\partial_x a_y-\partial_ya_x}{\pi} =-\frac{2}{2\pi}f_{xy}$ \cite{CPNdAdda, CPNWitten, PolyakovBook}. Note, since the flux of $a$ is tied to a conserved charge, monopole fluctuations which change the flux in units of $2\pi$  are disallowed on symmetry grounds, unlike in the usual  CP$^1$ model. Thus the dynamics of this model resembles that of the non-compact CP$^1$ theory \cite{MV04}, which explicitly disallows monopoles. 

The CP${}^1$ action takes the form
  \begin{multline}
      S[z] = \int d^3 \br \left\{ \frac{\Lambda}{g} |D_\mu z|^2 + \frac{2 e \mu}{2\pi c}  f_{xy} \right. \\ \left.
     +\frac{1}{2c} \int d^2\br' \frac{f_{xy}(\br)}{\pi}  V(r-r') \frac{f_{xy}(\br')}{\pi} \right\}
    \label{eq:LCP1g}
  \end{multline}
    where $D_\mu  = (\partial_\mu - i a_\mu) $. Note, we rescaled the (imaginary) time as $\tau = \frac{1}{c} r_z$, introduced the cutoff scale $\Lambda = 1/\sqrt{A_M}$, the coupling $g = \sqrt{\frac{J A_M}{2\R}}$, and velocity $c  = \frac{2}{\Lambda} \sqrt{2 J A_M \R}$.

In the following, we compute the phase diagram of the CP${}^1$ model at finite doping and interpret the results for MATBG. We use the large-$N$ approximation by extending to $N$ complex fields $(z_1,\,z_2,\dots z_N)$.  We start by reviewing the solution in the absence of doping and Coulomb interaction ($\lambda = \mu = V = 0$) which was discussed in the pioneering works of Refs.~\cite{CPNdAdda, CPNWitten, CPNArefeva}. In this limit, the model has two phases: (i) an ordered phase for $g < g_c = 4\pi$ characterized by a finite expectation value of $z$ with $a_\mu$ Higgsed and (ii) a disordered phase for $g > g_c = 4\pi$ where the $z$ variables are gapped and $a_\mu$ is free. The gap is given by $|\Delta| = \Lambda (1 - g_c/g)$ and the Lagrangian for the gauge field $a_\mu$ has the standard Maxwell form $\sim \frac{1}{|\Delta|} (\partial_\mu a_\nu - \partial_\nu a_\mu)^2$ with an additional Chern-Simons coupling to the background $\U(1)$ electromagnetic field $\frac{i e}{\pi} \epsilon_{\mu \nu \lambda} a_\mu \partial_\mu A_\lambda$. The disordered phase actually describes a paired superfluid which can be seen by integrating out the field $a_\mu$  by introducing a dual phase variable $\phi$ which can be identified with the phase of the superfluid order parameter
\begin{gather}
    S = \int d^2 \br d\tau \left\{ \frac{\rho_{\rm SC}}{2} (\partial_i \phi - 2e A_i)^2 + \frac{\chi_{\rm SC}}{2} (\partial_\tau \phi - 2e A_\tau)^2 \right\} \nonumber \\
\rho_{\rm SC} = \frac{3|\Delta|c}{\pi N}, \quad \chi_{\rm SC} = \frac{3|\Delta|}{\pi c N}
\label{eq:RhoChi}
\end{gather}
The phase stiffness $\rho_{SC}$ yields the temperature scale for superconductivity, using the standard formula $T_c = \frac{\pi \rho_{\rm SC}}{2 k_B}$ \cite{NelsonKosterlitz}. Let us now turn to producing  the superconductor by varying the chemical potential, rather than the coupling $g$. 

\begin{figure*}
    \centering
    \includegraphics[width = 0.75\textwidth]{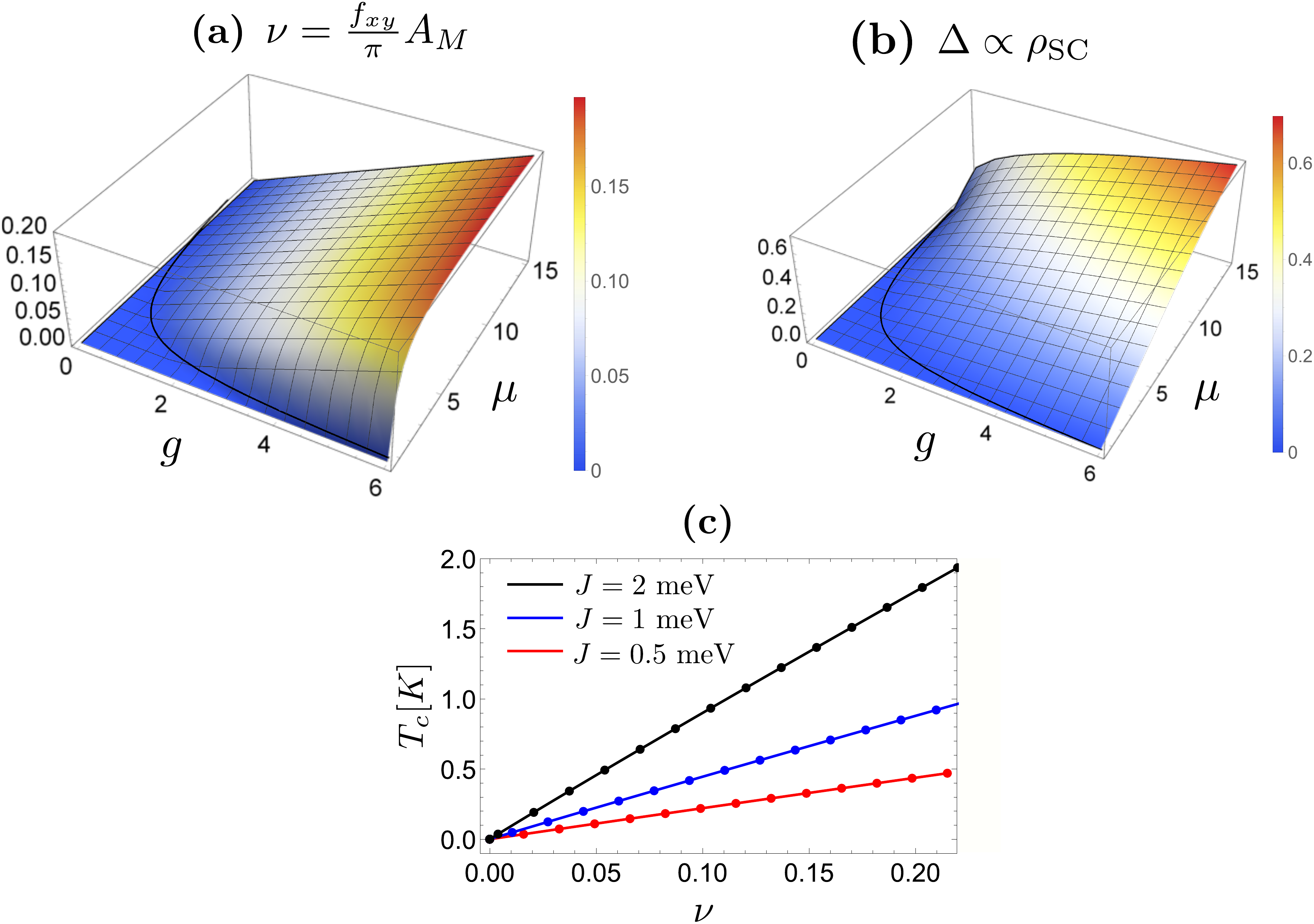}
\caption{{\bf Large-N phase diagram for the doped CP${}^1$ model}: (a) plot of the charge 2e skyrmion filling fraction $\nu$ which is directly related to the flux $f_{xy}$ of the gauge field $a_\mu$ and (b) the gap for the spinon field $\Delta$  which is directly proportional to the superfluid stiffness (cf.~Eq.~\ref{eq:RhoChi}) in the superconducting phase plotted as a function of the chemical potential $\mu$ and the dimensionless coupling $g$. The energies $\mu$ and $\Delta$ are measured in units of the energy cutoff $\Lambda c$. The black line represents the phase boundary between the insulator and the superconductor. In (c), we plot an estimate for superconducting $T_c$ in the CP${}^{N-1}$ model, computed using the relation: $\frac{3\pi c |\Delta|}{2 \pi N k_B}$ while setting $N = 2$  and $\rho_{\rm ps} = 1$ meV as a function of doping $\nu$ for different values of $J$.}
    \label{fig:PhaseDiagram} 
\end{figure*}

For nonzero chemical potential $\mu$, the phase diagram can be obtained numerically by solving the mean-field self-consistency equation (see supplemental material) as a function of $g$ and $\mu$  as shown in Fig.~\ref{fig:PhaseDiagram}. Note, a finite value $|\mu_c| = \frac{\pi c \Lambda}{g} (1 - g/g_c)$ is needed to introduce skyrmions in the ordered phase $g < g_c$. This value reduces to the elastic skyrmion energy $4\pi \R$ for $g \ll g_c$ which serves as a check on the CP${}^N$ theory. For larger $\mu$, the skyrmion density $\nu$ is obtained numerically. For any finite skyrmion density $\nu$, the effective field theory takes the form (\ref{eq:RhoChi}) describing a superconductor whose phase stiffness can be computed numerically as shown in Fig.~\ref{fig:PhaseDiagram}. In the dilute limit $\nu \ll 1$ and for $g \ll g_c$,  $\Delta$ can be expressed in terms of the doping $\nu$ leading to
\beq
    \rho_{\rm SC}  \approx \frac{3 \Lambda c \nu g}{4\pi N} = \frac{3 J A_M \nu}{2\pi N} , \quad \chi_{\rm SC} \approx \frac{3 \Lambda g \nu}{4\pi N c} = \frac{1}{N} \frac{3 \nu}{16 \pi \R A_M}
\eeq
Interestingly, the superfluid stiffness (charge compressibility) of the superconductor is inversely proportional to the pseudospin compressibility (psuedoespin stiffness) of the zero doping insulator with the proportionality constant being, up to numerical prefactors, just the filling $\nu$. This leads to $T_c \sim J \nu$ in agreement with Eq.~\ref{eq:Tc}. The phase diagram of the chemical potential tuned CP$^1$ theory with realistic parameters and screened Coulomb interaction should be also be accessible in future studies using the Monte Carlo technique\cite{Babaev}.

To fix the pairing symmetry of the superconductor  consider the symmetries  of the operator $ \Delta$ that creates flux: $\nabla \times a = 2\pi$, which corresponds to the Cooper pair creation operator. Since flux is left invariant by  pseudospin rotation, $ \Delta$ is expected to be a pseudospin singlet. Hence the Cooper pair operator must be one of: $\Delta_\mu(\bk) = g_\mu(\bk) c_{-\bk}^T \eta_y\gamma_\mu c_{\bk} $ which gives four possible pairings. Reverting to the picture of skyrmion-anti-skyrmion pairing, we note that pairs reside in opposite Chern sectors, which eliminates two of the four pairing channels leaving us with $\Delta_{x,y} \propto \eta_y\gamma_{x,y}$. Of these, the first is antisymmetric in internal indices leading to even pairing $g_x(-\bk) = g_x(\bk)$ whereas the second is symmetric in internal indices leading to odd pairing $g_y(-\bk) = -g_y(\bk)$. Further evaluating the rotation quantum number for these pairings \cite{DQCPPRB,Shift, Monopole} (see supplemental material for details), one sees that the natural zero angular momentum skyrmion corresponds to  $\Delta_x \propto \eta_y\gamma_x  = \tau_y $ , while the other $\Delta_y \propto \eta_y \gamma_y = \tau_x \sigma_z$ option corresponds to a nonzero (odd) angular momentum.

Separately, it is worth noting that the pairing channels $\Delta_{x,y}$ are also the ones that correspond to the maximal gap in the presence of the insulating pseudospin antiferromagnetic background. This is readily seen by checking that the corresponding matrices anticommute in the Nambu basis, (see Supplementary materials) thus the insulating and superconducting gaps add in quadrature when they are simultaneously present.  Furthermore, taking into account the kinetic part of the Hamiltonian $h$, in Eq. \ref{Eqn:h}, it is found to anticommute with $\Delta_x$, leading to a bigger gap compared to $\Delta_y$ which commutes with $h$.

\section{Schematic phase diagram and duality between the insulator and superconductor}

    \begin{figure}
\includegraphics[width=0.45 \textwidth]{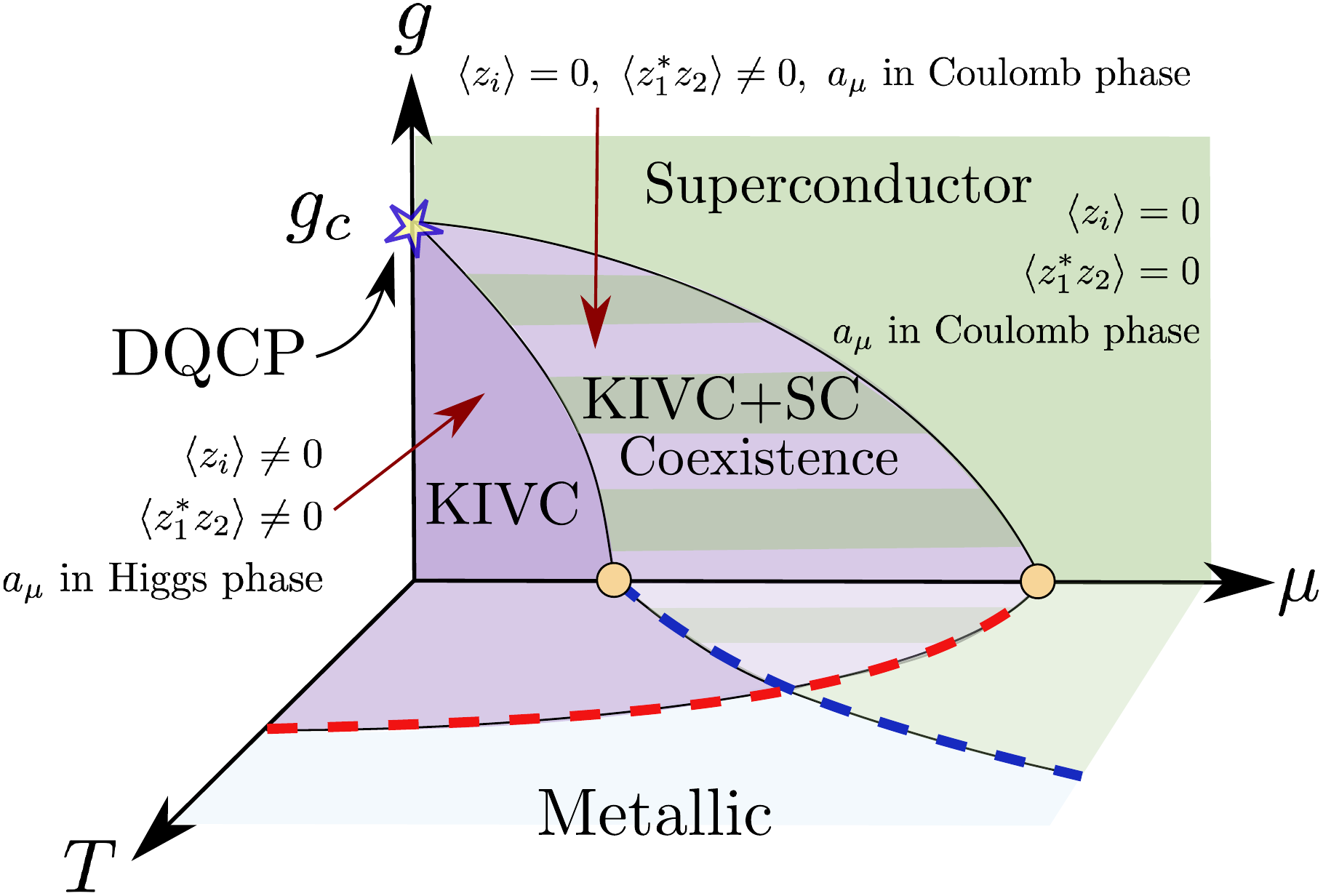}
\caption{Schematic phase diagram in the limit of short screening length. We will be mainly concerned with chemical potential driven transitions at small $g<g_c$. The dotted blue (red) curve at nonzero $T$ indicates the BKT transition out of the superconductor (KIVC) with quasi-long range order. On increasing $g$ at $T = 0$ and $\mu = 0$, a direct transition between the KIVC and SC states (deconfined criticality) is potentially allowed, which then splits into two ordinary quantum critical points (yellow circles) with a region of co-existence when $\mu \neq 0$. }
\label{fig:ScPhasediagram}
\end{figure}

The large $N$ mean field analysis identified two phases: a K-IVC insulator and a superconductor. There is however a third possibility: a phase where both orders coexist. Such phase is possible when superconductivity is driven by tuning the chemical potential since the condensation of skyrmions in this case does not disorder the phase of the insulator XY order parameter. Thus, for finite $N$, we expect a phase diagram which schematically has the form of Fig.~\ref{fig:ScPhasediagram} where the three different phases can be identified using the expectation values $\langle z_i \rangle$ and $\langle z_1^* z_2 \rangle$ as follows \cite{MV04, DQCPScience, DQCPPRB}:

 \emph{1. Pure insulator:} In the absence of doping, at small $g$  below some critical coupling $g_c$,  we have the Higgs phase $\langle z_i\rangle \neq 0$ which, with the expected easy plane anisotropy ($\lambda>0$), favors equal amplitudes for $z_{1,2}$, takes the form 
     \beq
     \langle z_i\rangle =  r_0 e^{i\theta_i}.
     \eeq
     This is nothing but  the K-IVC phase, with K-IVC order parameter $\langle z_1^*z_2\rangle =r_0^2 e^{i(\theta_1-\theta_2)}$. The gapless photon of $a$ is absent due to the Higgs mechanism. This indicates that density and current fluctuations are suppressed,  implying an electrical insulator. This can be seen by explicitly integrating out $a$ in the Higgs phase which gives: $ {\mathcal L} = \frac{(2e)^2}{2\kappa}(\epsilon_{\mu\nu\lambda}\partial_\mu A_\lambda)^2$ with $\kappa = 16\pi^2 r_0^2\Lambda/g$, as expected for a insulating dielectric. On the other hand, fluctuations of $\theta$ in the K-IVC phase lead to a gapless Goldstone mode.
     
     \emph{2. Coexistence phase:} Next, raising the chemical potential $\mu$ corresponds to applying a magnetic field to the Higgs condensate. Beyond  a critical value of $\mu$,  vortices of the Higgs condensate form to accommodate flux. The cheapest such configuration is a simultaneous vortex in both $z_1$ and $z_2$, which has a finite energy cost and carries charge $2e$ (this is nothing but the $\bn$ skyrmion). Condensing these vortices kills the Higgs condensate in $\langle z_i \rangle=0$, which restores $a$ to its Coulomb phase. This is the superfluid and the photon of $a$ is simply the superfluid Goldstone mode. Nevertheless the K-IVC order parameter is not destroyed since the phase of both $z_1,\,z_2$ wind around such a vortex, their relative phase $z_1^*z_2$ remains well defined, so $\langle z_1^*z_2 \rangle \neq 0$ . This is the coexistence phase.  The phase transition from the insulator to the coexistence phase with doping is in the  superfluid-insulator transition driven by the chemical potential, with dynamical exponent $z=2$ \cite{sachdev_2011}.
     
     \emph{3. Superconductor:} Finally, on further raising $\mu$, the amplitude of the KIVC order is reduced, potentially driving a transition when the K-IVC order disappears. Beyond this we simply have the superfluid order since $a$ remains in the Coulomb phase. This transition is expected to be in the 3D XY universality class.

     The relationship between the K-IVC insulator and the superconductor can be highlighted by the following duality. We start by writing the K-IVC and the superconducting order parameters as $\langle c^\dagger_{K\sigma} c_{K'\sigma'} \rangle = \Delta_{\rm KIVC} \sigma^y_{\sigma\sigma'}$ and $\langle c^\dagger_{\tau\sigma} c^\dagger_{\tau'\sigma'} \rangle = \Delta_{\rm SC} \delta_{\sigma,\sigma'}\tau^y_{\tau \tau'}$, respectively,  where $\Delta_{\rm IVC}=|\Delta_{\rm IVC}|e^{i\vartheta}$ and $\Delta_{\rm SC}=|\Delta_{\rm SC}|e^{i\varphi}$ are complex numbers. The two are related by the following  particle-hole transformation that acts in only one of the valleys:
\begin{eqnarray}
c_{K,\sigma} &\mapsto& c_{K,\sigma}, \qquad c_{K',\sigma} \mapsto -i \sigma^y_{\sigma,\sigma'} c^\dagger_{K',\sigma'}\\
\hat{n}_c &\leftrightarrow &  \hat{n}_v, \,\,\,\,\qquad\varphi \leftrightarrow \vartheta
\end{eqnarray}
This duality transformation exchanges the total $\U(1)_c$ charge conservation (of total charge $\hat{n}_c$) and the $\U(1)_v$ valley-charge conservation (of valley charge $\hat{n}_v$) which is consistent with the fact that the KIVC (SC) order breaks (preserves) $\U(1)_v$ but preserves (breaks) $\U(1)_c$.  Furthermore, this transformation interchanges the phases $\vartheta \leftrightarrow \phi$ of the K-IVC and Superconductor, conjugate to the charges $n_v$ and $n_c$ which are interchanged.

In fact, it is helpful to organize the order parameters into a five component object $n_i$:
$n_1+in_2=\Delta_{\rm IVC}$ and $n_4+in_5=\Delta_{\rm SC}$ while $n_3$ represents the valley-Hall order. Under the duality it is  readily verified  from the fermion transformation that $n_3\leftrightarrow n_3$, so the duality acts as a rotation in SO(5). In fact combining the SO(3) pseudospin rotations of $(n_1,\, n_2,\,n_3)$ and the $U(1)_c$ rotation of $(n_4,\,n_5)$ with this duality, one obtains the full SO(5) group.

This duality immediately fixes properties of the superconductor, based on what we know about the K-IVC insulator. 
For example,  vortices in the order parameters are interchanged by duality. 
Here, vortices of K-IVC have `cores' that contain the valley-Hall order $n_3$,  giving rise to merons, or two component vortices with opposite direction of $n_3$ order in their cores. These carry electric charge $\pm e$ for opposite $n_3$ configurations of the same vorticity. Under the duality, these are mapped to superconducting vortices, which also have a two component structure, with $n_3$ core that now determines their valley-charge. Denoting the K-IVC and SC vortex creation fields by $(w_1,\,w_2)$ and $(z_1,\,z_2)$, their properties are summarized in Table \ref{tab:duality}.

 \begin{table}[]
     \centering
     \begin{tabular}{|c|c|c|c|c}
     \hline \hline
        Vortex & Gauge Field & Charge ($n_c,\,n_v$) & Order Parameter\\
        \hline
        K-IVC: $w_{1,\,2}$ & $\tilde{a}$ & ($\pm1$, 0) & $w_1^*w_2 = \Delta_{\rm SC}$\\ 
        \hline
        SC: $z_{1,\,2}$ & ${a}$ & (0, $\pm1$) & $z_1^*z_2 = \Delta_{\rm KIVC}$\\ 
        \hline \hline
     \end{tabular}
     \caption{Summary of properties of K-IVC vortex fields(merons) in the first row, charge quantum numbers and relation to superconductor order parameter. Second row, the corresponding properties for superconducting vortices obtained from duality. The superconductor vortex fields is equivalent to the CP${}^1$ representation of K-IVC order.}
     \label{tab:duality}
 \end{table}
 
 \section{$\SO(5)$ sigma model and WZW term}
 In the previous sections, we have mainly focused on the transition into the superconductor starting deep inside the K-IVC state by tuning the chemical potential, treated within the CP$_1$ theory. However, following Refs. \cite{SenthilFisher, GroverSenthil, LeeSachdev}, it is instructive to derive an equivalent effective field theory which deals with the insulator and superconductor on equal footing. Such a field theoretic description is known to include a topological WZW term. In the following, we will outline this derivation in the context of magic angle graphene including the effect of the chemical potential.

To derive universal aspects of the field theory, such as the presence of a topological term, it is sufficient to adopt a convenient starting point where we take a  dispersion with Dirac points for the nearly flat bands with a spontaneously induced single particle gap (Dirac mass) that is much smaller than the dispersion. The dispersion can then be linearized close to the moir\'e Dirac points $K_M$ and $K'_M$ where we expect a gap to be induced via a Dirac mass term. To deal with the insulating and superconducting states on equal footing, we introduce the Nambu basis defined as 
\beq
\chi^T_{\bk} = (\psi_{K_M,\bk}^T, \psi_{K'_M,-\bk}^\dagger) 
\eeq
Let us now introduce the Pauli matrices $\rho_{x,y,z}$ which act within the two-dimensional Nambu space in $\chi_\bk$. The Hamiltonian now takes the form 
\beq
\H_D = k_x \gamma_x \rho_z + k_y \gamma_y + \mathcal M, 
\label{eq:HD}
\eeq
The mass $\mathcal M$ is a matrix in $\gamma$, $\eta$, and $\rho$ spaces containing both the K-IVC and the superconducting  order parameters, as well as  the valley Hall order which is a part of the antiferromagnetic manifold. All these correspond to anticommuting mass terms for the Dirac equation and hence can be written as: $\mathcal M = \sum_{i=1}^5 n_i \Gamma_i$ where the $\SO(5)$ order parameter $\hat n$ is defined as $(\bn, \Re \Delta_{\rm SC}, \Im \Delta_{\rm SC})$. The corresponding orders and the matrices $\Gamma_i$ are shown in the Table \ref{Table:Gamma}.

\begin{table}[h!]
\centering
\begin{tabular}{ |c||c|c|c| c|c|} 
 \hline
 Order & $\Re \Delta_{\rm IVC}$ & $\Im \Delta_{\rm IVC}$ & $\Delta_{\rm VH}$ & $\Re \Delta_{\rm SC}$ & $\Im \Delta_{\rm SC}$\\ 
 \hline\hline
 $\hat{n}$ & $n_1$ & $n_2$ & $n_3$ &  $n_4$ & $n_5$ \\ 
 \hline
 $\Gamma$ & $\gamma_z \eta_x \rho_z$  & $\gamma_z \eta_y$ &$\gamma_z \eta_z \rho_z$ & $\eta_y \gamma_x \rho_y $ & $\eta_y \gamma_x \rho_x$\\ 
 \hline
\end{tabular}
\caption{SO(5) Order parameters and Dirac mass terms }
\label{Table:Gamma}
\end{table}

We note that the massive Dirac Hamiltonian (\ref{eq:HD}) is invariant under the particle-hole symmetry $\P = \gamma_z \rho_y \K$ (i.e. $\{\H_D, \P\} = 0$) where $\K$ is complex conjugation. As a result, it belongs to symmetry class C of the Altland-Zirnbauer classification \cite{AltlandZirnbauer}. The mass term $\mathcal M$ parametrizes the symplectic Grassmanian manifold $\frac{\Sp(4n)}{\Sp(2n) \times \Sp(2n)}$ (our convention here is that $Sp(2n)$ constitutes $2n \times 2n$ symplectic matrices). For our case, $n=1$ and $\frac{\Sp(4)}{\Sp(2) \times \Sp(2)}$ is isomorphic to the 4-sphere parametrized by the 5-dimensional unit vector $\hat n$. The topology of the symplectic Grassmanian $\pi_4\left(\frac{\Sp(4n)}{\Sp(2n) \times \Sp(2n)}\right) = \Z$ is what allows for the existence of a WZW term in the action \cite{Kitaev09, Schnyder09, Ryu10}.

Following the standard procedure by integrating out the fermions and performing the gradient expansion (see supplemental material for details), we derive the following effective theory
\beq
S = S_{\rm WZW} + \int d\tau d^2 \br {\mathcal L}[\hat n] 
\label{eq:SSO5}
\eeq
where ${\mathcal L}[\hat n] $ is given by
\begin{multline}
   {\mathcal L}[\hat n] = \frac{\tilde \rho}{2} (\nabla \hat n)^2 + \frac{\tilde \chi}{2} (\partial_\tau \hat n)^2 + (g - g_c + 2\tilde \chi \mu^2) (\bn^2 - |\Delta|^2) \\ + \lambda n_3^2 + 2\tilde \chi \mu \Delta^\dagger \partial_\tau \Delta - \frac{\mu}{2\pi} \bn \cdot (\partial_x \bn \times \partial_y \bn )
   \label{eq:LSO5}
\end{multline}
We see that the chemical potential enters the action in three different places. First it couples to the topological density of the vector $\bn$ representing the insulator. Note that since $\bn$ now does not have a fixed length, this term is not quantized. Second, the chemical potential couples linearly to the first derivative of the superconducting gap and quadratically to its magnitude. The latter coupling affects the effective potential, favoring superconductivity at finite doping. We note that both couplings were considered before in the context of $\SO(5)$ theories for cuprates Ref.~\cite{ZhangSO5, DemlerSO5} although the topological term which we turn to next was absent in that context.  

$S_{\rm WZW}$ is the well-known Wess-Zumino-Witten term which can be written by introducing an auxillary integration variable as
\beq
S_{\rm WZW} = \frac{3i}{4\pi} \int_0^1 du \int d^2 \br d\tau \epsilon^{abcde} n_a \partial_u n_b \partial_\tau n_c \partial_x n_d \partial_y n_e
\eeq

At $\mu = 0$, this theory has the same form as the theory describing the transition between an antiferromagnet and a valence-bond-solid \cite{DQCPScience, DQCPPRB,Sandvik2007,SenthilFisher}. In this case, the vector $\hat n$ is always either fully in the superconducting ($12$) phase or the insulating ($345$) phase, i.e. there is no coexistence regime. At $g = g_c, \lambda = 0$, the model has been conjectured to possess an emergent $\SO(5)$ symmetry \cite{Nahum15,WNMXS17}, reduced to an $SO(4)$ symmetry in the easy plane limit. The existence of a WZW term implies that one cannot simultaneously disorder both the superconductor and the K-IVC/Valley Hall order without inducing either a gapless critical point (deconfined criticality as shown in Figure \ref{fig:ScPhasediagram}) or topological order. 

\section{Role of Spin Quantum Number}

The spinless model discussed in this paper is directly relevant to the vicinity of the half-filling insulator (at fillings $\nu = \pm (2+\epsilon)$), where charge neutrality corresponds to $\nu = 0$. Achieving an insulator at half-filling requires polarizing a flavor (eg. spin)  \emph{and}  opening a gap at the Dirac point, presumably through developing a pseudospin quantum Hall ferromagnet (with $\bn_+ = -\bn_-$). Spin polarization means that for each valley, $K$ and $K'$, only one spin species is filled which may correspond to a simple spin ferromagnet or a spin-valley locked state with spin anti-aligned in opposite valleys. The main assumption leading to the reduction to a spinless problem is that spin polarization takes place at a higher energy scale than the scale relevant for superconductivity. This is strongly suggested by the cascade picture of Refs.~\cite{CascadeShahal, CascadeYazdani} which identified a relatively high temperature scale where flavor polarization takes place, before the development of insulating or superconducting behavior at lower temperatures. As a result, one spin flavor is completely frozen at the scales relevant for superconductivity allowing us to focus on the spinless problem. We note that most elements of our discussion remain valid even without the assumption of a frozen spin flavor -- pseudospin skyrmions still carry charge and pair due to antiferromagnetic coupling -- but the ground state manifold has a more complicated structure and allows for other competing charged excitations.

\begin{figure}
    \centering
    \includegraphics[width = 0.45\textwidth]{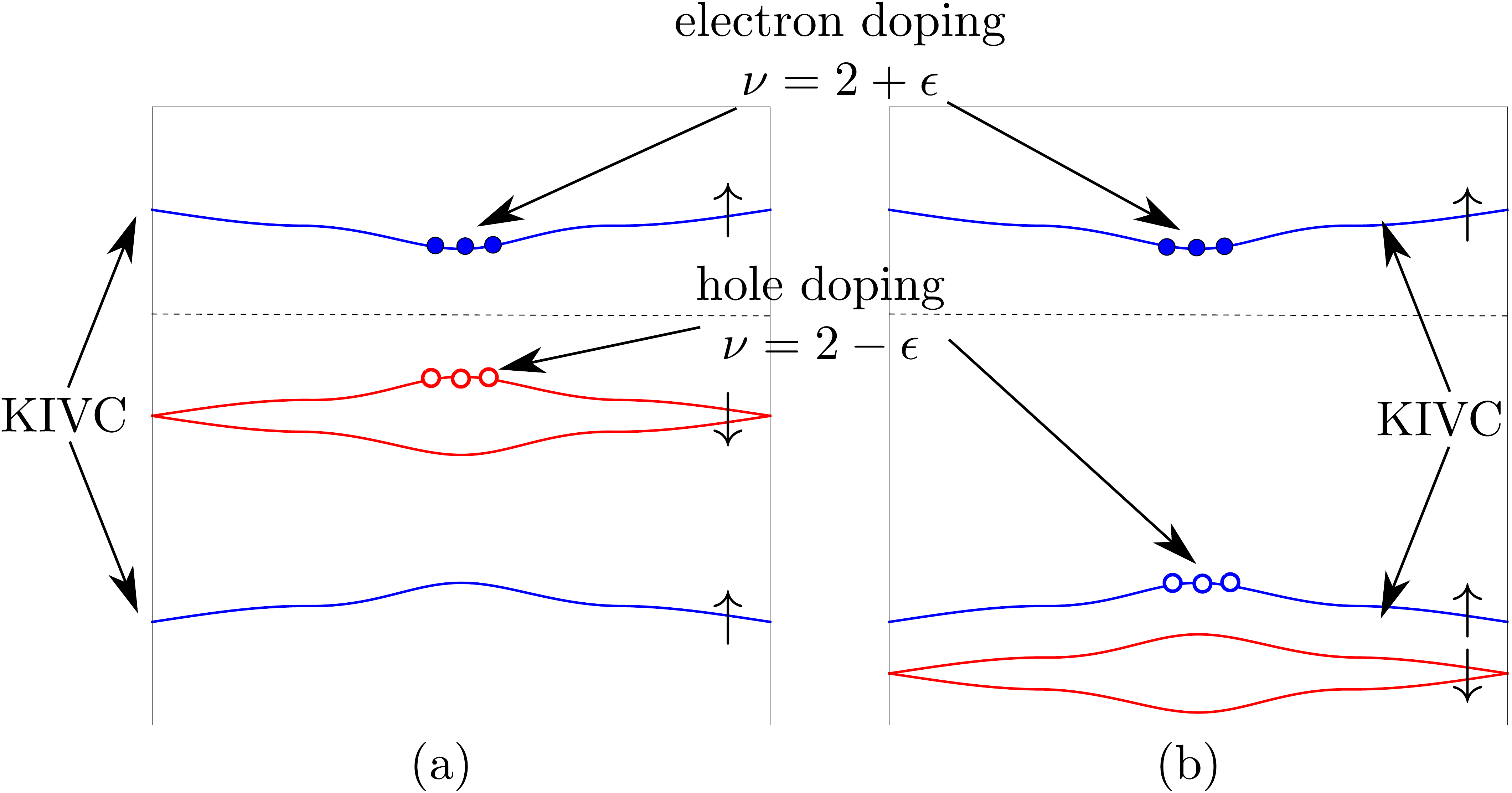}
    \caption{{\bf MATBG close to half-filling $\nu=2$}: Two possible scenarios for the spin-polarized insulating state at half-filling: (a) the opposite spin band lies inside the  gap of the psuedospin antiferromagnet (taken to be the K-IVC order here) or (b) the opposite spin band lies outside this gap. In both scenarios, doped electrons go into the upper K-IVC band. On the other hand, doped holes go into the opposite spin band for (a) but  into the same spin K-IVC band for (b). We note here that for simplicity, we considered the case where flavor polarization corresponds to full spin polarization. Another possible flavor polarized state would be the spin-valley locked state for which the same considerations apply.}
    \label{fig:HFDoping}
\end{figure}

It is worth noting that while the spinless model is particle-hole symmetric, we do not expect particle-hole symmetry in the real system in the vicinity of half-filling. To see this, note that  there are two scenarios for such an insulator shown schematically in Fig.~\ref{fig:HFDoping} where the opposite spin band lies outside or inside the Dirac gap. In the first scenario (a), electron (hole) doping at $\nu = 2 (-2)$ resembles doping a spinless version of the charge neutrality state whereas hole (electron) doping resembles doping a spinless version of the full (empty) band structure whereas in the second scenario (b), there is no difference between particle and hole doping at half-filling and both reduce to a spinless version of the doped state near charge neutrality. The first scenario is strongly supported by the quantum oscillation measurements where the Landau fans are only observed at $\nu = \pm (2 + \epsilon)$ but not at $\nu = \pm (2 - \epsilon)$ with a degeneracy that is half that observed at charge neutrality \cite{PabloSC, Dean-Young, efetov}. Further support of this picture is provided by the cascade transition picture based on recent compressibility \cite{CascadeShahal} as well as STM \cite{CascadeYazdani} data, which suggests that the $\nu = \pm (2 + \epsilon)$ state is a flavor polarized version of the one near charge neutrality.

The pairing symmetry favored by skyrmion condensation (in the spinless model) is an intervalley singlet, with same sublattice pairing i.e. $ \hat{\Delta} = \psi^\dagger_{\sigma\tau}\tau^y_{\tau'\tau}\psi^\dagger_{\sigma\tau'}$. In order to embed the spinless model within spinful TBG at $\nu=2$, we need to polarize a flavor as explained above. If the polarized flavor is just spin, then we are left with pairing between equal spins. On the other hand, if the polarized flavor is more complex, for example a spin valley locked combination, the corresponding spin structure of the Cooper pair is also more involved.  A general discussion of the spinful model is contained in the supplementary material.

\section{Discussion}

Let us compare and contrast the present work with previous theoretical discussions of skyrmion superconductivity. Starting with  Dirac fermions,   \cite{AbanovWiegmann,GroverSenthil} pointed out that  charge 2e skyrmions can arise when  interactions spontaneously create topological insulators from  symmetry breaking. Further, if these  skyrmion textures are the lowest energy charge excitations,  refs. \cite{GroverSenthil, Assaad,Palumbo2015} argued that doping could lead to superconductivity.  
 Here, we have examined a microscopic model with predominantly repulsive interactions and shown that it exhibits a related topological ground state with  low energy charge 2e skyrmions.  Note, restricting to repulsion is necessary to establishing an electronic mechanism of superconductivity.  To the best of our knowledge,  this collection of properties  has not  been established before \cite{Rachel} in a microscopic model with repulsive interactions.  A distinction from the previous Dirac models is that the 2e skyrmion here is  is actually a composite of a skyrmion anti-skyrmion pair. Furthermore, our model,  is motivated by magic angle graphene which seems to naturally incorporate the requisite physics. We also have calculated the energetics of skyrmions and shown that in a range of parameters they are the lowest energy charge excitations. On doping skyrmions we obtain a superconductor, whose transition temperature we calculate.      In our theory an essential role is played by $J$, which induces pairing between opposite skyrmion textures, and arises from electron tunneling between Chern sectors. In particular, we provide a scenario where pairing can be achieved in the absence of any extrinsic attractive pairing mechanism e.g. phonons. In addition, our mechanism explains how the large Coulomb repulsion can be evaded without screening or retardation allowing for the smaller pseudospin antiferromagnetic coupling $J \sim h^2/U \sim 1$ meV to overcome the  larger Coulomb repulsion as summarized in Figure \ref{fig:ChargedSkx}.  

The role of the coupling $J$ as the source of pairing can account for the absence of superconductivity in MATBG samples which are aligned with the hexagonal Boron Nitride (hBN) substrate. The latter generates a sublattice potential which breaks $C_2 \T$ symmetry by inducing an energy gap of about $15-20$ meV between the bands polarized on sublattices A and B \cite{sharpe2019emergent, Bultinck19, ZhangSenthil}. This acts as an opposite Zeeman field in the opposite Chern sectors (a Zeeman field for the vector $\bn$) which (a) shrinks the charge $2e$ skyrmions and makes them energetically less favorable and (b) introduces an
 energy detuning between previously resonant bands suppresses the tunneling term responsible for the coupling $J$. This mechanism is also absent in other moir\'e systems lacking $C_{2}$ symmetry such as twisted double bilayer graphene \cite{TDBGexp2019, PabloTDBG, IOP_TDBG} and ABC graphene on hBN \cite{FengWang-TLGonBN-SC} which can also be viewed as consisting of two quantum Hall systems with opposite Chern numbers. In addition to the suppression of $J$ due to broken $C_2$ symmetry, the direct tunneling $h$ is also strongly suppressed in these systems since the two opposite Chern sectors reside in opposite valleys and are thus very weakly coupled due to valley-charge conservation. Apart from pristine MATBG samples, the other platform that retains the relevant $C_2 \T$ symmetry is twisted trilayer (and multilayer) graphene with alternating twist as described in Ref.~\cite{Khalaf2019}. In addition to having the same symmetries as MATBG, it was shown \cite{Khalaf2019} that the wavefunctions of these systems can be mapped exactly to those of MATBG making them very attractive candidates for psuedospin antiferromagnetic order and skyrmion mediated superconductivity,  and are promising platforms for future theoretical and experimental studies. 

{We have noted the key role played by $C_2$ symmetry, which is important to defining a degenerate pusedospin.  
A natural question that arises is whether disorder, which is expected to to locally break $C_2$, would play an antagonistic role. However, it is important to note that in contrast to the global  breaking of $C_2$, the symmetry is not broken on average by disorder. For our mechanism, we only need $C_2$ to hold on distances comparable to the moir\'e scale. Disorder that preserves $C_2$ on average and is sufficiently smooth on the moir\'e scale is therefore not expected to significantly affect our conclusions. }

It is worth emphasizing that the superconducting mechanism we  discussed here is applicable to a much wider class of systems than just magic angle graphene.   The key requirements are a pair of quantum Hall (or Chern) ferromagnets, related by time reversal symmetry, and coupled to one another by tunneling. This leads, via a superexchange process, to an antiferromagnetic coupling $J$  which is maximized when tunneling connects states at the same energy. 

Let us mention a few other promising systems that incorporate all these ingredients. First, we have already briefly mentioned the multilayer graphene platform as described in Ref.~\cite{Khalaf2019} where each successive layer is twisted by an alternating angle  (i.e. $\theta,-\theta,\theta\dots$), and where alternating layers, which are at the same angle, are in registry.  These systems have shifted magic angles but otherwise retain all essential symmetries and are promising platforms for realizing skyrmion superconductivity. For the simplest alternating angle trilayer case, an additional ingredient is a  coexisting Dirac node which intersects the flat bands, but which is not expected to significantly alter the physics due to its rapid dispersion~\cite{Khalaf2019}.  
A second route involves strain induced Landau levels in graphene which have been observed \cite{Crommie2010}, with opposite Chern number in opposite valleys. Here spin plays the  role of our pseudospin, although translation symmetry blocks tunneling between opposite valleys.  Activating tunneling with an appropriate translation symmetry breaking order (eg. a substrate with charge density wave order), can help realize the skyrmion superconductor on doping. Other potential realizations include any spinful valley-Chern insulator, e.g. twisted double bilayer graphene, where again the crucial additional ingredient absent in current platforms is the presence of a source of translation symmetry breaking that opens up intervalley tunneling. More generally, 2D materials with electronic bands harboring the same fragile topology as magic angle graphene\cite{Po2018,Po2018faithful,ahn2018failure,Song},  will be interesting future candidates if they can be also brought into the  correlated regime. Alternately, consider a pair of Haldane models with opposite Chern numbers,  with spin degeneracy. Here spins play the role of our pseudospin. We note that such a model, dubbed the "shift insulator" has  been recently studied in the context of crystalline topology \cite{Shift}. Further, adding repulsive interactions and tunneling between the opposite Chern sectors  should realize the physics studied here when placed near quarter filling. Pursuing materials realization of these models, such as bilayers of anomalous quantum Hall ferromagnets, will be an interesting direction for future work.

{Finally, let us briefly discuss  our proposal in light of  recent experiments on MATBG superconductivity. First, although we have chosen to discuss our scenario in terms of doping a parent psuedospin-ordered insulator, the presence of the insulating gap is not essential. While an insulating state serves as proxy for order, it is by no means necessary, since an ordered metallic phase can also obtain at integer filling,  due to low-energy electron-hole pockets. The essential features of the skyrmion pairing mechanism however remain intact, although a theoretical description must now include  fermions in addition to the bosonic $\bn$ degrees of freedom.  Hence, our results are consistent with recent experimental reports of superconductivity in the absence of correlated insulating behavior \cite{YoungScreening, EfetovScreening} - an interesting question there is whether psuedospin order is present in the metallic state near integer filling. We note that in the  calculation of Sec.~4  we find a superconducting phase with doping in the absence of long range psuedospin order. Thus strictly, the only real requirement is short-ranged psuedospin order in which skyrmions can be  defined. 
Second, an important energy scale in our theory is $J \sim t^2/U$ which controls the strength of superconductivity.  This may provide a conceptual framework to help design better superconductors. For example, it is helpful to remember that in the interacting problem, not just  $U$ but also  $t$ increase with increasing the Coulomb scale, since the flat band dispersion is mostly generated by the interaction as discussed in \cite{KIVCpaper}. }

The discovery of superconductivity  in MATBG came as a  bolt from the blue, and it would be surprising if an explanation only invoked conventional ingredients. Instead, we believe this observation calls for an unusual mechanism like the skyrmion pairing one described here, which relies on  characteristics  unique to MATBG. More generally, we have identified a new mechanism for superconductivity from repulsion, and it will be interesting to explore in the future a variety of other settings that realize the essential ingredients of Chern ferromagnets that are tunnel-coupled to their time reversed conjugates. 

\emph{Note}--- We would like to draw the reader’s attention to Refs.~\cite{SkDMRG, Christos} on related topics, that appeared after this wok was first released.

\section{Acknowledgments} 
We thank Shang Liu for a related previous collaboration and S. Sachdev for helpful feedback. SC thanks V. Lohani for helpful discussions. AV was supported by a Simons Investigator award and by the Simons Collaboration on Ultra-Quantum Matter, which is a grant from the Simons Foundation (651440, AV). EK was supported by a Simons Investigator Fellowship, by NSF-DMR 1411343, and by the German National Academy of Sciences Leopoldina through grant LPDS 2018-02 Leopoldina fellowship.  SC acknowledges support from the ERC synergy grant UQUAM via E. Altman. MZ was supported by the Director, Office of Science, Office of Basic Energy Sciences, Materials Sciences and Engineering Division of the U.S. Department of Energy under contract no. DE-AC02-05-CH11231 (van der Waals heterostructures program, KCWF16)

\pagebreak
\widetext
\begin{center}

\textbf{\large Supplemental material:\\ Charged Skyrmions and Topological Origin of Superconductivity in Magic Angle Graphene}
\end{center}
\setcounter{equation}{0}
\setcounter{figure}{0}
\setcounter{table}{0}
\setcounter{page}{1}
\makeatletter
\renewcommand{\theequation}{S\arabic{equation}}
\renewcommand{\thefigure}{S\arabic{figure}}
\renewcommand{\bibnumfmt}[1]{[S#1]}
\setcounter{section}{0}


\section{Derivation of the effective field theory}
\label{App:NLSM}
In this section, we present the derivation of the effective field theory describing the low-lying excitations (spin waves and skyrmions) for a pair of tunnel-coupled $\U(n)$ quantum hall system related by $C_2 \T$ symmetry. This describes the manifold of low-energy states in TBG close to charge neutrality $n=2$ and half-filling $n=1$ \cite{KIVCpaper}. Although we focus in the main text on the spinless ($n=1$) case, our derivation will be done for arbitrary $n$. 

The Hamiltonian is given by (see Ref.~\cite{KIVCpaper} for details)
\begin{gather}
\H_{\rm eff} = h + \H_{\rm int}, \qquad h = \sum_\bk c_\bk^\dagger h(\bk) c_\bk, \qquad \H_{\rm int}  = \frac{1}{2A} \sum_\bq V_\bq \delta \rho_{-\bq} \delta \rho_{\bq} \\
\delta \rho_{\bq} = \rho_{\bq} - \bar \rho_{\bq},  \qquad \rho_{\bq} = \sum_\bk c_\bk^\dagger \Lambda_{\bq}(\bk) c_{\bk + \bq}, \qquad \bar \rho_{\bq} = \frac{1}{2}\sum_{\bk,\bG} \delta_{\bq,\bG} \tr \Lambda_{\bG}(\bk)
\end{gather}
Here, $\bk$ denotes a momentum in the first Brillouin zone, $\bq$ denotes a generic (unrestricted) momentum, and $\bG$ denotes a reciprocal lattice momentum. In the following, we will use the basis of Chern sector and pseudospin defined in the main text given by
\beq
\gamma_{x,y,z} = (\sigma_x, \sigma_y \tau_z, \sigma_z \tau_z), \qquad \eta_{x,y,z} = (\sigma_x \tau_x, \sigma_x \tau_y, \tau_z)
\label{GammaEta}
\eeq
In this basis, the Chern number is given by $C = \gamma_z$, the tunneling term is
\beq
h(\bk) = h_x(\bk) \gamma_x + h_y(\bk) \gamma_y
\eeq
The density operator $\rho_\bq$ (as well as the average density $\bar \rho_\bq$) can be split into a chiral symmetric piece and a chiral symmetry-breaking piece with the latter being only non-vanishing away from the chiral limit \cite{Tarnopolsky, KIVCpaper}. The main approximation involved in this work is to neglect the effect of the chiral symmetry breaking form factor which was shown to be relatively small in Ref.~\cite{KIVCpaper}. The form factor then reduces to the simple expression
\beq
\Lambda_\bq(\bk) = F_\bq(\bk) e^{i \phi_\bq(\bk) \gamma_z}
\eeq

We begin by considering translationally symmetric Slater determinant states at half-filling (which describes charge neutrality for $n=2$ and the spinless model for $n=1$) which can be described in terms of the projection operator $P(\bk)$
\beq
P(\bk) = \frac{1}{2}(1 + Q(\bk)), \qquad Q^2 = 1, \qquad \tr Q = 0
\eeq
The states which minimize $\H_{\rm int}$ are Slater determinant states characterized by the $\bk$-independent order parameter $Q$ commuting with $\gamma_z$. A fixed ground state $|\Psi_0 \rangle$ is specified by a given $Q_0^2 = 1$, $\tr Q_0 = 0$ and $[Q_0, \gamma_z] = 0$. Projected onto the manifold of ground state the effective energy functional has two contributions: (i) the tunneling term $h$ which breaks the $\U(2n) \times \U(2n)$ symmetry down to $\U(n)$, (ii) slowly varying fluctuations in the order parameter $Q_0$ that cost small energy.

In the spinless limit $n=1$ which we consider in this work, the matrix $Q$ can be written in terms of two unit vectors $\bn_\pm$ in the $\pm$ Chern sectors as
\beq
Q = \left(\begin{array}{cc}\bn_+ \cdot {\boldsymbol \eta} & 0 \\ 0 & \bn_-(\br) \cdot {\boldsymbol \eta} \end{array} \right)_{\gamma_z = \pm}
\label{Qpm}
\eeq

\subsection{Order parameter fluctuations}

The effective action associated with slow variations of the order parameter in space can be derived following Ref.~\cite{MoonMori} with the generalization to the SU(N) case from Ref.~\cite{Arovas99}. The manifold of low energy excitations contains the massless spin waves as well as charged topological excitations, skyrmions, which cost a finite energy . In computing the cost of slow variations of the order parameter, we will only consider the effect of $\H_{\rm int}$ since the other terms are smaller corrections.

The order parameter fluctuations are parametrized using the generators of the $\U(2n) \times \U(2n)$ symmetry group of the interaction term $\H_{\rm int}$ which we denote by $t^\mu$, $\mu = 1,\dots,4n^2$. These can be obtained from the generators of $\U(4n)$ by restricting to those which commute with $\gamma_z$. The effect of the order parameter fluctuations can be taken into account by writing
\beq
|\Psi(w)\rangle = e^{-i \hat O} |\Psi_0 \rangle, \qquad \hat O = \sum_{\mu} \int d^2 \br w^\mu(\br) \psi^\dagger(\br) t^\mu \psi(\br)
\eeq
Here the sum over $\mu$ goes over the generators of $\U(2n) \times \U(2n)$ \emph{excluding} those which leave the ground state $Q_0$ invariant. The latter corresponds to the subgroup $\U(n) \times \U(n) \times \U(n) \times \U(n)$. This means that we restrict the sum to $\{t^\mu, Q_0\} = 0$. $\psi(\br)$ is the creation operator for the microscopic electron at point $\br$, it can be written in terms of the operator $c_{\alpha,\bk}$ which creates a state in the flat band labelled by $\alpha$ (which for the spinless case corresponds to a combined valley-sublattice index $\alpha = (\tau,\sigma)$) at momentum $\bk$ as
\beq
\psi(\br) = \sum_{\alpha,\bk} \phi_{\alpha,\bk}(\br) c_{\alpha,\bk}, \qquad  \int d^2 \br \phi^\dagger_{\alpha,\bk}(\br) \phi_{\beta,\bk'}(\br) = \delta_{\alpha \beta} \delta_{\bk, \bk'}
\label{Psir}
\eeq
Here, $\phi_{\alpha,\bk}(\br)$ are the Bloch wavefunctions written in terms of the periodic wavefunctions $u_{\alpha,\bk}(\bk)$ as 
\beq
\phi_{\alpha,\bk}(\br) = e^{i \bk \cdot \br} u_{\alpha,\bk}(\br), 
\eeq
Introducing the Fourier transform
\beq
w^\mu(\br) = \sum_\bq e^{i \bq \cdot \br} w^\mu_\bq, 
\eeq
leads to
\beq
\hat O = \sum_{\bq,\mu} w^\mu_\bq \hat S^\mu_{-\bq},  \qquad \hat S^\mu_\bq = \sum_{\alpha,\beta,\bk} c_{\alpha,\bk}^\dagger c_{\beta,\bk + \bq} \langle u_{\alpha,\bk}| t^\mu| u_{\beta,\bk + \bq} \rangle
\label{Ohat}
\eeq
We can now rewrite the action of $t^\mu$ in terms of the corresponding projected operator $T^\mu$ defined as $t^\mu |u_{\alpha,\bk} \rangle = \sum_\beta T^\mu_{\alpha,\beta} |u_{\beta,\bk} \rangle$ leading to 
\beq
\hat S^\mu_\bq = \sum_{\alpha,\beta,\eta,\bk} c_{\alpha,\bk}^\dagger c_{\beta,\bk + \bq} [\Lambda_\bq(\bk)]_{\alpha,\eta} T^\mu_{\beta,\eta} = \sum_\bk c_\bk^\dagger \Lambda_\bq(\bk) [T^\mu]^T c_{\bk + \bq}, \qquad [T^\mu, \gamma_z] = 0, \qquad \{ T^\mu, Q_0 \} = 0
\eeq
The expression above assumes $\U(2n) \times \U(2n)$ is a good symmetry and includes only the the symmetric part of the form factor. Although the form (\ref{Ohat}) is more reminiscent of the standard derivation for the SU(2) spin skyrmions, we will find it more convenient to express $\hat O$ in the slightly different form
\beq
\hat O = \sum_\bq \hat O_{\bq}, \qquad \hat O_{\bq} = \sum_\bk c_\bk^\dagger \Lambda_\bq(\bk) w_{-\bq}^T c_{\bk + \bq}, \qquad w_\bq = \sum_\mu w^\mu_\bq T^\mu, \qquad [w_\bq, \gamma_z] = \{w_\bq, Q_0\} = 0
\label{Ohatq}
\eeq
Thus, $w_\bq$ is a hermitian matrix parametrizing the tangent space to the ground state manifold at $Q_0$.

In order to simplify the evaluation of the different terms in the expansion in $\hat O$, we will derive a few simple identities below. We will be considering fermion bilinear operators
\beq
\hat A = \sum_{\alpha, \beta} c_\alpha^\dagger A_{\alpha,\beta} c_\beta
\eeq
where we use the hat notation $\hat A$ for the second quantized operator corresponding to the first quantized operator $A$. Using the relation
\beq
[c_\alpha^\dagger c_\beta, c_\gamma^\dagger c_\delta] = \delta_{\beta \gamma} c^\dagger_\alpha c_\beta - \delta_{\alpha \delta} c^\dagger_\gamma c_\beta
\eeq
we get
\beq
[\hat A, \hat B] = \widehat{[A,B]} 
\eeq
The ground state expectation values can be evaluated using the identities
\beq
\langle \hat A \rangle = \Tr P_0^T A, \qquad \langle \hat A \hat B \rangle = \Tr A B P_0^T + \Tr A P_0^T \Tr B P_0^T - \Tr A P_0^T B P_0^T
\label{ExpVal}
\eeq
where
\beq
[P_0]_{\alpha,\beta} = \langle c_\alpha^\dagger c_\beta \rangle, \qquad P_0 = \frac{1}{2}(1 + Q_0)
\eeq
The traces in the above expressions go over all internal indices including momenta (we will use the symbol ``Tr'' with capital T to denote traces which involve momentum summation to distinguish them from those involving only flavor summation denoted by ``tr''). We notice that whenever $A$ or $B$ commutes with $P_0$, then the first and third terms in (\ref{ExpVal}) cancel leaving only the second term. Similarly, if $A$ or $B$ anticommute with $Q_0$, then the second and third terms vanish leaving only the first. This can be summarized as 
\beq
\langle \hat A \hat B \rangle = \begin{cases} 
\Tr A P_0^T \Tr B P_0^T &: [A,Q_0] = 0 \text{ or } [B,Q_0]=0 \\
\Tr A B P_0^T  &: \{A,Q_0\} = 0 \text{ or } \{B,Q_0\}=0 \\
0 &: \{A B, Q_0 \} = 0
\end{cases}
\label{ExpValSimp}
\eeq
In our notation, we can write the following
\beq
\rho_\bq = \hat \Lambda_\bq, \qquad [\Lambda_\bq]_{\bk,\bk'} = \delta_{\bk',\bk+\bq} \Lambda_\bq(\bk), \qquad O_\bq = \Lambda_\bq w_{-\bq}^T
\eeq

We note that since the order parameter manifold is isomorphic to the product of two Grassmanian manifolds $\U(2n)/\U(n) \times \U(n)$, each of these manifolds admits charged skyrmion textures since $\pi_2(\U(2n)/\U(n) \times \U(n)) = \Z$. The skyrmion charge can be evaluated directly using
\beq
\langle \Psi(w)| \delta \rho_\bq| \Psi(w) \rangle = -i\langle [\hat O, \rho_\bq] \rangle - \frac{1}{2} \langle [\hat O, [\hat O, \rho_\bq] \rangle
\eeq
The first order term vanishes since
\beq
\langle [\hat O, \rho_\bq] \rangle = \langle \widehat{[ O, \Lambda_\bq]} \rangle = \Tr P_0^T O \Lambda_\bq = 0
\eeq
which follows from the fact that $O \Lambda_\bq$ anticommutes with $Q_0$. The second order term can be evaluated as
\beq
\delta \rho_\bq^{(2)} = -\frac{1}{2} \langle [\hat O, [\hat O, \rho_\bq]] \rangle = -\frac{1}{2}\langle \widehat{[ O, [O, \Lambda_\bq]]} \rangle = \Tr P_0^T [ O, [O, \Lambda_\bq]] = -\frac{1}{2}\sum_{\bq'} \Tr (w_{\bq+\bq'} w_{-\bq'} Q_0)^T \Lambda_{\bq'} [\Lambda_{-\bq'-\bq}, \Lambda_\bq]
\label{drho2}
\eeq
Since the fluctuations $w_\bq$ are slowly varying in space, we can expand to leading order in small momenta $\bq$ and $\bq'$. To simplify the expansion, we note that
\beq
\Lambda_\bq(\bk) = 1 + i \bq \cdot \bA(\bk) + O(\bq^2), 
\eeq
where $\bA(\bk) = \diag(\bA^+(\bk), \bA^-(\bk))_{\gamma_z}$ with $\bA^\pm(\bk)$ denoting the Berry connection for the $\pm$ Chern sectors $\gamma_z = \pm 1$. This leads to the following identity
\beq
[\Lambda_\bq, \Lambda_{\bq'}] = (\bq \cdot \nabla) (i \bq' \cdot \bA) -  (\bq' \cdot \nabla) (i \bq \cdot \bA) + O(\bq^3) = i (\bq \wedge \bq') (\nabla \times \bA) + O(\bq^3) = i (\bq \wedge \bq') \Omega(\bk) + O(\bq^3)
\eeq
where $\Omega(\bk)$ denotes the Berry curvature.

Substituting in (\ref{drho2}) yields
\beq
\langle [\hat O, [\hat O, \rho_\bq] \rangle = i  \sum_{\bq'} (\bq' \wedge \bq ) \tr Q_0 w_{-\bq'} w_{\bq+\bq'} \sum_{\bk} \Omega(\bk) = \frac{i}{2\pi} \int d^2 \br \tr \gamma_z Q_0 \nabla_\br w(\br) \wedge \nabla_\br w(\br)
\eeq
where we used the fact that $\sum_\bk \Omega^\pm(\bk) = \frac{A}{(2\pi)^2} \int_{\rm BZ} d^2\bk \, \Omega^\pm(\bk) = \pm \frac{A}{2\pi}$. Thus, the charge density associated with a skyrmion is
\beq
\delta \rho(\br) = -\frac{i}{4\pi} \tr \gamma_z Q_0 \nabla_\br w(\br) \wedge \nabla_\br w(\br) =  -\frac{i}{4\pi} [\tr Q_{0,+} \nabla_\br w_+(\br) \wedge \nabla_\br w_+(\br) - \tr Q_{0,-} \nabla_\br w_-(\br) \wedge \nabla_\br w_-(\br)]
\label{drhor2}
\eeq
where we split $Q_0$ and $w(\br)$ into $\pm$ Chern sectors. Since this expression only depends on gradients of $w$, we can lift the assumption that $w$ is small used in the derivation and consider a general slowly varying deformation of the order parameter $Q(\br) = T(\br)^{-1} Q_0 T(\br)$ with $T(\br) = e^{-i w(\br)}$. We can then make the replacement $\nabla w(\br) \mapsto i (\nabla T) T^{-1}$. Substituting in (\ref{drhor2}) yields after a sequence of straightforward manipulations
\beq
\delta \rho(\br) = \frac{i}{16\pi} \epsilon_{ij} \tr \gamma_z Q(\br) \partial_i Q(\br)  \partial_j Q(\br) = \frac{i}{16\pi}\epsilon_{ij} [\tr Q_+(\br) \partial_i Q_+(\br)  \partial_j Q_+(\br) - \tr Q_-(\br) \partial_i Q_-(\br)  \partial_j Q_-(\br)]
\label{drhor}
\eeq
This expression says that a skyrmion with topoplogical winding $n$ is associated with charge $\pm n$ in the $\pm$ Chern sector.

In the spinless limit $n=1$, we can substitute the expression (\ref{Qpm})  in (\ref{drhor}), to get the more familiar expression
\beq
\delta \rho_\pm = \mp \frac{1}{8\pi} \int d^2 \br \epsilon_{ij} \bn_\pm(\br) \cdot  [\partial_i \bn_\pm(\br) \times \partial_j \bn_\pm(\br)]
\eeq

The effective Hamiltonian is obtained as
\beq
\langle \Psi(w)| \H| \Psi(w) \rangle = \langle \Psi_0| e^{i \hat O} \H e^{-i \hat O}| \Psi_0 \rangle = \langle \Psi_0| e^{i \ad_{\hat O}} \H | \Psi_0 \rangle = \langle \Psi_0| i \ad_{\hat O} \H - \frac{1}{2} \ad^2_{\hat O} \H + \dots | \Psi_0 \rangle, \qquad \ad_{\hat O} A = [\hat O,A]
\label{Hexp}
\eeq
where we used the factor that the Hamiltonian annihilates $|\Psi_0 \rangle$.  It will be useful to rewrite the Hamiltonian (\ref{Hexp}) as
\beq
\H = \frac{1}{2A} \sum_\bq V_\bq \rho_\bq \rho_{-\bq} - \frac{1}{A} \sum_\bG V_\bG \rho_{\bG} \bar \rho_{-\bG}
\label{HB}
\eeq
by separating the density-density part from the background part.

To evaluate the leading order term, we need to evaluate
\beq
\langle \ad_{\hat O} \rho_\bq \rho_{-\bq} \rangle  = \langle [\hat O,\rho_\bq \rho_{-\bq}] \rangle = \langle \widehat{[O, \Lambda_\bq]} \hat \Lambda_{-\bq} \rangle + \langle \hat \Lambda_\bq \widehat{[O, \Lambda_{-\bq}]} \rangle
\eeq
which follows from the standard identity of commutators
\beq
[A, B C] = [A, B] C + B [A, C] \quad \leftrightarrow \quad \ad_A B C = (\ad_A B) C + B \, \ad_A C
\label{Comm}
\eeq
Each term can be evaluated using (\ref{ExpVal}). From (\ref{ExpValSimp}), we can see that both terms vanish. In fact, this is the case for the expectation value of any product of operator containing an odd number of $\hat O$ operators. 

The second order term is evaluated by evaluating the expectation value
\beq
\langle \ad_{\hat O}^2\rho_\bq \rho_{-\bq} \rangle = \langle (\ad_{\hat O}^2\rho_\bq) \rho_{-\bq} \rangle + 2\langle (\ad_{\hat O}\rho_\bq) (\ad_{\hat O} \rho_{-\bq}) \rangle + \langle \rho_\bq \ad^2_{\hat O} \rho_{-\bq} \rangle
\label{ad2O}
\eeq
which can be obtained by repeatedly applying (\ref{Comm}). The first term yields
\beq
\langle (\ad_{\hat O}^2\rho_\bq) \rho_{-\bq} \rangle = \langle \widehat{[O,[O, \Lambda_\bq]]} \hat \Lambda_{-\bq} \rangle = \Tr P_0^T \Lambda_{-\bq} \Tr P_0^T [O, [O, \Lambda_\bq]] = \sum_\bG \bar \rho_{-\bG} \langle [\hat O, [\hat O, \rho_\bG]] \rangle
\eeq
which, together with the last term, cancels against the corresponding term in the background (second term in (\ref{HB})). The second term is
\beq
 \langle (\ad_{\hat O}\rho_\bq) (\ad_{\hat O} \rho_{-\bq}) \rangle = \langle \widehat{[ O,  \Lambda_\bq]} \widehat{[O, \Lambda_{-\bq}]} \rangle = \Tr P_0^T [ O,  \Lambda_\bq] [ O,  \Lambda_{-\bq}] = \sum_{\bq'} \Tr (w_{\bq'} w_{-\bq'} P_0)^T [\Lambda_{\bq'}, \Lambda_\bq] [\Lambda_{-\bq'}, \Lambda_{-\bq}]
\eeq
We notice that, unlike (\ref{drho2}) where both $\bq$ and $\bq'$ were small since the fluctuations are slowly varying in space, this expression only includes $w_{\pm \bq'}$ and as a result we should not assume that $\bq$ is small. Introducing the gauge invariant combination
\beq
M_\bk(\bq,\bq') = \Lambda_\bq(\bk) \Lambda_{\bq'}(\bk+\bq) - \Lambda_{\bq'}(\bk) \Lambda_\bq(\bk + \bq')
\eeq
We can write 
\beq
 \langle (\ad_{\hat O}\rho_\bq) (\ad_{\hat O} \rho_{-\bq}) \rangle = \frac{1}{2}  \sum_{\bk,\bq'} M_\bk(\bq,\bq') M_\bk(-\bq,-\bq') \tr w_{-\bq'} w_{\bq'}
\eeq
The corresponding energy is then given by
\beq
\delta E^{(2)} = -\frac{1}{2} \langle \ad_{\hat O}^2 \H \rangle =  \sum_{i,j=x,y} \rho_{ij} \int d^2 \br \tr \partial_i w(\br) \partial_j w(\br)
\eeq
where $\rho_{ij}$
\beq
\rho_{ij} = -\frac{1}{4 A^2} \frac{\partial^2}{\partial_{q'_i} \partial_{q'_j}} \sum_{\bk,\bq} V_q M_\bk(\bq,\bq') M_\bk(-\bq,-\bq') \Big|_{\bq'=0}
\eeq
Since we expect the continuum theory to be rotationally symmetric, we can assume $\rho_{xx} = \rho_{yy} = \R$ and $\rho_{xy} = \rho_{yx} = 0$ leading to the expresssion
\beq
\R = -\frac{1}{8 A^2} \nabla^2_{\bq'} \sum_{\bk,\bq} V_q M_\bk(\bq,\bq') M_\bk(-\bq,-\bq') \Big|_{\bq'=0}
\label{rhos}
\eeq
with the energy given by
\beq
\delta E^{(2)} = \frac{\R}{4} \int d^2 \br \tr [\nabla Q(\br)]^2 =  \frac{\R}{4} \int d^2 \br \tr\{ [\nabla Q_+(\br)]^2 + [\nabla Q_-(\br)]^2\}, 
\label{dE2}
\eeq

One possible simplification of (\ref{rhos}) was suggested in \cite{Chatterjee19} by assuming that the magnitude of the form factor $F_\bq(\bk)$ decays relatively quickly with increasing $|\bq|$ and depends weakly on $\bk$ such that the terms containing the derivative $\nabla_\bk F_\bq(\bk)$ can be ignored. This justifies expanding in small $\bq$ using $\Lambda_\bq(\bk) = F_\bq(\bk) [1 + i \bq \cdot \bA(\bk)]$  leading to
\beq
M_\bk(\bq,\bq') = i \epsilon_{ij} q_i q'_j F_\bq(\bk) \Omega(\bk)
\eeq
which gives
\beq
\R = \frac{1}{8 A^2} \sum_\bq V_\bq \bq^2 \sum_\bk F_\bq(\bk)^2 \Omega(\bk)^2
\eeq
in agreement with \cite{Chatterjee19}.

The next order term is the fourth order. This term is presumably smaller than the second order term since it has more derivatives. However, it has a part that becomes relevant if the potential $V(\br)$ is sufficiently long-ranged (weak or no screening limit). To see this, we write
\beq
\langle \ad_{\hat O}^4 \rho_\bq \rho_{-\bq} \rangle = \langle (\ad_{\hat O}^4 \rho_\bq) \rho_{-\bq} \rangle + 4 \langle (\ad_{\hat O}^3 \rho_\bq) (\ad_{\hat O} \rho_{-\bq}) \rangle + 6 \langle (\ad_{\hat O}^2 \rho_\bq) (\ad_{\hat O}^2 \rho_{-\bq}) \rangle + 4 \langle (\ad_{\hat O} \rho_\bq) (\ad_{\hat O}^3 \rho_{-\bq}) \rangle + \langle \rho_\bq (\ad_{\hat O}^4 \rho_{-\bq}) \rangle
\eeq
The first and last terms vanish against the background. The second and fourth terms include the product of two operators which each anticommute with $Q_0$ which implies that the ground state expectation value yields a single trace (cf.~Eq.~\ref{ExpValSimp}). As a result, the expression has the form $\sim w_{\bq_1} w_{\bq_2} w_{\bq_3} w_{-\bq_1-\bq_2-\bq_3}$ which enables taking the $\bq$ dependence out of the trace similar to the manipulations leading to (\ref{dE2}). The resulting expression is local in real space and contains at least four derivative terms. This term is obviously smaller than the gradient term (\ref{dE2}) and can be safely neglected. On the other hand, the third term is the product of two even terms under $Q_0$ which yields a product of Traces leading to the form $\sim w_{\bq_1} w_{-\bq - \bq_1} w_{\bq_2} w_{\bq - \bq_2}$. In real space, this couples to the gradients of the order parameter at different points as we will see below. This means that such term can be important for sufficiently long range interactions.

Using (\ref{ExpValSimp}), we find
\beq
\langle (\ad_{\hat O}^2 \rho_\bq) (\ad_{\hat O}^2 \rho_{-\bq}) \rangle = \Tr P_0^T [O, [O, \Lambda_\bq]] \Tr P_0^T [O, [O, \Lambda_{-\bq}]] = 4 \delta \rho_\bq \delta \rho_{-\bq}
\eeq
where we used (\ref{drho2}). The corresponding correction to the Hamiltonian is
\beq
\delta E^{(4)} = \frac{1}{2} \int d^2 \br d^2 \br' \delta \rho(\br) V(\br - \br') \delta \rho(\br')
\eeq

We can now summarize the results of this section as
\begin{gather}
\delta E = \langle \Psi(w)| \H - \mu \delta \rho |\Psi(w) \rangle = \frac{\R}{4} \int d^2 \br \tr [\nabla Q(\br)]^2 + \frac{1}{2} \int d^2 \br d^2 \br' \delta \rho(\br) V(\br - \br') \delta \rho(\br') - \mu e \int d^2 \br \delta \rho(\br), \\
\delta \rho(\br) = \frac{i}{16\pi} \epsilon_{ij} \tr \gamma_z Q(\br) \partial_i Q(\br)  \partial_j Q(\br) = \delta \rho_+(\br) + \delta \rho_-(\br), \qquad \delta \rho_\pm(\br) = \pm \frac{i}{16\pi} \epsilon_{ij} \tr Q_\pm(\br) \partial_i Q_\pm(\br)  \partial_j Q_\pm(\br) 
\end{gather}

\subsection{Effect of tunneling $h$}
The dispersion term $h$ is off-diagonal in the Chern basis $h \propto \gamma_{x,y}$. This means that 
\beq
\langle \ad_{\hat O}^n h \rangle = 0
\eeq
for any $n$. That is, the energy does not recerive any corrections to first order in $h$. This is easy to understand since $h$ creates a particle-hole pair between bands with opposite Chern number. Such excitations are massive and cannot be described by slow variations of the order parameter within the manifold of low-energy states. The effect of $h$ can be captured by allowing for massive excitations which take $Q_0$ out of the ground state manifold. In the following, we will restrict ourselves to massive excitations which can be created by $h$ which are given by
\beq
|\Psi_0\rangle \mapsto |\Psi(M)\rangle = e^{-i \hat M} |\Psi_0\rangle, \qquad \hat M = \sum_\bk c^\dagger_\bk M_\bk c_\bk, \quad \{M_\bk, \gamma_z\} = \{M_\bk,Q_0 \} = 0
\eeq
The energy associated with these fluctuations can be evaulated by writing a very similar expanstion to (\ref{Hexp}) with $\hat O$ replaced by $\hat M$. Similarly, the first order term in the expansion vanishes and the second order term can be expanded as in (\ref{ad2O}) with the first and last terms cancelling against the background. The resulting expression then simplifies to
\begin{align}
E^{(2)}_M &= -\frac{1}{4A}\sum_\bq V_\bq \langle \ad_{\hat M}^2 \rho_\bq \rho_{-\bq} \rangle =  -\frac{1}{4A}\sum_\bq V_\bq \Tr P_0^T [M, \Lambda_\bq] [M, \Lambda_{-\bq}] = \frac{1}{4A}\sum_\bq V_\bq \Tr P_0^T [M, \Lambda_\bq] [M, \Lambda_{-\bq}] \nonumber \\
&=\frac{1}{2A}\sum_{\bq,\bk} V_\bq F_\bq(\bk)^2 \tr P_0^T \{M_\bk^2 - M_\bk M_{\bk+\bq} [\Lambda^\dagger_\bq(\bk)]^2\} = \frac{1}{4A}\sum_{\bq,\bk} V_\bq F_\bq(\bk)^2 \tr \{M_\bk^2 - M_\bk M_{\bk+\bq} [\Lambda^\dagger_\bq(\bk)]^2\}
\label{E2M}
\end{align}
where we used the fact that $M_\bk$ anticommutes with $Q_0$ in the last line. We can parametrize $M_\bk$ as
\beq
M_\bk = \sum_\mu r^\mu (m_{\mu,\bk} \gamma_+ + m^*_{\mu,\bk} \gamma_-), \qquad Q_+ r^\mu Q_- = - r^\mu, \qquad \tr r^\mu r^\nu = \delta_{\mu,\nu}
\label{rmu}
\eeq
where $r^\mu$ are the generators of $\U(2n)$ satisfying $Q_+ r^\mu Q_- = - r^\mu$ (The map $M \mapsto Q_+ M Q_-$ is a linear map on the space of $2n \times 2n$ matrices which squares to the identity, so it induces a decomposition on the space of matrices into even and odd subspaces under this operator. $r^\mu$ are taked to be the generators of the odd subspace). Here, $m_{\mu,\bk}$ are complex functions of $\bk$. Substituting in (\ref{E2M}) yields
\beq
E^{(2)}_M = \frac{1}{2} \sum_{\bk,\bk',\mu} m^*_{\mu,\bk} R_{\bk,\bk'} m_{\mu,\bk'}, \qquad R_{\bk,\bk'} = \frac{1}{A}\sum_{\bq} V_\bq F_\bq(\bk)^2 \{\delta_{\bk,\bk'} - \delta_{\bk',\bk+ [\bq]}e^{2i \phi_\bq(\bk)}\}
\eeq
where $[\bq]$ denotes the component of $\bq$ in the first Brillouin zone. We now compute the leading correction in $h$ given by
\beq
E^{(1)}_h = -i \langle [\hat M, \hat h] \rangle = -i \Tr P_0^T [M, h] = -i \sum_\bk \tr Q_0 M_\bk h(\bk) = -i \sum_{\bk,\mu} (m_{\mu,\bk} z^*_\bk - m^*_{\mu,\bk} z_\bk) \tr Q_+ r^\mu, \quad z_\bk = h_x(\bk) + i h_y(\bk)
\eeq
where we used the properties of the generators (\ref{rmu}). Thus the effective Hamiltonian for the massive excitations $m_{\mu,\bk}$ is given by
\beq
\H_M = \frac{1}{2} \sum_{\bk,\bk',\mu} m^*_{\mu,\bk} R_{\bk,\bk'} m_{\mu,\bk'} -i \sum_{\bk,\mu} (m_{\mu,\bk} z^*_\bk - m^*_{\mu,\bk} z_\bk) \tr Q_+ r^\mu
\eeq
We can now integrate out the massive terms to get
\beq
\H_h = -2 \sum_{\bk,\bk'} z_\bk^* (R^{-1})_{\bk,\bk'}  z_\bk \sum_\mu \tr Q_+ r^\mu \tr Q_+ r^\mu 
\label{Hh}
\eeq
To evaluate the product of traces summed over the generators $r^\mu$, we use the Fierze identity for the generators of $\U(N)$ given by
\beq
\sum_\mu t^\mu_{\alpha \beta} t^\mu_{\gamma \delta} = \delta_{\alpha \delta} \delta_{\beta \gamma}
\eeq
which when restricted to generators satisfying (\ref{rmu}) yields
\beq
\sum_\mu \tr A r^\mu \tr B r^\mu = \frac{1}{4}[\tr A B + \tr Q_+ A Q_- Q_+ B Q_- - \tr Q_+ A Q_- B - \tr A Q_+ B Q_-]
\eeq
Substituting in (\ref{Hh}) then yields
\beq
\H_h = -A J (\tr 1 - \tr Q_+ Q_-) = A J \tr Q_+ Q_- + \text{const.} = J \int d^2 \br \tr Q_+(\br) Q_-(\br), \qquad J = \frac{1}{A} \sum_{\bk,\bk'} z_\bk^* (R^{-1})_{\bk,\bk'}  z_\bk
\eeq
This result means that integrating out the massive modes yields an antiferromagnetic coupling between the $+$ and $-$ Chern sectors which favors $Q_+ = - Q_-$.

\subsection{Dynamical term}
To obtain the full effective action, we need also to include the dynamical term obtained by allowing the fluctuations $\hat O$ to be (imaginary) time-dependent
\beq
\langle \Psi(w)| \partial_\tau| \Psi(w) \rangle = -\frac{1}{2}\langle \ad^2_{\hat O} \partial_\tau \rangle = -\frac{1}{2} \langle [\hat O(\tau),\partial_\tau \hat O(\tau)] \rangle =- \frac{N}{2} \tr P_0^T [w_0^T(\tau), \partial_\tau w^T_0(\tau)] = -\frac{N}{2} \tr Q_0 w_0 \partial_\tau w_0(\tau)
\eeq
Writing that $Q(\br,\tau) = T^{-1}(\br,\tau) Q_0 T(\br,\tau)$, we notice that this term matches the leading term in the expansion of $\frac{1}{2} \tr T^{-1} Q_0 \partial_\tau T$. This is in fact the well-known one-dimensional Wess-Zumino term which is actually the unique expression that satisfies the required symmetries. Let us write it as
\begin{align}
S_{\tau} &= \frac{1}{2} \sum_\bR \int_0^\beta d\tau  \tr \{T_+(\bR)^{-1} Q_{0,+} \partial_\tau T_+(\bR) +  T_-^{-1}(\bR) Q_{0,-} \partial_\tau T_-(\bR)\} \nonumber \\ &= \frac{1}{2A_M} \int d^2 \br d\tau  \tr \{T_+(\br)^{-1} Q_{0,+} \partial_\tau T_+(\br) +  T_-(\br)^{-1} Q_{0,-} \partial_\tau T_-(\br)\}
\label{St}
\end{align}
It is well-known that this term cannot be written in a gauge invariant way in terms of $Q = T^{-1} Q_0 T$. To see this, we notice that the operator $T$ has a gauge freedom $T(\br,\tau) \mapsto K(\br,\tau) T(\br,\tau)$ for any matrix $K(\br,\tau)$ which commutes with $Q_0$. The matrix $T$ can be any matrix in $\G = S(\U(2n) \times \U(2n))$. The matrix $Q$ on the other hand parametrizes the ground state manifold which is given by the coset space $\G/\K$ where $\K$ is the subgroup of $\G$ which commutes with $Q_0$ (which is isomorphic to $\U(n) \times \U(n) \times \U(n) \times \U(n)$). All other terms in the action are manifestly gauge invariant since they are written explicitly in terms of $Q$. The Wess-Zumino term (\ref{St}) on the other hand can only be written in a gauge invariant way by introducing an auxiliary dimension which we will not do here. Its transformation under gauge transformations $T(\br,\tau) \mapsto K(\br,\tau) T(\br,\tau)$ is given by (for simplicity, consider the $+$ sector and drop the subscript and consider a fixed position $\br = \bR_0$)
\begin{align}
S_\tau &= \frac{1}{2} \int_0^\beta d\tau \tr  T^{-1} Q_{0} \partial_\tau T \mapsto \frac{1}{2} \int_0^\beta d\tau \tr  T^{-1} K^{-1} Q_{0} \partial_\tau K T \nonumber \\ &= S_\tau +\frac{1}{2} \int_0^\beta d\tau \, \partial_\tau \tr Q_{0} \ln K = S_\tau +\frac{1}{2} \tr Q_{0} (\ln K[\beta] - \ln K[0]) 
\end{align}
Since $Q_0$ and $K$ commutes, we can simultanuously diagonalize them. The eigenvalues of $Q_0$ are $\pm$ whereas the eigenvalues of $\ln K[\beta] - \ln K[0]$ is an integer multiple of $4\pi i$ (since $\det K = 1$) due to periodic boundary conditions, which implies that the extra term is an integer multiple of $2\pi i$.

We can write this term in the familiar form in the spinless limit by taking $Q_0 = \eta_z$ and $T = e^{-\frac{i}{2} \phi \eta_z} e^{\frac{i}{2} \theta \eta_x} e^{\frac{i}{2} \phi \eta_z}$. Substiting in (\ref{St}) yields 
\beq
S_\tau = -\frac{i}{2} \int_0^\beta d\tau \, (1 - \cos \theta) \partial_\tau \phi
\label{Stn}
\eeq

\subsection{Full effective action}
The full effective action is given by
\beq
\label{SQ}
S[Q] = \int d^2 \br d\tau \left\{\frac{1}{2A_M} \tr T^{-1} Q_0 \partial_\tau T + \frac{\R}{4} \tr [\nabla Q(\br)]^2 + \frac{J}{4} \tr (Q \gamma_x)^2 - \mu e \delta \rho(\br) + \frac{1}{2} \int  d^2 \br' \delta \rho(\br) V(\br - \br') \delta \rho(\br')  \right\}
\eeq
In the spinless limit where $Q$ is given by (\ref{Qpm}), this leads to the field theory (4) in the main text.

\subsection{Deviation from perfect sublattice polarization}
We will now briefly investigate the effect of imperfect sublattice polarization. This corresponds to considering the sublattice off-diagonal part of the form factor which can be generally written as $\tilde \Lambda_\bq(\bk) = \gamma_x \eta_z \tilde F_\bq(\bk) e^{i \tilde \Phi_\bq(\bk) \gamma_z}$. This leads to the following contribution to the action
\beq
E_\lambda = \frac{1}{2A} \sum_\bq V_\bq \langle \hat {\tilde \Lambda}_{\bq} \hat {\tilde \Lambda}_{-\bq} \rangle = \frac{\lambda A}{8} \tr [Q, \gamma_x \eta_z][Q, \gamma_x \eta_z]^\dagger = -\frac{\lambda A}{4} \tr (Q \gamma_x \eta_z)^2 + \text{const.} = -\frac{\lambda}{4} \int d^2 \br  \tr (Q(\br) \gamma_x \eta_z)^2 + \text{const.} 
\eeq
with $\lambda$ given by
\beq
\lambda = \frac{1}{2A^2} \sum_{\bk,\bq} V_\bq F_{A,\bq}(\bk)^2
\eeq
This reduces to the expression
\beq
E_\lambda = -\lambda (n_{+,z} n_{-,z} - \bn_{+,xy} \cdot \bn_{-,xy})
\eeq

\section{Effects of dynamical fluctuations on skyrmions}
\label{app:fluc}
In this section, we discuss the estimate for the effective mass of a skyrmion antiskyrmion pair in a first quantized description of skyrmions. In the main text, we considered the Lagrangian of a skyrmion-antiskyrmion pair with a frozen shape, i.e. where only the location of the skyrmions is a dynamical variable. Here, we critically examine the validity of this assumption and also derive the effective "spring constant" $k$ used in Eq.~(6) of the main text. 

Our strategy for considering dynamical degrees of freedom of the skyrmion uses a somewhat crude model for physical transparency and analytical amenability, where the skyrmion is treated as a magnetic bubble with a nearly circular domain wall. Outside the domain wall, the spins are approximately pointing up whereas inside, they point down, and along the domain wall the spins lie in the x-y plane, i.e, $n_z(\br) = 0$ defines the domain wall. Therefore, we can parametrize the domain wall by the following collective coordinates: $r(\phi)$ denoting the radius as a function of the azimuthal angle $\phi$, (for a static bubble $r = R$) and $\Phi(\phi)$ which denotes the in-plane spin fluctuations on the domain wall. Our goal is to find an effective action for the collective coordinates, and then integrate out the spin fluctuations $\Phi(\phi)$ to generate the dynamics of $r(\phi)$, the harmonics of which corresponds to motion/deformation of the bubble.

Following Ref.~\onlinecite{Oleg2012}, we write down the action for a circular domain wall of static radius $R$, denoting the angular momentum density by $g$.
\beq
S = \int dt L , \text{where } L = R \int d\phi \, \mathcal{L}(r, \Phi), \text{ with } \mathcal{L}(r, \Phi) = - g \Phi  \dot r - \frac{\kappa}{2}(\Phi - \Phi_{eq})^2
\eeq
where $\Phi_{eq}$ denotes the equilibrium configuration. As before, our strategy is to integrate over $\phi$ first to obtain coupled equations of motion for $r(\phi)$ and $\Phi(\phi)$, and then integrate out $\Phi(\phi)$ to generate the effective dynamics for $r$. For this purpose, we decompose the fields into circular harmonics:
\beq
r(\phi) = R + \sum_{m = -\infty}^{\infty} r_m \, e^{i m \phi}, ~~ \Phi(\phi) = \phi + \phi_1 + \sum_{m = -\infty}^{\infty} \Phi_m \, e^{i m \phi}
\label{eq:rphi}
\eeq
In Eq.~(\ref{eq:rphi}), $r_m = \overline{r}_{-m}$ denote circular harmonics of $r$. For example, $r_0$ parametrizes the breathing mode, $r_{\pm 1} = (X \mp i Y)/2$ correspond to the center-of-mass motion of the bubble, $r_{\pm 2}$ parametrize elliptical deformations, and so on. Similarly, $\Phi_m = \overline{\Phi}_{-m}$, and the shift by constant $\phi_1$ is motivated by the observation that in equilibrium, the spins prefer to point at a given angle to the domain wall \cite{Oleg2012}. For example, for Bloch domain walls, $\phi_1 = \pm \pi/2$, so we have $\Phi_{eq}(\phi) = \phi \pm \pi/2 - \frac{1}{r} \frac{\partial r}{\partial \phi} \approx \phi \pm \pi/2  - \frac{1}{R} \frac{\partial r}{\partial \phi} $. However, our results will be insensitive to the the precise value of the angle $\phi_1$. Finally, $\kappa \sim \R W$ parametrizes the spin stiffness, $W$ being the width of the domain wall. Plugging this into the Lagrangian $L$, we find that (upto total time-derivatives):
\beq
L = 2 \pi R \left[ \left(- g r_0 \dot \Phi_0 - \frac{\kappa}{2} \Phi_0^2 \right) + \sum_{m = 1}^{\infty}  (-g) (\dot r_m \overline{\Phi}_{m} +  \overline{ \dot r}_m \Phi_m) - \kappa \bigg| \Phi_m - \frac{i m r_m}{R}\bigg|^2 \right]
\eeq
Now, we can integrate out the $(\Phi_m, \bar{\Phi}_m)$ modes to generate the effective action for the skyrmion coordinates:
\beqarray
L_{eff} &=& \left( \frac{\pi R g^2}{\kappa} \right) \dot{r}_0^2 + \sum_{m = 1}^{\infty} \left( \frac{2\pi R g^2}{\kappa} \right) |\dot r_m|^2 + 2 \pi i m g (\overline{ \dot{r}}_m r_m - \dot{r}_m \overline{r}_m)  \nn
& = & \frac{M_s}{2} \dot {\bR}^2 - \pi g (\bR \times \dot{\bR}) \cdot \vechat{z} + \ldots
\eeqarray
where $M_s = \frac{\pi R g^2}{\kappa}$ denotes the effective mass of the skyrmion, and the ellipsis denotes the $|m| \neq 1$ terms which also scale with the skyrmion mass. We see that the skyrmion mass $M_s$ is directly proportional to the linear size $R$ of the skyrmion, and inversely proportional to the stiffness $\kappa$ (a larger stiffness makes fluctuations expensive and suppresses the kinetic energy term of the skyrmion). Therefore, for a small-size skyrmion or a large spin-stiffness, the dynamical fluctuations are likely to be suppressed. On the contrary, the Berry phase term is independent of the skyrmion-size.

Next, we discuss the effective action for the skyrmion-antiskyrmion pair from opposite Chern sectors. On inclusion of the inertial term $M_s$ in addition to the Berry phase term \cite{Stone}, the action for the pair is given by (with $G = 2 \pi \hbar/A_M$):
\beq
S[\bR_+ , \bR_-] = \int dt L_{sas}, \text{ where } L_{sas} = \frac{G}{2} [(\bR_+ \times \dot{\bR}_+) - (\bR_- \times \dot{\bR}_-) ] \cdot \vechat{z} + \frac{M_s}{2} (\dot{\bR}_+^2 + \dot{\bR}_-^2 ) - \mathcal{E}(|\bR_+ - \bR_-|) \nn
\eeq
Moving to center of mass coordinates $\bR_s = (\bR_+ + \bR_-)/2$ and relative coordinates $\bR_d = \bR_+ - \bR_-$, a rewriting of the Lagrangian yields:
\beq
 L_{sas} =  G (\bR_d \times \dot{\bR}_s) \cdot \vechat{z} + M \dot {\bR}_s^2 + \frac{M}{4} \dot {\bR}_d^2 - \mathcal{E}(|\bR_d|)
\eeq
The canonical momenta, and hence the Hamiltonian of the pair, are therefore given by:
\beqarray
\bP_s &=& \frac{\partial L}{\partial  \dot {\bR}_s} =2 M \dot {\bR}_s + G( \vechat{z} \times  \bR_d), ~~~ \bP_d = \frac{\partial L}{\partial  \dot {\bR}_d} = \frac{M_s}{2}  \dot {\bR}_d \\
H_{sas} &=& \bP_s \cdot \dot{\bR}_s +  \bP_d \cdot \dot{\bR}_d - L = M_s \dot {\bR}_s^2 +  \frac{M_s}{4}\dot {\bR}_d^2 +  \mathcal{E}(|\bR_d|) \nn
& = & \frac{1}{4M_s} ( \bP_s - G \vechat{z} \times \bR_d )^2 + \frac{1}{M_s} \bP_d^2 + \mathcal{E}(|\bR_d|) 
\eeqarray
In order to make further analytical progress, one needs to resort the $M_s \rightarrow 0$ limit, which we analyzed in the main text, and is a good approximation when the size of single (charge $e$) skyrmions is small. In this limit, we have $ \bP_s = G \vechat{z} \times \bR_d$ and $\bP_d = 0$ strictly enforced, so that we can replace $\mathcal{E}(|\bR_d|)$ by $\mathcal{E}\left(\frac{|\bP_s|}{G} \right)$. Therefore, the mass of the pair can be derived by expanding $\mathcal{E}$ about $\bR_d = 0$: it is given by:
\beq
M_{\tp} = \frac{G^2}{\mathcal{E}^{\prime \prime}(0)}
\eeq
What is left now is to determine the pair potential $\mathcal{E}(|\bR_d|)$. For this purpose, we start with a representative KIVC ground state $\bn_+(\br) = -\bn_-(\br) = (1,0,0)$ for the action in Eq.~(5) in the main text (at $\mu = 0$); such a ground state would be chosen by  anisotropies favoring ordering in the $x$-$y$ plane that we have neglected in this work for simplicity. We consider skyrmionic (antiskyrmionic) textures in the $C = + $ (-) sector with a relavtive core separation of $\bR_d$, such that $\bn_\pm(\br) = \pm \bn_{sk}(\br \pm \bR_d/2)$. Here, $\bn_{sk}(\br)$ is a skyrmionic texture with unit winding for a unit-vector field in two spatial dimensions, given by:
\beq
\bn_{sk}(\br) = (\cos(\Theta(\br))\sin(\Theta(\br)) \cos(\Phi(\br)), \sin(\Theta(\br)) \sin(\Phi(\br))), ~ \text{ satisfying } \frac{1}{4\pi} \int d^2r \, \bn_{sk} \cdot ( \partial_x \bn_{sk} \times \partial_y \bn_{sk} ) = 1
\label{eq:nSk}
\eeq

Since the elastic energy of this pair depends only on the winding number, it is independent of the relative core-separation $\bR_d$. The exchange energy proportional to $J$ increases, while the Coulomb interaction energy decreases as the charged skyrmions are separated. As a function of separation $|\bR_d|$, the change of pair potential is given by:
\beqarray
\mathcal{E}(|\bR_d|) - \mathcal{E}(0) &=& J \int d^2r \, (1 + \bn_+(\br)\cdot \bn_{-}(\br)) + \int \frac{d^2q}{(2\pi)^2} \, V_{|\bq|} \delta\rho_\bq  \delta\rho_{-\bq} ( e^{i \bq \cdot \bR_d} - 1) \nn
&=& J \int d^2r \, (1 - \bn_{sk}(\br - \bR_d/2)\cdot \bn_{sk}(\br + \bR_d/2)) + \frac{1}{4 \pi \epsilon}\int_0^\infty dq (\delta \rho_q)^2 (J_0(q |\bR_d|) - 1)
\label{eq:EnDiff}
\eeqarray
where we have noted that $\delta \rho_{\bq} = \int d^2r \, \delta \rho(\br)e^{i \bq \cdot \br}$ is only dependent on $|\bq|$ for cylindrically symmetric charge distributions, and used $V_{|\bq|} = 1/(2 \epsilon |\bq|)$ as the Fourier transform of the (unscreened) Coulomb potential for two-dimensional motion. The pair mass, therefore, may simply be computed by Taylor expanding the RHS of Eq.~(\ref{eq:EnDiff}) to second order in $\bR_d$, which yields:
\beqarray
\mathcal{E}(|\bR_d|) - \mathcal{E}(0) &=& \left[ \frac{J}{4} \int d^2r (\mathbf{\nabla} \vechat{n})^2 - \frac{1}{16\pi \epsilon} \int_0^\infty dq (\delta \rho_q)^2 q^2 \right] |\bR_d|^2 + \text{O}(|\bR_d|^4) \nn
\implies \mathcal{E}^{\prime \prime}(0) &=& \frac{J}{2} \int d^2r (\mathbf{\nabla} \vechat{n})^2 - \frac{1}{8\pi \epsilon} \int_0^\infty dq (\delta \rho_q)^2 q^2
\eeqarray
Next, we compute this correction for the Belavin-Polyakov ansatz \cite{BP75} that corresponds to $\Theta(\br)= 2 \tan^{-1}(R/r)$, and $\Phi(\br)= \phi$ in the texture defined in Eq.~(\ref{eq:nSk}). For this ansatz $\delta \rho_q = e (q R) K_1(qR)$, and it follows that:
\beq
\mathcal{E}^{\prime \prime}(0) = 4 \pi J - \frac{45 \pi e^2}{2^{12} \epsilon R^3}
\eeq
For large skyrmions, the correction from the Coulomb term is insignificant and can be neglected, leading to $k = 4 \pi J$ as quoted in Eq.~(6) of the main text. For smaller skyrmions, the Coulomb term can reduce $\mathcal{E}^{\prime \prime}(0)$ substantially leading to an enhancement of the pair effective mass $M_\tp$ and a subsequent reduction to $T_c$. 

\section{Integrating out the ferromagnetic fluctuations}
\label{app:IntFerro}
In this section, we explain the procedure to integrate out ferromagnetic spin fluctuations leading to an effective theory. 
In terms of the matrix order parameter $Q$, ferromagnetic fluctuations denote deviations from the perfect antiferromagnetic alignment $Q_+ = -Q_-$ which can be parametrized as
\beq
Q(W) = \left(\begin{array}{cc} \tilde T^{-1} e^{\frac{i}{2} W} \tilde Q_0 e^{- \frac{i}{2} W} \tilde T & 0 \\ 0 &  -\tilde T^{-1} e^{-\frac{i}{2} W} \tilde Q_0 e^{\frac{i}{2} W} \tilde T \end{array} \right), \qquad \{W, \tilde Q_0\} = 0, 
\eeq
We assume that both $W$ and its gradients as well as the gradients of $T$ are small. To simplify the notation in the following, we will introduce the variables
\beq
\tilde Q = \tilde T^{-1} \tilde Q_0 \tilde T, \qquad A_\mu = (\partial_\mu \tilde T) \tilde T^{-1}
\eeq
Since the action is expressed in terms of $Q(W)$, it is invariant under the gauge transformations 
\beq
\tilde T \mapsto K\tilde T, \qquad W \mapsto K W K^{-1}, \qquad A_\mu \mapsto K A_\mu K^{-1} + (\partial_\mu K) K^{-1}, \quad \text{ for } \quad [K,Q_0] = 0
\eeq
Substituting in the different terms in the action, we get 
\begin{gather}
\frac{\R}{4} \tr [\partial_i Q(W)]^2 = \frac{\R}{2} \tr [\partial_i \tilde Q]^2 - \frac{\R}{8} \tr [Q_0, \partial_i W - [A_i, W]]^2,  \label{Swgrad} \\
\frac{1}{2} \tr Q_0 (\partial_\tau T(W)) T^{-1}(W) = -i\tr \tilde Q_0 W A_0 - \frac{i}{2} \tr \tilde Q_0 \partial_\tau W
\label{Swt}\\
\frac{J}{4} \tr (Q(W) \gamma_x)^2 = J \tr W^2 + \text{const.} 
\label{SwJ}\\
\delta \rho(\br) = \frac{i}{8\pi}\epsilon^{ij} \tr \tilde Q \partial_i \tilde Q  \partial_j \tilde Q + \frac{i}{8\pi}\epsilon^{i j} \tr \tilde Q_0 \partial_i W  \partial_j W
\label{Swrho}
\end{gather}
We notice that the second term in (\ref{Swt}) is a total derivative which can be ignored since $W(\beta) = W(0)$. We also notice that the second term in (\ref{Swrho}) is a total derivative which vanishes if $W$ is smooth and vanishes quickly enough at infinity (which is simply the statement that smooth deformations cannot change the topological index) so we can also neglect this term. As a result, the action for the ferromagnetic fluctuations $W$ has the form
\beq
S_W = \int d\tau \int d^2 r \left\{J \tr W^2 - \frac{\R}{8} \tr [\tilde Q_0, \partial_i W - [A_i, W]]^2 - \frac{i}{A_M} \tr \tilde Q_0 W A_0 \right\}
\label{SW}
\eeq

To proceed, we restrict ourselves to the spinless limit $n=1$, where we can take $\tilde Q_0 = \eta_z$ and write
\beq
W = \left(\begin{array}{cc} 0 & w \\ w^* & 0 \end{array} \right), \qquad A_\mu = A_{0,\mu} + \sum_{i=x,y,z} A_{i,\mu} \eta_i, \qquad A_{i,\mu} = \frac{1}{2} \tr A_\mu \eta_i
\eeq
Substituting in (\ref{SW}) yields
\beq
S_W = \int d\tau \int d^2 r \left\{2J w^* w + \R [(\partial_i + 2A_{z,i}) w^*]  [(\partial_i - 2A_{z,i}) w] + \frac{1}{A_M} (w A_{xy,0}^* - w^* A_{xy,0}) \right\}, \qquad A_{xy,\mu} = A_{x,\mu} - i A_{y,\mu}
\label{Sw}
\eeq
We notice that under the gauge transformation generated by $K = e^{i \frac{\phi}{2} \eta_z}$, the $w$ and $A$ variables transform as
\beq
w \mapsto e^{i \phi} w, \qquad A_{z,\mu} \mapsto A_{z,\mu} + \frac{i}{2} \partial_\mu \phi, \qquad A_{xy,\mu} \mapsto e^{i \phi} A_{xy,\mu}
\eeq
which leaves (\ref{Sw}) invariant.

We can now easily integrate the $w$ variable to get
\beq
S_{\rm eff} = \frac{1}{A_M^2} \int d\tau d^2\br A_{xy,0}^* [2J - \R (\partial_i - 2A_{z,i})^2]^{-1} A_{xy,0} 
\label{Seff}
\eeq
Let us now introduce the CP${}^1$ representation defined from the $2 \times 2$ $Q$ matrix via
\beq
\tilde Q_\pm = \pm \tilde T^{-1} \tilde Q_0 \tilde T = \pm \bn \cdot {\boldsymbol \eta}, \qquad \bn = z^\dagger {\boldsymbol \eta} z, \qquad z^T = (z_1, z_2), \qquad z^\dagger z = 1
\eeq
It is straightforward to verify that
\beq
\tilde T = \left(\begin{array}{c} z^\dagger \\ i z^T \sigma_y \end{array}\right), \qquad A_\mu = \left(\begin{array}{cc} \partial_\mu z^\dagger z & i \partial_\mu z^\dagger \sigma_y z^* \\ -i \partial_\mu z^T \sigma_y z & z^\dagger \partial_\mu z \end{array}\right) \quad \Rightarrow \quad A_{z,\mu} = -z^\dagger \partial_\mu z, \qquad A_{xy,\mu} = -i z^\dagger \sigma_y \partial_\mu z^* 
\eeq
Introducing the gauge field and covariant derivate
\beq
a_\mu = i z^\dagger \partial_\mu z, \qquad D_\mu = \partial_\mu - i a_\mu
\eeq
we can write
\beq
A_{z,\mu} = -i a_\mu, \qquad A_{xy,\mu} = -i z^\dagger \sigma_y (D_\mu z)^*
\eeq
where we used the fact that $z^\dagger \sigma_y z^* = 0$.
Substituting in (\ref{Seff}) yields
\beq
S_{\rm eff} =- \frac{1}{A_M^2} \int d\tau d^2\br z^T \sigma_y (D_0 z) [2J - \R (\partial_i + 2i a_\mu)^2]^{-1} z^\dagger \sigma_y (D_0 z)^*  
\label{Sz}
\eeq
In the following, we will be only interested in terms which has at most two gradients. This means we can neglect the second term in the denominator. The resulting expression can be simplified by using the identity
\beq
(a^\dagger \sigma_y b^*) (c^T \sigma_y d) = (a^\dagger d) (b^\dagger c) - (a^\dagger c)(b^\dagger d)
\eeq
for any 2D complex vectors $a$, $b$, $c$ and $d$. In addition, we can use the following identities
\beq
z^\dagger D_\mu z = (D_\mu z)^\dagger z = 0, \qquad z^\dagger D_\mu D_\nu z + (D_\mu z)^\dagger D_\nu z = 0, 
\eeq
Substituting in (\ref{Sz}) yields
\beq
S_{\rm eff} = \frac{1}{A_M^2} \int d\tau d^2\br \frac{1}{2J} |D_0 z|^2
\eeq
The full CP${}^1$ action is then given by Eq.~8 in the main text. 

\section{Large $N$ phase diagram}
\label{app:LargeN}
Our purpose in this appendix is to provide the details for the calculation of the large $N$ phase diagram for the CP${}^1$ model whose Lagrangian is repeated here for completeness
  \beq
        S[z] = \int d^3 \br \left\{ \frac{\Lambda}{g} |D_\mu z|^2 + \frac{2 e \mu}{c} \frac{\epsilon_{ij}}{2\pi} \partial_i a_j \right. \\ \left. 
     +\frac{1}{2c} \int d^2\br' \frac{\epsilon_{ij}}{\pi} (\partial_i a_j)_{\br} V(r-r') \frac{\epsilon_{lk}}{\pi} (\partial_l a_k)_{\br'} \right\}
    \label{eq:LCP1gApp}
    \eeq
    with $g$ and $c$ defined as
     \beq
 g  = \sqrt{\frac{J A_M}{2\R}}, \qquad c =  \frac{2}{\Lambda} \sqrt{2 J A_M \R}
 \eeq
    
    \subsection{Zero doping}
    Let's first review the solution for the case of $ \mu = 0$ \cite{CPNdAdda, CPNWitten, CPNArefeva, PolyakovBook}. In this case, we can start by writing
    \beq
    \frac{\Lambda}{g} |D_\mu z|^2 =  \frac{\Lambda}{g}(|\partial_\mu z|^2 - a_\mu^2)
    \eeq
    and decouple the quadratic term in $a_\mu$ via a Hubbard-Stratonovich transformation as
    \beq
    e^{\frac{\Lambda}{g} a_\mu^2} \propto \int d\alpha_\mu e^{-\frac{\Lambda}{g}(\alpha_\mu^2 - 2 a_\mu \alpha_\mu)}
    \eeq
    Upon substituting in (\ref{Sz}), this leads to
    \beq
    \L = \frac{\Lambda}{g} |\D_\mu z|^2, \qquad \D_\mu = \partial_\mu - i \alpha_\mu
    \eeq
    Since $\alpha_\mu$ enters the field theory in the same way as $a_\mu$, in the following, we will relabel $\alpha_\mu \mapsto a_\mu$. The constraint $z^\dagger z = 1$ can be included by writing the integral representation of the delta function
    \beq
    \delta(z^\dagger z - 1) = \int d\kappa \, e^{i \kappa (z^\dagger z - 1)}
    \label{eq:delta}
    \eeq
    leading to
    \beq
    \L = \frac{\Lambda}{g} \{|\D_\mu z|^2 + \Delta^2 (z^\dagger z - 1) \}, \qquad \Delta^2 = i \kappa \frac{g}{\Lambda}
    \eeq
    We now introduce the large $N$ limit by promoting $z$ to an $N$-component vector in CP${}^{N-1}$ satisfying the constraint $z^\dagger z = N$. The Lagrangian becomes
    \beq
    \L = \frac{\Lambda}{g} \left\{ \sum_{i=1}^{N} z_i^\dagger [- \D^2 + \Delta^2 ] z_i - \Delta^2 N \right\}
    \eeq
    Integrating out the $z$ variables leads to
    \beq
    S = N \left\{\tr \ln  \left[- \D^2 + \Delta^2 \right] - \frac{\Lambda}{g} \int d^3 \br \Delta^2  \right\}
    \eeq
    Variation with respect to $\Delta^2$ and $a_\mu$ yield
    \begin{gather}
    \frac{\Lambda}{g} = G_0(\Delta,\br) , \nonumber \\
     - iG_\mu(\Delta,\br) - \frac{\epsilon_{0 \mu \nu} \epsilon_{ij} }{\pi^2 c} \int d^2 \br' V(\br - \br')  \partial_\nu \partial_i a_j(\br')= 0
    \label{eq:saddle}
    \end{gather}
    where $G_0$ is the Green's function at equal points 
    \beq 
    G_0(\br) = G(\br, \br), \qquad G_\mu(\br) = \frac{\partial}{\partial r_\mu} G(\br, \br')\Big|_{\br' = \br},\qquad G(\br,\br') = \left(-\D^2 + \Delta^2 \right)^{-1}_{\br,\br'}
    \eeq
    and the derivative in $\partial_\mu G_0$ is taken relative to the first index. 
    Since $\mu = 0$, we can consider saddle points when there is no magnetic field, i.e. $\nabla \times a = 0$, so we can take $a_\mu = 0$. In this case, the second saddle point equation is automatically satisfied whereas the first one becomes
    \beq
    G_0(\Delta, \br) = G_0(\Delta) = \int \frac{d^3 \bk}{(2\pi)^3} \frac{1}{\bk^2 + \Delta^2} = \frac{\Lambda}{g}, 
    \label{G0g}
    \eeq
    We notice here that the Green's function at equal points is independent of $\br$. The integral can be regularized as follows
    \beq
    G_0(\Delta) =  \int \frac{d^3 \bk}{(2\pi)^3} \left\{ \frac{1}{\bk^2 + \Delta^2} - \frac{1}{\bk^2 + \Lambda^2}\right\} = \frac{1}{4\pi}\{\Lambda - |\Delta| \}
    \eeq
    As a result, we see that a self consistent solution to (\ref{G0g}) is only possible for $g > g_c = 4\pi$ yielding 
    \beq
    |\Delta| = \Lambda \left( 1 - \frac{4\pi}{g} \right) = 4\pi \Lambda (g_c^{-1} - g^{-1})
    \label{eq:Delta}
    \eeq
    This gives a mass for the $z$ fields which implies a finite correlation length for the spins, i.e. the phase is spin-disordered. Furthermore, expanding the action in variations in $a_\mu$ yields \cite{PolyakovBook, CPNArefeva}
    \begin{gather}
    S = \frac{N}{2} \sum_\bq a_\mu(\bq) \Gamma_{\mu \nu}(\bq) a_\nu(-\bq) \\  \Gamma_{\mu \nu}(\bq) = \int \frac{d^3 \bk}{(2\pi)^3} \left\{ \frac{2 \delta_{\mu \nu}}{\bk^2 + \Delta^2} - \frac{(2 k_\mu + q_\mu)(2 k_\nu + q_\nu) }{(\bk^2 + \Delta^2)((\bk + \bq)^2 + \Delta^2)} \right\} = \frac{1}{24\pi |\Delta|} (q^2 \delta_{\mu \nu} - q_\mu q_\nu)
    \label{eq:Gamma}
    \end{gather}
     which gives the Maxwell term 
     \beq
     \L_{\rm Maxwell} = \frac{N}{96 \pi |\Delta|} f_{\mu \nu}^2, \qquad f_{\mu \nu} = \partial_\mu a_\nu - \partial_\nu a_\mu
     \eeq
     To see how this is related to the phase stiffness of the superconductor, we first include the coupling to the electromagnetic gauge field $A$ which takes the form $\frac{2 i e }{2\pi}J_\mu A_\mu$ where the current operator $J_\mu$ is given by $J_\mu = \epsilon^{\mu \nu \lambda} \partial_\nu a_\lambda$. The Lagrangian for $a_\mu$ takes the form
     \beq
     \L = \frac{N}{48 \pi |\Delta|} J_{\mu}^2 + \frac{2 i e}{2\pi}J_\mu A_\mu
     \eeq
     where $J_\mu$ satisfies the current conservation $\partial_\mu J_\mu = 0$ which can be included in the Lagrangian using the delta function representation (\ref{eq:delta}) leading to
     \beq
     \L = \frac{N}{48 \pi |\Delta|} J_{\mu}^2 + \frac{2 i e}{2\pi}J_\mu A_\mu + \frac{i}{2\pi} \phi \partial_\mu J_\mu = \frac{N}{48 \pi |\Delta|} J_{\mu}^2 + \frac{i}{2\pi} J_\mu (\partial_\mu \phi - 2 e A_\mu) 
     \eeq
     We can now integrate out the $J_\mu$ variable leading to
     \beq
     \L = \frac{3|\Delta|}{\pi N} (\partial_\mu \phi - 2 e A_\mu)^2 
     \eeq
     Identifying the phase $\phi$ with the superconducting phase, we can write the corresponding action
     \beq
     S = \int d^2 \br d\tau \left\{\frac{\rho_{\rm SC}}{2} (\partial_i \phi - 2 e A_i)^2 + \frac{\chi_{\rm SC}}{2} (\partial_\tau \phi - 2 e  A_\tau)^2 \right\}, \qquad \rho_{\rm SC} = \frac{3|\Delta|c}{\pi N} \qquad \chi_{\rm SC} = \frac{3|\Delta|}{\pi c N}
     \label{eq:SCrhochi}
     \eeq
    
    We note here that it is also possible to access the spin-ordered phase by slightly modifying the procedure above, integrating out the variables $z_2,\dots,z_N$ but leaving $z_1$. In this case, we can similarly derive a set of saddle points \cite{AffleckBrezin, CPNArefeva} with a saddle point solution $\Delta = 0$ and
    \beq
    |z_1|^2 = 1 - \frac{g}{\Lambda}G_0(0) = 1 - \frac{g}{4\pi}
    \eeq
    so that such solution is only valid for $g < g_c$. This is the spin-ordered phase since $a_\mu$ field is in the Higgs phase.

    \subsection{Finite doping $\mu \neq 0$}
    Let us now reintroduce the chemical potential term $\mu$. Although the chemical potential $\mu$ does not enter the saddle point equations (\ref{eq:saddle}), its effect is to make it energetically favorable to introduce a finite magnetic flux of the field $a_\mu$ given by $b = \nabla \times a$. The spatial distribution of the flux $b$ is determined by solving the second equation in (\ref{eq:saddle}). In the following, we will assume the flux $b$ is spatially uniform. We will show at the end that for parameters of interest in MATBG, this assumption is true to a very good approximation. Since $a_\mu$ corresponds to the Skyrmion density in the original problem, the $b$ and the cyclotron frequency $w_c$ can be written in terms of the filling as 
    \beq
    b = \nabla \times a = \frac{\nu \pi}{A_M} = \pi \nu \Lambda^2 \qquad w_c = 2b = 2\pi \nu \Lambda^2
    \eeq
     with $\nu$ being the filling fraction. In the presence of a magnetic field, the spectrum of the operator $-\D^2 + \Delta^2$ corresponds to the Landau level spectrum 
     \beq
     \epsilon_{n,k_z} = k_z^2 + w_c(n+1/2) + \Delta^2, \quad \text{ with degeneracy } \frac{1}{2}\nu \Lambda^2 A
     \eeq
     The Green's function $G_0(\Delta,\br)$ is modified to
    \beq
    G_0(\Delta,\nu, \br) =  G_0(\Delta,\nu) = \frac{1}{2} \nu \Lambda^2 \sum_n \int \frac{dk_z}{2\pi} \left\{ \frac{1}{k_z^2 + 2\pi \nu \Lambda^2 (n + 1/2) + \Delta^2} - \frac{1}{k_z^2 + 2\pi \nu \Lambda^2 (n + 1/2) + \Lambda^2}\right\}
    \eeq
    where we used the completeness of the wavefunctions within each Landau level to eliminate the $\br$ dependence. This integral converges for $\Delta^2 > -\pi \nu \Lambda^2$ and can be evaluated as
    \beq
    G_0(\Delta, \nu) = \frac{ \Lambda}{4} \sqrt{\frac{\nu}{2\pi}}\left\{ \zeta\left(\frac{1}{2}, \frac{1}{2} + \frac{\Delta^2}{2\pi \nu \Lambda^2} \right) - \zeta\left(\frac{1}{2}, \frac{1}{2} + \frac{1}{2\pi \nu} \right) \right\},
    \label{eq:G0nu}
    \eeq
    where $\zeta(s,q)$ is the Hurwitz zeta function defined as \cite{NIST}
    \beq
    \zeta(s,q) = \sum_{n=0}^\infty \frac{1}{(q + n)^s}
    \eeq
    For $\nu \neq 0$, the saddle point equation (\ref{eq:saddle})  is always satisfied since $G_0(\Delta,\nu)$ is unbounded from above (we can choose $\Delta^2$ to be as close as we want to $-\pi \nu \Lambda^2$) with the caveat being that the mass $\Delta^2$ would have to be negative whenever $g < \Lambda/G_0(0,\nu)$. This analysis implies that the purely ordered phase where $z$'s are gapless and $a_\mu$ is Higgsed is immediately lost for any finite doping. This arises since doping forces magnetic flux into the system. In principle, we can have a vortex lattice or a coexistence phase as discussed in the main text. However, our current mean field scheme does not allow for any coexistence since non-zero doping immediately implies the loss of spin order $z = 0$. For the action (\ref{eq:LCP1gApp}), the doping is controlled through the chemical potential. The optimal doping at a given chemical potential is obtained by finding the minimum of the action as a function of doping $\nu$. To write the action for finite doping $\nu$, let's first consider the first term (in the following, we will consider the rescaled action $S = N V \tilde S$)
    \beq
    \tilde S_1 = \frac{1}{V} \tr \ln (-\D_\mu^\dagger \D_\mu + \Delta^2) 
    \eeq
    We notice that the second derivative of this action with respect to $\Delta^2$ is finite and can be evaluated as
    \begin{align}
    \frac{\partial^2 \tilde S_1}{\partial (\Delta^2)^2} &=-\frac{1}{2} \nu \Lambda^2 \sum_n \int \frac{dk_z}{2\pi} \frac{1}{(k_z^2 + \omega_c (n + 1/2) + \Delta^2)^2} = -\frac{1}{8} \nu \Lambda^2 \sum_n  \frac{1}{( \omega_c (n + 1/2) + \Delta^2)^{3/2}} \nonumber \\ &= -\frac{1}{16 \pi  \Lambda \sqrt{2\pi \nu}} \zeta\left(\frac{3}{2}, \frac{1}{2} + \frac{\Delta^2}{2\pi \nu \Lambda^2}\right)
    \end{align}
    This can be integrated to give a finite action with the help of two counter terms for the mass and free energy renormalization \cite{AffleckBrezin}
    \beq
    \tilde S_1[\Delta, \nu] = \nu \Lambda^3 \sqrt{\frac{\nu}{2\pi}}\zeta\left(-\frac{1}{2}, \frac{1}{2} + \frac{\Delta^2}{2\pi \nu \Lambda^2}\right)+ A(\nu) \Delta^2 + C(\nu)
    \eeq
    $A(\nu)$ is fixed by the requirement that the first derivative of the action with respect to $\Delta^2$ is equal to $G_0(\Delta,\nu)$ given in (\ref{eq:G0nu}) while $C(\nu)$ is fixed by the requirement the the free energy vanishes for $\Delta = 0$ which yields
    \beq
    A(\nu) = -\frac{ \Lambda}{4} \sqrt{\frac{\nu}{2\pi}} \zeta\left(\frac{1}{2}, \frac{1}{2} + \frac{1}{2\pi \nu} \right), \qquad C(\nu) = -\nu \Lambda^3 \sqrt{\frac{\nu}{2\pi}}\zeta\left(-\frac{1}{2}, \frac{1}{2}\right)
    \eeq
    Collecting everything gives
    \beq
    \tilde S = \frac{S}{N V \Lambda^3} = \sqrt{\frac{\nu}{2\pi}} \left\{ \nu \zeta\left(-\frac{1}{2}, \frac{1}{2} + \frac{\Delta^2}{2\pi \nu \Lambda^2}\right) -\frac{1}{4} \zeta\left(\frac{1}{2}, \frac{1}{2} + \frac{1}{2\pi \nu} \right) \frac{\Delta^2}{\Lambda^2} - \nu \zeta\left(-\frac{1}{2}, \frac{1}{2}\right) \right\}  + \left\{- \frac{\Delta^2}{g \Lambda^2} + \tilde \mu  \nu +  \frac{\nu^2 \tilde V_0}{2} \right\}
    \eeq
    where 
    \beq
    \tilde V_0 = \frac{V_0}{c}, \qquad V_0 = \Lambda^2 \int d^2 \br V(\br) = \frac{1}{A_M} \int d^2 \br V(\br) = \frac{e^2 d}{2 \epsilon \epsilon_0 A_M}, \qquad \tilde \mu = \frac{\mu}{c \Lambda}
    \eeq
    $\Delta^2$ can be eliminated by solving the saddle point equation
    \beq
    1 = \frac{g}{\Lambda} G_0(\Delta, \nu) = \frac{g}{4} \sqrt{\frac{\nu}{2\pi}}\left\{ \zeta\left(\frac{1}{2}, \frac{1}{2} + \frac{\Delta^2}{2\pi \nu \Lambda^2} \right) - \zeta\left(\frac{1}{2}, \frac{1}{2} + \frac{1}{2\pi \nu} \right) \right\}, 
    \label{eq:SCgap}
    \eeq
     The gap $\Delta$ can be obtained by solving Eq.~\ref{eq:SCgap} in the limit $g \sqrt{\nu} \ll 1$. Writing
    \beq
    \Delta^2 = \Lambda^2[-\pi \nu + \nu^2 g^2 a^2  + O(\nu^3 g^4)]
    \label{eq:Delta2}
    \eeq
    We then use the asymptotic expressions
    \beq
    \zeta(s, x\ll 1) \approx x^{-s} + O(x^{-s+1}), \qquad \zeta(s, x\gg 1) = \frac{x^{1-s}}{s-1} + \frac{1}{2} x^{-s} + O(x^{-1-s})
    \label{eq:ZetaAsymp}
    \eeq
    Substituting in (\ref{eq:SCgap}), we find
    \beq
    a = \frac{1}{4} 
    \eeq
    which is valid for $g \nu \ll 1, g < g_c$.
    
    The phase boundary between the spin ordered phase and the superconductor can be determined by looking for the point at which the spin ordered phase becomes unstable, i.e. when the derivative of the Free energy relative to the doping is negative. Using (\ref{eq:Delta2}) and (\ref{eq:ZetaAsymp}) we find that
    \beq
    \frac{\partial \tilde S}{\partial \nu} \Big|_{\nu = 0} = -\frac{1}{4} + \frac{\pi}{g} +\tilde \mu = 0 \quad \implies \quad \mu_c = -\frac{\pi c \Lambda}{g} \left(1 - \frac{g}{g_c} \right) = -4\pi \rho_s \left(1 - \frac{g}{g_c} \right)
    \eeq
    This gives the phase boundary between the spin ordered state and the superconductor. This expression reduces to the energy of a single skyrmion $4\pi \rho_s$ in the $\SU(2)$ limit when $g \ll g_c$.

    We can also estimate the stiffness and compressibility of the superconductor similar to the derivation leading to Eq.~\ref{eq:SCrhochi}. In general, this requires recomputing (\ref{eq:Gamma}) using the propagator at finite doping. Such evaluation is in general complicated since momentum conservation is lost and we need to use the Landau level wavefunctions. In the following, we will assume we are in the limit $\nu \ll 1$ so that the discrete sum in $G_0(\Delta,\nu)$ can be transformed into an integral. This is done by writing
    \beq
    \frac{1}{2} \nu \Lambda^2 \sum_n f(2\pi \nu \Lambda^2 n) = \int \frac{d\epsilon}{4\pi} f(\epsilon) = \int \frac{d^2 \bk}{(2\pi)^2} f(\bk^2)
    \eeq
    This allows us to evaluate the integral of (\ref{eq:Gamma}) with the main difference being the replacement $\Delta \mapsto \tilde \Delta = \sqrt{\Delta^2 + \pi \nu \Lambda^2}$ leading to
    \beq
    \rho_{\rm SC} \approx \frac{3c }{\pi N} |\tilde \Delta|, \qquad \chi_{\rm SC} \approx \frac{3 }{\pi N c} |\tilde \Delta|, \qquad \nu \ll 1
    \eeq
    Using the expressions (\ref{eq:Delta2}) which are valid for $g \sqrt{\nu} \ll 1$, we get
    \beq
    \rho_{\rm SC} \approx \frac{3 \Lambda c \nu g}{4\pi N}  = \frac{1}{N } \frac{3 J \nu}{2\pi} , \qquad \chi_{\rm SC}  \approx \frac{3 \Lambda g \nu}{4\pi N c} = \frac{1}{N} \frac{3 \nu}{16 \pi \rho_s A_M} , \qquad g \sqrt{\nu} \ll 1
    \eeq
    Remarkably, the stiffness and compressibility of the superconductor are inversely proportional to the compressibility and the stiffness of the insulator, respectively.
    
    At the end, let us revisit the justification for assuming a uniform magnetic flux $b$. To assess this assumption, let us write
    \beq
    b(\br) = b + \delta b(\br), \qquad b = \pi \nu \Lambda^2
    \eeq
    Substituting in the second equation in (\ref{eq:saddle}), we see that the constant part $b$ drops in the second term. We can estimate the magnitude of the second term to be $~\frac{V_0}{c \Lambda^2} \epsilon_{ij} \partial_j \delta b(\br)$. The first term $G_\mu$ has dimensions inverse length square and its magnitude can be estimated using the magnetic length $l_B \sim \frac{1}{\Lambda \sqrt{\nu}}$ to be $G_\mu \sim \Lambda^2 \nu \sim b$. Thus, we find
    \beq
    \frac{1}{\Lambda} \epsilon_{ij} \partial_j \delta b(\br) \sim \frac{\Lambda c}{V_0} b
    \eeq
    For parameters relevant to MATBG, we find that $\Lambda c \sim 1$ meV while $V_0 \sim 30-120$ meV for the realistic values of the screening parameter $d = 10 - 40$ nm. Thus, the variation in the magnetic flux on the Moir\'e scale is neglegible compared to the value of the flux. This justifies our assumption of uniform flux.
    
    \section{Pairing symmetry for the skyrmion superconductor}
    In this section, we will discuss the pairing symmetry for the skyrmion superconductor. We will begin by reviewing the different symmetries of MATBG and their implementation in the flat band projected sublattice-polarized basis following Ref.~\cite{KIVCpaper}. We will then discuss how these symmetries act on the CP${}^1$ variables and correspondingly on the monopole operator generating a $2\pi$ flux of the gauge field $a_\mu$. Finally, we will compare this to the action of symmetries on the different pairing channels.
    
    \subsection{Action of symmetries in the flat bands}

To compute the quantum numbers of the skyrmions under the different symmetries, it is instructive to start by reviewing the different symmetries of TBG. First, let us start by the microscopic symmteries acting on the microscopic annihilation operators $f_\bk$ which is a vector in valley ($\tau$), sublattice ($\sigma$), and layer ($\mu$). These can be divided into spatial and local symmetries. The former are given by
\beq
C_3 f_\bk C_3^{-1} = e^{i\frac{\pi}{3} \sigma_z \tau_z} f_{O_3 \bk}, \qquad C_2 f_\bk C_2^{-1} = \sigma_x \tau_x f_{-\bk}, \qquad M_y f_\bk M_y^{-1} = \sigma_x \mu_x f_{m_y \bk}
\eeq
where $m_y (k_x, k_y) = (k_x, -k_y)$ and $O_3 \bk$ corresponds to counter clockwise rotation by $2\pi/3$. The local symmetries are
\beq
\T f_\bk \T^{-1} = \tau_x f_{-\bk}, \qquad \P f_\bk \P^{-1} = \sigma_x \mu_y f^\dagger_{-\bk}, \qquad \S f_\bk \S^{-1} = \sigma_z f^\dagger_{\bk}, \qquad U_v f_\bk U_v^\dagger = e^{i \phi \tau_z} f_\bk
\eeq
We notice that the unitary particle hole symmetry $\P$ is only a symmetry of the continuum model for small angles and the sublattice $\S$ only holds in the so-called chiral limit \cite{Tarnopolsky}. Here, following Ref.~\cite{KIVCpaper}, we will assume all these symmetries are good symmetries which is equivalent to the sublattice polarized approximation discussed in the main text. This approximation relies on the fact that the wavefunctions in the flat bands still have significant sublattice polarization. The combination of $\S$, $\P$, and $\T$ symmetries implies the existence of a $\Z_2$ unitary symmetry $R = i\S \P \T = \sigma_y \tau_x \mu_y$ which anticommutes with $\tau_z$, the generator of $U_v$. The resulting problem has an enlarged $\SU(2)$ symmetry generated by $\tau_z$, $R = \sigma_y \tau_x \mu_y$, and $i \tau_z R = \sigma_y \tau_y \mu_y$. This generates the  $\SU(2)$ pseudospin rotation symmetry considered in the main text.

We then project the operator $f_\bk$ to the flat band operator $c_\bk$ which is now labeled by sublattice index $\sigma$ and valley index $\tau$. In general, the representation of different symmetries depends on the gauge choice for the flat band wavefunctions $u_{\sigma,\tau}(\bk)$. Once a gauge is fixed for one valley and sublattice, we can fix the gauge in the other valley and sublattice using two of the symmetries which relates different valleys and sublattices. Once this choice is fixed, we no longer have a gauge freedom and the representation of the remaining symmetries will in general have complicated $\bk$-dependence. Since we want a simple $\bk$-independent representation for the generators of $\SU(2)$ pseudospin rotation, we will use $\T$ and $\P$ symmetries to fix the gauge leading to the simple expressions $\T = \tau_x \K$ and $\P = \tau_z \sigma_y \K$. In this gauge, the generators of $\SU(2)$ pseudospin rotations are $\sigma_x \tau_x$, $\sigma_x \tau_y$ and $\tau_z$. The remaining symmetries $C_2$, $C_3$, and $M_y$ will generally have more complicated $\bk$-dependent forms in this gauge. To see this, we will find it easier to define $C_2' = i \tau_z R C_2 = \sigma_z \tau_z \mu_y$  which is diagonal in valley and sublattice and satisfies 
\beq
C_2'^2 = 1, \quad [\T,C_2'] = 0, \quad [R, C_2'] = 0
\label{C2pIdentities}
\eeq
Its action on the flat bands is generally given by
\beq
C_2' u_{\sigma,\tau}(\bk) = e^{i \theta_{\sigma,\tau}(\bk)} u_{\sigma,\tau}(-\bk)
\eeq
Using (\ref{C2pIdentities}) implies
\beq
C_{2\bk}' = e^{i \theta(\bk)} \sigma_0 \tau_0, \qquad \theta(-\bk) = -\theta(\bk)
\eeq
We note that we cannot set $\theta(\bk)$ to zero everywhere due to the band topology which requires the product of parity eigenvalues $e^{i\theta(\bk)}$ at TRIMs to be $-1$ \cite{Bernevig2012}.

Similarly, $C_3$ symmetry is also diagonal in valley and sublattice so it has the form
\beq
C_3 u_{\sigma,\tau}(\bk) = e^{i \phi_{\sigma,\tau}(\bk)} u_{\sigma,\tau}(O_3 \bk)
\eeq
Similar to the case of $C_2'$,  $\phi_{\sigma,\tau}(\bk)$ has to be $\bk$-dependent so that the product of $e^{i \phi_{\sigma,\tau}(\bk)}$ at $C_3$-symmetric points is compatible with the Chern number \cite{Bernevig2012}. Commutation with $C_2' \T$, $R$ and $C_2'$ implies 
\beq
C_{3\bk} = e^{i \sigma_z \tau_z \phi(\bk)}, \qquad \phi(-\bk) = \phi(\bk), 
\eeq
Finally, $M_y$ exchanges sublattice but not valley
\beq
M_y u_{\sigma,\tau}(\bk) = e^{i \alpha_{\sigma,\tau}(\bk)} u_{-\sigma,\tau}(m_y \bk)
\eeq
Commutation with $R$, $C_2 \T$, and $\T$ implies 
\beq
M_{y\bk} = \sigma_x e^{i \alpha(\bk) \sigma_z \tau_z}, \qquad \alpha(-\bk) = \alpha(\bk)
\eeq

These symmetries can also be represented in terms of the Chern-pseudospin basis defined in the main text (repeated in Eq.~\ref{GammaEta} of the supplemental mateiral)
\beq
{\boldsymbol \gamma} = (\sigma_x, \sigma_y \tau_z, \sigma_z \tau_z), \qquad {\boldsymbol \eta} = (\sigma_x \tau_x, \sigma_x \tau_y, \tau_z)
\eeq
We can then summarize the symmetries as follows
\begin{table}[h]
\begin{tabular}{c|c|c|c|c|c|c|c|c}
\hline \hline
basis &$\T$ & $\P$ & $\S$ & $\SU(2)$ generators $(\tau_z, R, i\tau_z R)$ & $C_2$ & $C_2' = i \tau_z R C_2$ & $C_3$ & $M_y$ \\
\hline
microscopic & $\tau_x \K$ & $\sigma_x \mu_y \K$ & $\sigma_z$ & $(\tau_z, \sigma_y \tau_x \mu_y, \sigma_y \tau_y \mu_y)$ & $\sigma_x \tau_x$ & $\sigma_z \tau_z \mu_y$ & $e^{i \frac{\pi}{3} \sigma_z \tau_z}$ & $\sigma_x \mu_x$\\
\hline
projected $(\sigma,\tau)$ & $\tau_x \K$ & $\tau_z \sigma_y \K$ & $\sigma_z$ & $(\tau_z, \tau_y \sigma_x, \tau_x \sigma_x)$ & $\sigma_x \tau_x e^{i \theta(\bk)}$ & $e^{i \theta(\bk)}$ & $e^{i \phi(\bk) \sigma_z \tau_z}$ & $\sigma_x e^{i \alpha(\bk) \sigma_z \tau_z}$ \\
\hline
projected $(\gamma,\eta)$ & $\gamma_x \eta_x \K$ & $\gamma_y \K$ & $\gamma_z \eta_z$ & $(\eta_z, \eta_y, \eta_x)$ & $\eta_x e^{i \theta(\bk)}$ & $e^{i \theta(\bk)}$ & $e^{i \phi(\bk) \gamma_z}$ & $\gamma_x e^{i \alpha(\bk) \gamma_z}$\\
\hline \hline
\end{tabular}
\end{table}

\subsection{Action of symmetries on the CP${}^1$ variables and the monopole operators}

Using the results of table above, we can deduce the action of the different symmetries on the pseudospin vectors $\bn_\pm$ as well as the CP${}^1$ variables $z$ and the corresponding gauge flux $f_{xy} = \frac{1}{\pi} \epsilon_{ij} \partial_i a_j$. The $\SU(2)$ pseudospin rotation acts on the $\bn_\pm$ as an ${\rm SO}(3)$ rotation and it acts naturally as an $\SU(2)$ rotation on $z$. For the remaining symmetries, we will find it convenient to multiply them sometimes with a pseudospin rotation such that they either leave the pseudospin invariant and flip its direction. The action of the remaining symmetries is given in the Table below
\begin{table}[h]
\begin{tabular}{c|c|c|c|c}
\hline \hline
field & $\T' = i \eta_z \T = \eta_y \gamma_x \K$ & $M_y$ & $C_2'$ & $C_3$ \\
\hline
$\bn_\pm(\br)$ & $-\bn_\mp(\br)$ & $\bn_\mp(m_y \br)$ & $\bn_\pm(-\br)$ & $\bn_\pm(O_3 \br) $\\
\hline
$z(\br)$ & $z^*(\br)$ & $z(m_y \br)$ & $z(-\br)$ & $z(O_3 \br)$ \\
\hline
$f_{xy}(\br)$ & $f_{xy}(\br)$ & $-f_{xy}(m_y \br)$ & $f_{xy}(-\br)$ & $f_{xy}(O_3 \br)$ \\
\hline \hline
\end{tabular}
\end{table}

Here, we have also included the Kramers time-reversal symmetry $\T'$ discussed in Ref.~\cite{KIVCpaper}. The cooper pair creation operator can be associated with the monopole operator $ \Delta$ which creates a $2\pi$ flux. Since this operator is a scalar, it transforms as a singlet under $\SU(2)$ spin rotation \cite{DQCPPRB, Shift, Monopole}. In addition, it transforms as a one-dimensional representation under the group of spatial symmetries $G$ generated by $\{C_2', C_3, M_y\}$. Notice here that we use $C_2'$ rather than $C_2$ since only the former commutes with the generators of the $\SU(2)$ pseudospin rotation enabling us to decompose the symmetry representation into a spatial part and pseudospin part.

Within the CP${}^1$ model described in terms of the variable $\bn = \bn_+ = -\bn_-$, we cannot narrow down the properties of the monopole operator further. However, it is possible to determine its angular momentum quantum numbers under $C_2'$ and $C_3$ if we go back to the original $\bn_\pm$ variable and assumed that the monopole operator creates a $2\pi$ flux in one Chern sector and a $-2\pi$ flux in the other sector, related to it by Kramers time-reversal symmetry. Denoting the flux creation operators in the $\pm$ Chern sectors by $\Delta_\pm$, we can write $\Delta_- = \T' \Delta_+ \T'^{-1}$. Denoting the angular momentum quantum numbers of $\Delta_+$ under $C_2'$ and $C_3$ by $L_2$ and $L_3$, respectively, we find that
\begin{gather}
C_2' \Delta_+ C_2'^{-1} = e^{i \pi L_2} \Delta_+ \, \implies \, C_2' \Delta_- C_2'^{-1} = C_2' \T' \Delta_+ \T'^{-1} C_2'^{-1} = \T' C_2' \Delta_+ C_2'^{-1} \T'^{-1} = \T' e^{i \pi L_2} \Delta_+ \T'^{-1} =  e^{-i \pi L_2} \Delta_- \nonumber \\
C_3 \Delta_+ C_3^{-1} = e^{i \frac{2\pi}{3} L_3} \Delta_+ \, \implies \, C_3 \Delta_- C_3^{-1} = C_3 \T' \Delta_+ \T'^{-1} C_3^{-1} = \T' C_3 \Delta_+ C_3^{-1} \T'^{-1} = \T' e^{i \frac{2\pi}{3} L_3} \Delta_+ \T'^{-1} =  e^{-i \frac{2\pi}{3} L_3} \Delta_-
\end{gather}
Here, we used $\T'$ commutes with $C_2'$ and $C_3$. As a result, the monopole operator $\Delta \sim \Delta_+ \Delta_-$ has zero angular momentum under both $C_2'$ and $C_3$. In summary, $\Delta$ is invariant under the action of $\SU(2)$ pseudospin rotations and the group of spatial symmetries $G$.

\subsection{Transformation properties of the superconducting pairing channels}
To relate the skyrmion superconductor to a weak pairing BCS superconductor, we investigate the properties of pairing functions in momentum space under the different symmetries in what follows. We restrict ourselves to pairing between time-reversal related momenta which has the form
\beq
\Delta_\bk = c_\bk c_{-\bk}^T
\eeq
which behaves under unitary transformations as
\beq
c_\bk \mapsto U_\bk c_{O \bk} \qquad \implies \qquad \Delta_\bk \mapsto U_\bk \Delta_{O \bk} U_{-\bk}^T
\eeq
To behave as a singlet under pseudospin $\SU(2)$ rotation $\Delta_\bk$ has to be proportional to $\eta_y$. This means that the candidate pairing channels are
\beq
\Delta_\mu(\bk) = g_\mu(\bk) \eta_y \gamma_\mu, \quad \mu=0,x,y,z \qquad g_{0,x,z}(-\bk) = g_{0,x,z}(\bk), \quad g_y(-\bk) = - g_y(\bk)
\eeq
where we used the fact that the full pairing function is antisymmetric by construction $\Delta(\bk)^T = -\Delta(-\bk)$.

To distinguish these pairing channels, we use the transformation properties under the group of spatial symmetries $G$. The symmetry transformations properties under $C_3$ and $M_y$ takes the form 
\beq
C_3 \Delta(\bk) C_3^{-1} = e^{i \phi(\bk) \gamma_z} \Delta(O_3 \bk) e^{i \phi(\bk) \gamma_z}, \qquad M_y \Delta(\bk) M_y^{-1} = \gamma_x e^{i \alpha(\bk) \gamma_z} \Delta(m_y \bk) e^{i \alpha(\bk) \gamma_z} \gamma_x
\eeq
From this symmetry action, it is immediately obvious that $\Delta_{x,y}$ transform as a one-dimensional representation of $G$. On the other hand, the transformation properties of $\Delta_{0,z}$ can be understood by defining instead the pairing channels
\beq
\Delta_\pm(\bk) = g_\pm(\bk) \eta_y \frac{1 \pm \gamma_z}{2}, \qquad g_\pm(-\bk) = g_\pm(\bk)
\eeq
The action of $C_3$ and $M_y$ on $\Delta_\pm$ is 
\beq
C_3 \Delta_\pm(\bk) C_3^{-1} = e^{\pm 2i \phi(\bk)} \Delta_\pm(O_3 \bk), \qquad M_y \Delta_\pm(\bk) M_y^{-1} = e^{\pm 2i \alpha(\bk)} \Delta_\mp(m_y \bk)
\eeq
Since the action of the symmetries is $\bk$ dependent, it is instructive to focus on high symmetry points. In this regard, we notice that the phase $\phi(\bk)$ has to be a multiple of $2\pi/3$ at all $C_3$ symmetric points: $\Gamma$, $K$, and $K'$ with $\phi(K) = \phi(K')$. Furthermore, due to the Chern number of the bands, the sum $\phi(K) + \phi(K') + \phi(\Gamma)$ has to be equal to $\pm 2\pi/3$ modulo $2\pi$ \cite{FGB}. This, in turn implies that either $\phi(\Gamma) \neq 0$ or $\phi(K) \neq 0$. In the first case, $\Delta_\pm(\Gamma)$ has nonzero and opposite angular momentum under $C_3$ and are interchanged by $M_y$. Thus, $\Delta_\pm(\Gamma)$ transform as a 2D irrep of $G$. In the second case, $\Delta_\pm(K)$ has opposite angular momentum under $C_3$ and are interchanged by the action of $M_x = C_2' M_y$ (which leaves the $K$ invariant). Thus, $\Delta_\pm(K)$ transforms as a 2D irrep under $G$. We conclude from this analysis that $\Delta_\pm(\bk)$ transforms as a 2D irrep under $G$ either at $\Gamma$ or at $K$ and as a result cannot correspond to the skyrmion superconductor which transforms as a 1D irrep under $G$.

This leaves the two pairing channels $\Delta_{x,y}$ discussed in the main text. These can be distinguished further by their behavior under $C_2'$ 
\beq
C_2' \Delta(\bk) C_2'^{-1} = e^{i \theta(\bk)} \Delta(-\bk) e^{i \theta(-\bk)} = \Delta(-\bk)
\eeq
Thus, $\Delta_y$ is odd under $C_2'$ whereas $\Delta_x$ is even under $C_2'$. Comparing with the transformation properties of the monopole operator, we deduce that the skyrmion superconductor can only be one of the channels $\Delta_x$. It should be emphasized, however, that the selection between the channels $\Delta_{x,y}$ relies on extra assumptions about the internal structure of the CP${}^1$ monopole operator in terms of the monopole operators in the $\bn_\pm$ variables.
\subsection{Anticommutation of insulating and superconducting gaps}
Here we discuss the relation between superconducting and insulating gaps, in particular we show  that the pairing channels $\Delta_{x,y}$ are also the ones that correspond to the maximal gap in the presence of the insulating pseudospin antiferromagnetic background. This is readily seen by checking that the corresponding matrices anticommute in the Nambu basis,  thus the insulating and superconducting gaps add in quadrature when they are simultaneously present.

To check the anticommutation, we perform a particle-hole transformation in the $C = -1$ sector only given by 
\beq
c_{+,\bk} \mapsto c_{+,\bk}, \qquad c_{-,\bk}^T \mapsto  c_{-,-\bk}^\dagger \eta_y
\eeq
Under this transformation, the gap functions for the pseudospin antiferromagnets $c_\bk^\dagger \gamma_z \eta_{x,y,z} c_\bk$ are invariant while the superconducting gap maps to the excitonic insulating gap 
\beq
\Delta_x(\bk) + \text{h.c.} \mapsto c_\bk^\dagger [\gamma_x \Re g_x(\bk)  + \gamma_y \Im g_x(\bk)] c_\bk, \quad \Delta_y(\bk) + \text{h.c.} \mapsto  c_\bk^\dagger [\gamma_y \Re g_y(\bk) - \gamma_x \Im g_y(\bk)] c_\bk
\eeq
which manifestly anticommutes with the insulating gap functions implying that the gaps will add in quadrature. Taking into account the kinetic part $h$, favors the pairing channel $\Delta_x$ which anticommutes with $h$ relative to $\Delta_y$ which commutes with $h$.

\section{Theory of the intertwined insulator and superconductor order parameters}
\label{Appendix:SO(5)}

In the following, our goal is to derive an effective field theory which deals with the superconducting and insulating order parameters on equal footing. Similar to Appendix \ref{App:NLSM}, we will perform a general number of flavors $n$ which reveals the general structure of the theory even though in this work we are focusing exclusively on the case $n=1$. As a result, we will employ a more general approach than the ones used in Ref.~\cite{RyuVishwanath} which uses the fact that the mass term has the form of a unit vector in $S^d$. Instead, our approach is similar to the derivations employed to derive effective theories in the contexts of disordered systems \cite{AltlandWeylPRL, AltlandWeylPRB, AltlandSimons}. In particular, we will follow the derivation in Ref.~\cite{AltlandWeylPRB} very closely.

 To derive an effective theory for the insulating and superconducting phases it is helpful to go close to the point $g=g_c$ where there is a perfect symmetry rotating the two into each other. We consider the weak coupling limit where the order parameter $M$ constitutes a small mass term to the Dirac equation. Although the relation of the parameters of the resulting theory to the microscopic parameters will not be valid for the strong coupling regime where $M$ is a lot larger than the bandwidth, we expect the form of the action, which is fixed by symmetries, to remain valid. To this end, we expand the non-interacting Hamiltonian in the vicinity of the Moir\'e Dirac points $K_M$ and $K'_M$ as
 \beq
 \H_D = v_F \sum_\bk [\psi^\dagger_{K_M,\bk} (k_x \gamma_x + k_y \gamma_y) \psi_{K_M, \bk} - \psi^\dagger_{K_M',\bk} (k_x \gamma_x + k_y \gamma_y) \psi_{K_M', \bk} ]
 \label{eq:HDF}
 \eeq
 The relative negative sign is fixed by time-reversal symmetry which has the form $\T = \gamma_x \eta_x \K$ and maps $K_M$ to $K_M'$. To deal with the insulator and superconductor on equal footing, we introduce the Nabmu basis defined as
 \beq
 \chi_\bk^T = (\psi^T_{K_M,\bk}, \psi^\dagger_{K'_M, -\bk})
 \eeq
 This form is inspired by the fact that the superconducting pairing takes place between states at opposite momenta and opposite Moir\'e valleys.
 
 We can now introduce the Pauli matrices $\rho_{x,y,z}$ which act in the Nambu space and rewrite the Dirac Hamiltonian (\ref{eq:HDF}) as
 \beq
 \H_{D} = v_F \sum_\bk \chi^\dagger_{\bk} (k_x \gamma_x \rho_z + k_y \gamma_y) \chi_{\bk} 
 \eeq
 By construction, the Nambu Hamiltonian $\H_{D}$ is invariant under the particle-hole transformation $\P = \gamma_z \rho_y \K$ which satisfies $\P^2 = -1$ (class C). The analysis can be further simplified by introducing the Pauli matrices
 \beq
{\boldsymbol \alpha} = (\gamma_x \rho_z, \gamma_y, \gamma_z \rho_z), \qquad {\boldsymbol \beta} = (\gamma_y \rho_y, \gamma_y \rho_x, \rho_z)
\label{eq:Msym}
\eeq
In this basis, the Dirac Hamiltonian has the form $\H_{D} \propto k_x \alpha_x + k_y \alpha_y$ and $\P = \alpha_x \beta_y \K$. The most general mass term for the Dirac Hamiltonian takes the form $\tilde M = \alpha_z M$ with 
 \beq
 \alpha_x \beta_y \tilde M^* \alpha_x \beta_y = - \tilde M \quad \implies \quad \beta_y M^* \beta_y = +M
 \eeq
 If we impose in addition the condition $M^2 = 1$, this means that $M$ parametrizes the symplectic Grassmanian manifold $\Sp(4n)/\Sp(2n) \times \Sp(2n)$ which for $n=1$ is isomorphic to $S^4$. In this case, we can write $M$ explicitly in terms of the vector $\bn$ represting the insulating phases and the complex number $\Delta$ representing the superconductor or alternatively an $\SO(5)$ vector $n$ as
 \beq
M = \left(\begin{array}{cc} \bn \cdot {\boldsymbol \eta} & \Delta \eta_y \\ \Delta^* \eta_y & \bn \cdot {\boldsymbol \eta}^T \end{array} \right)_\beta = \sum_{i=1}^5 n_i \Gamma_i, \qquad \hat \Gamma = (\eta_x, \eta_y \beta_z, \eta_z, \eta_y \beta_x, \eta_y \beta_y)
\label{eq:M1}
\eeq

The full massive Dirac action for for $\chi$ becomes
 \beq
S =  \int d\tau d^2\br \chi^\dagger[\partial_\tau - i \alpha_x \partial_x - i \alpha_y \partial_y - \mu \beta_z + \alpha_z m M] \chi
 \eeq 
 where we have included the chemical potential $\mu$. We now define $\bar \chi = \chi^\dagger \alpha_z$ to get
 \beq
 S = \int d\tau d^2\br \bar \psi[\tilde \alpha_\nu \partial_\nu - \mu \tilde \alpha_z \beta_z + m M] \tilde \psi, \qquad \tilde {\boldsymbol \alpha} = \alpha_z (\alpha_0, i\alpha_x, i\alpha_y) = (\alpha_z, -\alpha_y, \alpha_x)
 \eeq
 We can now integrate out the fermions to get an effective action of the matrix field $M$ leading to
 \beq
 S_{\rm eff} = - \Tr \ln [\tilde \alpha_\nu \partial_\nu - \mu \tilde \alpha_z \beta_z + m M]
 \eeq
 This expression is UV-divergent and needs to be regularized before any further manipulation. In the following, we follow Refs.~\cite{AltlandSimons, AltlandWeylPRL, AltlandWeylPRB} by subtracting off the following action
 \beq
 S_0 = - \lim_{\epsilon \rightarrow 0^+} \Tr \ln [\tilde \alpha_\nu \partial_\nu - \mu \tilde \alpha_z \beta_z + \epsilon M]
 \eeq
 In the limit $\epsilon \rightarrow 0^+$, the $M$-dependence of this action drops and it contributes an inessential constant. On the other hand, for large momenta, this term cancels the UV-divergence of $S_{\rm eff}$. We can now use the condition $M^2 = 1$ to write 
 \beq
 M(x,\tau) = U^\dagger(x,\tau) \Lambda U(x,\tau), \qquad \Lambda^2 = 1, \qquad \tr \Lambda = 0
 \eeq
 where $\Lambda$ is just a constant matrix which can then be substituted in the gradient expansion. Since the action is now UV-finite, we can use the cyclic property of the trace to get 
 \beq
 S_{\rm eff} = - \Tr \ln [\tilde \alpha_\nu \partial_\nu + \tilde \alpha_\nu A_\nu  - \mu \tilde \alpha_z R + m \Lambda] - S_0, \qquad A_\nu = U\partial_\nu U^\dagger = - \partial_\nu U U^\dagger, \qquad R = U\beta_z U^\dagger
 \eeq
 We can now expand the action in powers of $\mu$ and the gradients $A_\mu$
 \beq
  S_{\rm eff} = - \Tr \ln [G_0^{-1}(m) - \Sigma] - S_0 = -\Tr \ln G_0^{-1} - S_0 + \sum_{n=1}^\infty \frac{1}{n} \Tr (G_0(m) \Sigma)^n
 \eeq
 with $G_0$ and $\Sigma$ given by
 \beq
 G_0(k) = \frac{1}{i k_\nu \tilde \alpha_\nu + m \Lambda} = \frac{-i k_\nu \tilde \alpha_\nu + m \Lambda}{k^2 + m^2}, \qquad \Sigma = \tilde \alpha_\nu \tilde A_\nu, \qquad \tilde A_\nu = A_\nu  - \mu \, \delta_{\nu,0} R
 \eeq
 The quadratic term in the action can be written as
 \beq
 S_2 = \frac{1}{2} \sum_{\bk, \bq} \frac{1}{(\bk^2 + m^2)((\bk + \bq)^2 + m^2)} \left\{ m^2 \tr \ta_\mu \ta_\nu \tr \Lambda \tA_\mu(\bq) \Lambda \tA_\nu(-\bq) - k_\rho (k_\lambda + q_\lambda) \tr \ta_\mu \tilde \alpha_\rho \ta_\nu \ta_\lambda \tr \tA_\mu(\bq) \tA_\nu(-\bq) \right\}
 \eeq
 In the long wavelength limit, we can set $\bq = 0$. In this case, we can use the symmetry of the integral under $\bk \mapsto -\bk$ and the rotation symmetry to reduces $k_\rho k_\lambda = \delta_{\rho \lambda} \bk^2/3$. In addition, the traces over products of $\ta$ can be evaluated as 
 \beq
 \tr \ta_\mu \ta_\nu = 2 \delta_{\mu \nu}, \qquad  \sum_\rho \tr \ta_\mu \ta_\rho \ta_\nu \ta_\rho = -2 \delta_{\mu \nu}
 \eeq
 Evaluating the momentum integrals in $S_2$ leads to
 \beq
     S_2 = \frac{V}{2\pi^2} \int dk k^2 \frac{1}{(k^2 + m^2)^2} \left\{ m^2 \tr \Lambda \tilde A_\nu \Lambda \tilde A_\nu + \frac{1}{3} k^2 \tr \tA_\nu \tA_\nu \right\} = \frac{m V}{8\pi} \tr [(\Lambda \tA_\nu)^2 - \tA_\nu^2] = \frac{m V}{16\pi} \tr [\Lambda, \tA_\nu]^2
 \eeq
 Here, in evaluating the momentum integral, we have subtracted off a term from $S_0$ to cancel the UV divergence of the second term. The result can be written in terms of the mass matrix $M$ as
 \beq
 S_2 = \int d^2 \br d\tau \frac{\rho_\nu}{8} \tr [\Lambda, \tilde A_\nu]^2 = \int d^2 \br d\tau  \frac{\rho_\nu}{8} \tr (\D_\nu M)^2, \qquad \D_\nu M = \partial_\nu M - \mu \delta_{\nu,0} [\beta_z, M]
 \eeq
 where
 \beq
 \rho_i = \tilde \rho = \frac{m}{2 \pi}, \qquad \rho_\tau = \tilde \chi = \frac{m}{2 \pi v_F^2}
 \eeq
 
 For $n=1$, this can be written explicitly by substituting (\ref{eq:M1}) leading to
 \beq
 S_{2} = \int d^2 \br d\tau  \left\{\frac{\tilde \rho}{2} [(\nabla \bn)^2 +  |\nabla \Delta|^2] +\frac{\tilde \chi}{2} [(\partial_\tau \bn)^2 + |\partial_\tau \Delta|^2] + 2 \tilde \chi \mu \Delta^* \partial_\tau \Delta - 2 \tilde \chi \mu^2 |\Delta|^2 \right\}
 \eeq
 The last term implies that the chemical potential always disfavors the insulator. 
 
We notice that $A_\mu$ is of the same order as the momentum $q_\mu$. Thus, the third order term has two contributions. First there is the term $\sim A^3$
 \begin{align}
     S_{3,A^3} &= \frac{1}{3} \sum_\bk \frac{1}{(\bk^2 + m^2)^3} \left\{ m^3 \tr \ta_\mu \ta_\nu \ta_\lambda \tr \Lambda \tA_\mu \Lambda \tA_\nu \Lambda \tA_\lambda - 3k_\rho k_\delta \, m \, \tr \ta_\mu \ta_\rho \ta_\nu \ta_\delta \ta_\lambda \tr \Lambda \tA_\mu \tA_\nu \tA_\lambda \right\} \nonumber \\
     &= \frac{2i \epsilon_{\mu \nu \lambda} m}{3} \sum_\bk \frac{1}{(\bk^2 + m^2)^3} \left\{  m^2 \tr \Lambda \tA_\mu \Lambda \tA_\nu \Lambda \tA_\lambda + k^2  \tr \Lambda \tA_\mu \tA_\nu \tA_\lambda \right\}
 \end{align}
 Here, both momentum integrals are UV finite and there are no corrections coming from $S_0$ and we have used
 \beq
 \tr \ta_\mu \ta_\nu \ta_\lambda = 2 i \epsilon_{\mu \nu \lambda}
 \eeq
 Evaluating the momentum integrals yields
 \beq
 S_{3,A^3} = \frac{i}{48 \pi} \int d^2 \br d\tau \epsilon_{\mu \nu \lambda} \{\tr \Lambda \tA_\mu \Lambda \tA_\nu \Lambda \tA_\lambda + 3 \tr \Lambda \tA_\mu \tA_\nu \tA_\lambda \}
 \eeq
 Second, there is the term proportional to $q A^2$ which is obtained from the second order term in the gradient expansion of the action
 \begin{align}
 S_{3,q A^2} &= -\frac{i m}{2} \sum_{\bk,\bq} \frac{q_\lambda}{(k^2 + m^2)^2} \tr \ta_\mu \ta_\lambda \ta_\nu \tr \Lambda \tA_\mu(\bq) \tA_\nu(-\bq) = -\frac{V \epsilon_{\mu \nu \lambda}}{8\pi} \sum_\bq q_\lambda \tr \Lambda \tA_\mu(\bq) \tA_\nu(-\bq) \nonumber \\
 &= -\frac{i \epsilon_{\mu \nu \lambda}}{8\pi} \int d^2 \br d\tau \tr \Lambda \tA_\mu \partial_\lambda \tA_\nu
 \end{align}
 
 The part not containing $\mu$ is obtained by replacing $\tilde A_\mu$ by $A_\mu$. Using $\partial_\nu A_\mu = A_\mu A_\nu - (\partial_\mu \partial_\nu U) U^\dagger$, it can be written as
 \beq
 S_3 = \frac{i}{48 \pi} \int d^2 \br d\tau \epsilon_{\mu \nu \lambda} \{\tr \Lambda A_\mu \Lambda A_\nu \Lambda A_\lambda - 3 \tr \Lambda A_\mu A_\nu A_\lambda \}
 \eeq
 This part has the form of a Chern-Simons term in the non-Abelian gauge field $A_\mu$ and it is not manifestly gauge invariant. In fact, under a gauge transformation $U \mapsto K U$, with $[K, \Lambda] = 0$, it changes by a total derivative + a topological contribution associated with large gauge transformations which cancels against a corresponding contribution from the regulator $S_0$ \cite{RedlichPRL84, RedlichPRD84} (see Ref.~\cite{AltlandWeylPRB} for a detailed discussion). As a result, this term is only gauge invariant on a manifold without a boundary and it cannot be written explicitly in terms of $M$ without introducing an extra dimension. To write it in the more familiar form, we promote $A_\mu$ to depend on the extra variable $u$ and notice that
 \beq
 \epsilon_{\mu \nu \lambda \rho} \partial_\rho [\tr \Lambda A_\mu \Lambda A_\nu \Lambda A_\lambda - 3 \tr \Lambda A_\mu A_\nu A_\lambda] = -3\epsilon_{\mu \nu \lambda \rho} [\tr \Lambda A_\mu \Lambda A_\nu \Lambda A_\lambda A_\rho - \tr \Lambda A_\mu A_\nu A_\lambda A_\rho] = \frac{3}{8} \epsilon_{\mu \nu \lambda \rho} \tr M \partial_\mu M \partial_\nu M \partial_\lambda M \partial_\rho M
 \eeq
 which means that we can write
 \beq
S_{3,0} = \frac{i}{128\pi}   \int_0^1 du \int d^2 \br d\tau \epsilon_{\mu \nu \lambda \rho} \tr M \partial_\mu M \partial_\nu M \partial_\lambda M \partial_\rho M
 \eeq
 For the spinless model where $M = n_a \Gamma_a$ with $a=0,\dots,4$, this reduces to
 \begin{align}
 S_{3,0} &= \frac{i}{32\pi} \int_0^1 du \int d^2 \br d\tau \epsilon_{\mu \nu \lambda \rho} \epsilon_{abcde} n_a \partial_\mu n_b \partial_\nu n_c \partial_\lambda n_d \partial_\rho n_e \nonumber \\
 &= 2\pi i \frac{3}{8\pi^2} \int_0^1 du \int d^2 \br d\tau \epsilon_{abcde} n_a \partial_u n_b \partial_\tau n_c \partial_x n_d \partial_y n_e
 \end{align}
 which is the well-known WZW term. Finally, we can extract the linear term in $\mu$ in $S_3$ which is given by
 \beq
 S_{3,\mu} = \frac{i \mu}{16 \pi} \int d^2 \br d\tau \epsilon_{ij} [\tr \Lambda A_i \Lambda A_j \Lambda R - \tr \Lambda A_i A_j R - \tr \Lambda A_i R A_j - \tr \Lambda R A_i A_j] 
 \label{eq:S3mu}
 \eeq
 Now we notice that $\partial_\mu R = [R, A_\mu]$ which means that the term $\epsilon_{ij} \tr \Lambda A_i A_j R$ is a total derivative since
 \beq
 \epsilon_{ij} \partial_j \tr \Lambda A_i R = \epsilon_{ij} \tr \Lambda (A_i A_j R + A_i R A_j - A_i A_j R) = \epsilon_{ij} \tr \Lambda A_i R A_j
 \eeq
 As a result, we can rewrite (\ref{eq:S3mu})  as
\beq
 S_{3,\mu} = \frac{i \mu}{16 \pi} \int d^2 \br d\tau \epsilon_{ij} \tr R \Lambda [\Lambda, A_i] [\Lambda, A_j]   = \frac{i \mu}{16 \pi} \int d^2 \br d\tau \epsilon_{ij} \tr \beta_z M \partial_i M \partial_j M 
 \eeq
For $n=1$, we can substitute (\ref{eq:M1}) in the above expression yielding 
 \beq
 S_{3,\mu} = \frac{\mu}{4\pi} \int d^2 \br d\tau \epsilon_{ij} \bn \cdot (\partial_i \bn \times \partial_j \bn)
 \eeq
 Combining all the terms yields the effective action (18) in the main text.

\end{document}